%% file: main.tex
\documentclass[a4,center,fleqn, nocrop]{NAR}
\input{sections/packages}
\input{sections/commands}

\copyrightyear{2023}
\pubdate{20 01 2023}
\pubyear{2023}
\jvolume{5}
\jissue{1}
\shortjname{lqad}
\artid{004}

\newcites{supp}{Supplementary References}
\newcites{rev}{Revision References}

\begin{document}

\reversemarginpar

\title{\ltitle}

\author{%
Can Firtina\,$^{1,*}$,
Jisung Park\,$^{1,2}$,
Mohammed Alser\,$^{1}$,
Jeremie S. Kim\,$^{1}$,
Damla Senol Cali\,$^{3}$,\\
Taha Shahroodi\,$^{4}$,
Nika Mansouri Ghiasi\,$^{1}$,
Gagandeep Singh\,$^{1}$,
Konstantinos Kanellopoulos\,$^{1}$,
Can Alkan\,$^{5}$,
and Onur Mutlu\,$^{1,}$%
\footnote{To whom correspondence should be addressed. Email: firtinac@ethz.ch omutlu@ethz.ch}
}

\address{%
$^{1}$ETH Zurich
$^{2}$POSTECH
$^{3}$Carnegie Mellon University
$^{4}$TU Delft
and
$^{5}$Bilkent University}


\maketitle

\input{sections/0_abstract}
\input{sections/1_introduction}
\input{sections/2_methods}
\input{sections/3_evaluation}
\input{sections/4_discussion}
\input{sections/5_conclusion}

\section{Data Availability}
We make the real and simulated datasets we use available on the Zenodo website. For the human genome dataset, we download the data from the EBI website. We provide the accession numbers of all the public datasets we use in Table~\ref{tab:dataset}. We provide all the scripts 1) with the Zenodo links to download real and simulated datasets and 2) to fully reproduce our results and figures. These scripts are available at \href{https://github.com/CMU-SAFARI/BLEND/tree/master/test}{https://github.com/CMU-SAFARI/BLEND/tree/master/test}, which also includes instructions to use these scripts. The source code of \proposal is available at \release. For easy installation, we also make \proposal available in Docker (\texttt{firtinac/blend}) and bioconda (\texttt{blend-bio}).

\section{Acknowledgements}

\canf{We thank the SAFARI Research Group members for their valuable feedback and the stimulating intellectual and scholarly environment they provide. SAFARI Research Group acknowledges the generous gifts of our industrial partners, including Intel and VMware. We are grateful for the detailed comments that Kristoffer Sahlin provided, which improved our mechanism and the manuscript greatly.}

\section{Funding}

We acknowledge the generous gifts of our industrial partners, including Intel and VMware. This work is also partially supported by the European Union’s Horizon programme for research and innovation [101047160 - BioPIM].

\subsubsection{Conflict of interest statement.} None declared.
\newpage
\balance
\bibliographystyle{IEEEtran}

{\small \bibliography{main}}

\input{sections/supp}

\end{document}

%% file: sections/packages.tex
\usepackage{xspace}
\usepackage{tikz}
\usepackage{xcolor}
\usepackage{soul}
\usepackage{multicol}
\usepackage{multirow}
\usepackage{booktabs}
\usepackage{fancyhdr}
\usepackage{pdfpages}
\usepackage[nomessages]{fp}
\usepackage{clipboard}

\usepackage{amsmath,amssymb,amsfonts}
\usepackage{algorithmic}
\usepackage{graphicx}
\usepackage{textcomp}
\usepackage{balance}
\usepackage{mathptmx} 
\usepackage[normalem]{ulem}
\usepackage{url}
\usepackage{marginnote}
\usepackage{enumitem}
\usepackage{ifthen}
\usepackage{caption}
\usepackage{listings}
\usepackage{setspace}
\usepackage{float}
\usepackage[us,12hr]{datetime}
\usepackage[textwidth=1cm, colorinlistoftodos,prependcaption,textsize=small]{todonotes}

\usepackage{xargs}
\usepackage{titletoc}
\usepackage[all]{nowidow}
\usepackage[framemethod=tikz]{mdframed}
\usepackage[most]{tcolorbox}
\tcbuselibrary{breakable}
\tcbuselibrary{hooks}

\usepackage{mathtools}

\usepackage{courier}
\usepackage[italic]{mathastext}

\usepackage{siunitx}
\usepackage{algorithm}
\usepackage{arydshln} 

\usepackage{pifont}
\usepackage{footmisc}
\usepackage{hhline} 
\usepackage[compact]{titlesec} 

\usepackage[bookmarks=true,breaklinks=true,colorlinks,linkcolor=black,citecolor=blue,urlcolor=black]{hyperref}

%% file: sections/commands.tex
\definecolor{denim}{rgb}{0.08, 0.38, 0.74}
\definecolor{darkolivegreen}{rgb}{0.33, 0.42, 0.18}
\definecolor{dgreen}{rgb}{0.00, 0.75, 0.00}
\definecolor{darkpink}{rgb}{0.88, 0.28, 0.54}
\definecolor{forestgreen}{rgb}{0.0, 0.27, 0.13}
\definecolor{amber}{rgb}{1.0, 0.49, 0.0}
\definecolor{lightyellow}{rgb}{0.980, 0.956, 0.623}
\definecolor{lightblue}{rgb}{0.980, 0.956, 0.623}

\newcommand{\release}{\href{https://github.com/CMU-SAFARI/BLEND}{https://github.com/CMU-SAFARI/BLEND}\xspace}

\newcommand\rev[1]{{\color{black}{#1}}}
\newcommand\canf[1]{{\color{black}{#1}}}

\newcommand\proposal{BLEND\xspace}
\newcommand\ltitle{BLEND: A Fast, Memory-Efficient, and Accurate Mechanism to Find Fuzzy Seed Matches in Genome Analysis\xspace}

\newcommand\ovmaxseed{$27.3\times$\xspace}

\newcommand\avgovpM{$19.3\times$\xspace}
\newcommand\avgovpHM{$40.3\times$\xspace}
\newcommand\ovpM{$83.9\times$\xspace}
\newcommand\movpM{$2.4\times$\xspace}

\newcommand\avgovmM{$3.8\times$\xspace}
\newcommand\avgovmHM{$7.2\times$\xspace}
\newcommand\ovmM{$14.1\times$\xspace}
\newcommand\movmM{$0.9\times$\xspace}

\newcommand\avgovpMH{$808.2\times$\xspace}
\newcommand\avgovpHMH{$1580.0\times$\xspace}
\newcommand\ovpMH{$4367.8\times$\xspace}
\newcommand\movpMH{$28.4\times$\xspace}

\newcommand\avgovmMH{$127.8\times$\xspace}
\newcommand\avgovmHMH{$214.0\times$\xspace}
\newcommand\ovmMH{$234.7\times$\xspace}
\newcommand\movmMH{$36.0\times$\xspace}

\newcommand\avgrmpM{$1.7\times$\xspace}

\newcommand\rmpM{$4.1\times$\xspace}
\newcommand\mrmpM{$0.8\times$\xspace}

\newcommand\avgrmmM{$1.0\times$\xspace}

\newcommand\rmmM{$1.1\times$\xspace}
\newcommand\mrmmM{$0.5\times$\xspace}

\newcommand\avgrmpL{$6.8\times$\xspace}

\newcommand\rmpL{$18.6\times$\xspace}
\newcommand\mrmpL{$1.2\times$\xspace}

\newcommand\avgrmmL{$0.6\times$\xspace}

\newcommand\rmmL{$1.0\times$\xspace}
\newcommand\mrmmL{$0.3\times$\xspace}

\newcommand\avgrmpW{$4.3\times$\xspace}

\newcommand\rmpW{$9.9\times$\xspace}
\newcommand\mrmpW{$1.1\times$\xspace}

\newcommand\avgrmmW{$1.5\times$\xspace}

\newcommand\rmmW{$4.1\times$\xspace}
\newcommand\mrmmW{$0.9\times$\xspace}

\newcommand\avgrmpS{$13.3\times$\xspace}

\newcommand\rmpS{$29.8\times$\xspace}
\newcommand\mrmpS{$1.4\times$\xspace}

\newcommand\avgrmmS{$1.6\times$\xspace}

\newcommand\rmmS{$4.2\times$\xspace}
\newcommand\mrmmS{$0.2\times$\xspace}

\let\oldmarginnote\marginnote

\renewcommand{\marginnote}[2][rectangle,draw,fill=blue!40,rounded corners]{%
    \oldmarginnote{%
    \tikz \node at (0,0) [#1]{#2};}%
}

\titlespacing*{\section}{0pt}{0ex}{0ex}
\titlespacing*{\subsection}{0pt}{0ex}{0ex}
\titlespacing*{\subsubsection}{0pt}{0ex}{0ex}
\titlespacing\section{1pt}{5pt plus 0.5pt minus 4pt}{5pt plus 0.5pt minus 4pt}
\titlespacing\subsection{1pt}{5pt plus 0.5pt minus 3pt}{5pt plus 0.5pt minus 3pt}
\titlespacing\subsubsection{1pt}{5pt plus 0.5pt minus 2pt}{5pt plus 0.5pt minus 2pt}

\makeatletter
\g@addto@macro{\normalsize}{%
  \setlength{\abovedisplayskip}{2pt plus 1pt minus 1pt}
  \setlength{\belowdisplayskip}{2pt plus 1pt minus 1pt}
  \setlength{\abovedisplayshortskip}{0pt}
  \setlength{\belowdisplayshortskip}{0pt}
  \setlength{\intextsep}{2pt plus 1pt minus 1pt}
  \setlength{\textfloatsep}{3pt plus 1pt minus 1pt}
  \setlength{\dbltextfloatsep}{3pt plus 1pt minus 1pt}
  \setlength{\skip\footins}{4pt plus 1pt minus 1pt}}
\makeatother

\newcommand{\circlednumber}[1]{\raisebox{0.1pt}{
  \protect\tikz[baseline=(myanchor.base)]{
  \protect\node[circle,fill=.,inner sep=0.8pt] (myanchor) {\color{-.}\scriptsize \textbf{#1}};}%
}}

\expandafter\def\expandafter\UrlBreaks\expandafter{\UrlBreaks
  \do\a\do\b\do\c\do\d\do\e\do\f\do\g\do\h\do\i\do\j
  \do\k\do\l\do\m\do\n\do\o\do\p\do\q\do\r\do\s\do\t
  \do\u\do\v\do\w\do\x\do\y\do\z\do\A\do\B\do\C\do\D
  \do\E\do\F\do\G\do\H\do\I\do\J\do\K\do\L\do\M\do\N
  \do\O\do\P\do\Q\do\R\do\S\do\T\do\U\do\V\do\W\do\X
  \do\Y\do\Z}

\newcommand{\squishlist}{
 \begin{list}{$\circ$}
  { \setlength{\itemsep}{0pt}
     \setlength{\parsep}{0pt}
     \setlength{\topsep}{0pt}
     \setlength{\partopsep}{0pt}
     \setlength{\leftmargin}{1em}
     \setlength{\labelwidth}{1em}
     \setlength{\labelsep}{0.5em} } }

\newcommand{\squishsublist}{
\begin{list}{$\rightarrow$}
 { \setlength{\itemsep}{0pt}
    \setlength{\parsep}{0pt}
    \setlength{\topsep}{-10em}
    \setlength{\partopsep}{-3pt}
    \setlength{\leftmargin}{1em}
    \setlength{\labelwidth}{1em}
    \setlength{\labelsep}{0.5em} } }

\newenvironment{squishitem}
{\begin{itemize}[leftmargin=*]
  \setlength{\itemsep}{0em}
  \setlength{\parskip}{0pt}
  \setlength{\parsep}{0pt}
  \setlength{\topsep}{3pt}
  \setlength{\partopsep}{0pt}}
{\end{itemize}}

\newcommand{\squishend}{\end{list}}
  
\let\oldmarginnote\marginnote

\renewcommand{\marginnote}[2][rectangle,draw,fill=blue!40,rounded corners]{%
    \oldmarginnote{%
    \tikz \node at (0,0) [#1]{#2};}%
}

\mdfdefinestyle{graybox}{
    splittopskip=0,%
    splitbottomskip=0,%
    frametitleaboveskip=0,
    frametitlebelowskip=0,
    skipabove=0,%
    skipbelow=0,%
    leftmargin=0,%
    rightmargin=0,%
    innertopmargin=0mm,%
    innerbottommargin=0mm,%
    roundcorner=1mm,%
    backgroundcolor=lightblue,
    hidealllines=true}
    
\mdfdefinestyle{graybox2}{
    splittopskip=0,%
    splitbottomskip=0,%
    frametitleaboveskip=0,
    frametitlebelowskip=0,
    skipabove=0,%
    skipbelow=0,%
    leftmargin=0,%
    rightmargin=0,%
    innertopmargin=2mm,%
    innerbottommargin=2mm,%
    roundcorner=2mm,%
    backgroundcolor=lightblue,
    hidealllines=true}

\newcommandx{\changev}[2][1=]{\todo[ linecolor=blue,backgroundcolor=blue!25,bordercolor=blue,#1,size=\scriptsize]{#2}}

\let\oldmarginnote\marginnote

\renewcommand{\marginnote}[2][rectangle,draw,fill=blue!40,rounded corners]{%
        \oldmarginnote{%
        \tikz \node at (0,0) [#1]{#2};}%
        }
        
\newcommand{\boxbegin} {
	\begin{tcolorbox}[enhanced, frame hidden, colback=gray!50, breakable]
}

\newcommand{\boxend} {
	\end{tcolorbox}
}

\newcommand{\yboxbegin} {
	\begin{tcolorbox}[breakable, enhanced, frame hidden, colback=yellow!50]
}

\newcommand{\yboxend} {
	\end{tcolorbox}
}

\mdfdefinestyle{graybox}{
    splittopskip=0,%
    splitbottomskip=0,%
    frametitleaboveskip=0,
    frametitlebelowskip=0,
    skipabove=0,%
    skipbelow=0,%
    leftmargin=0,%
    rightmargin=0,%
    innertopmargin=0mm,%
    innerbottommargin=0mm,%
    roundcorner=1mm,%
    backgroundcolor=lightblue,
    hidealllines=true}
    
\mdfdefinestyle{graybox2} {
    splittopskip=0,%
    splitbottomskip=0,%
    frametitleaboveskip=0,
    frametitlebelowskip=0,
    skipabove=0,%
    skipbelow=0,%
    leftmargin=0,%
    rightmargin=0,%
    innertopmargin=2mm,%
    innerbottommargin=2mm,%
    roundcorner=2mm,%
    backgroundcolor=lightblue,
    hidealllines=true}

\newcommand{\bboxbegin}{
    \begin{mdframed}[style=graybox]
}

\newcommand{\bboxend}{
    \end{mdframed}
}

\newcommand{\yyboxbegin}{
    \begin{mdframed}[style=graybox2]
}

\newcommand{\yyboxend}{
    \end{mdframed}
}

\let\oldtableofcontents\tableofcontents
\renewcommand\tableofcontents{
  \oldtableofcontents
  \clearpage
}

\makeindex

%% file: sections/0_abstract.tex
\begin{abstract}
Generating the hash values of short subsequences, called seeds, enables quickly identifying similarities between genomic sequences by matching seeds with a single lookup of their hash values. However, these hash values can be used only for finding exact-matching seeds as the conventional hashing methods assign distinct hash values for different seeds, including highly similar seeds. Finding only exact-matching seeds causes either 1)~increasing the use of the costly sequence alignment or 2)~limited sensitivity.

We introduce \emph{\proposal}, the first efficient and accurate mechanism that can identify \emph{both} exact-matching and highly similar seeds with a single lookup of their hash values, called fuzzy seed matches. \proposal 1) utilizes a technique called SimHash, that can generate the same hash value for similar sets, and 2) provides the proper mechanisms for using seeds as sets with the SimHash technique to find fuzzy seed matches efficiently.

We show the benefits of \proposal when used in read overlapping and read mapping. For read overlapping, \proposal is faster by \movpM-\ovpM (on average \avgovpM), has a lower memory footprint by \movmM-\ovmM (on average \avgovmM), and finds higher quality overlaps leading to accurate \emph{de novo} assemblies than the state-of-the-art tool, minimap2. For read mapping, \proposal is faster by \mrmpM-\rmpM (on average \avgrmpM) than minimap2. Source code is available at \release.
\end{abstract}

%% file: sections/1_introduction.tex
\section{Introduction} \label{sec:introduction}
High-throughput sequencing (HTS) technologies have revolutionized the field of genomics due to their ability to produce millions of nucleotide sequences at a relatively low cost~\cite{shendure_dna_2017}.
Although HTS technologies are key enablers of almost \emph{all} genomics studies~\cite{aynaud_multiplexed_2021, logsdon_long-read_2020, mantere_long-read_2019, friedman_genome-wide_2019, merker_long-read_2018, alkan_genome_2011}, HTS technology-provided data comes with two key shortcomings.
First, HTS technologies sequence short fragments of genome sequences. These short fragments are called \emph{reads}, which cover only a smaller region of a genome and contain from about one hundred up to a million bases depending on the technology~\cite{shendure_dna_2017}.
Second, HTS technologies can misinterpret signals during sequencing and thus provide reads that contain \emph{sequencing errors}~\cite{Goodwin2016}.
The average frequency of sequencing errors in a read highly varies from 0.1\% up to 15\% depending on the HTS technology~\cite{stoler_sequencing_2021, zhang_comprehensive_2020, hon_highly_2020, ma_analysis_2019, senol_cali_nanopore_2019}.
To address the shortcomings of HTS technologies, various computational approaches must be taken to process the reads into meaningful information accurately and efficiently. 
These include 1)~read mapping~\cite{li_minimap_2016, li_minimap2_2018, canzar_short_2017, kim_airlift_2021, kim_fastremap_2022}, 2)~\emph{de novo} assembly~\cite{ekim_minimizer-space_2021, cheng_haplotype-resolved_2021, robertson_novo_2010}, 3)~read classification in metagenomic studies~\cite{meyer2021critical, lapierre2020metalign, wood_improved_2019}, 4)~correcting sequencing errors~\cite{firtina_apollo_2020, vaser_fast_2017, loman_complete_2015}.

\begin{figure}[tbh]
  \centering
  \includegraphics[width=0.95\linewidth]{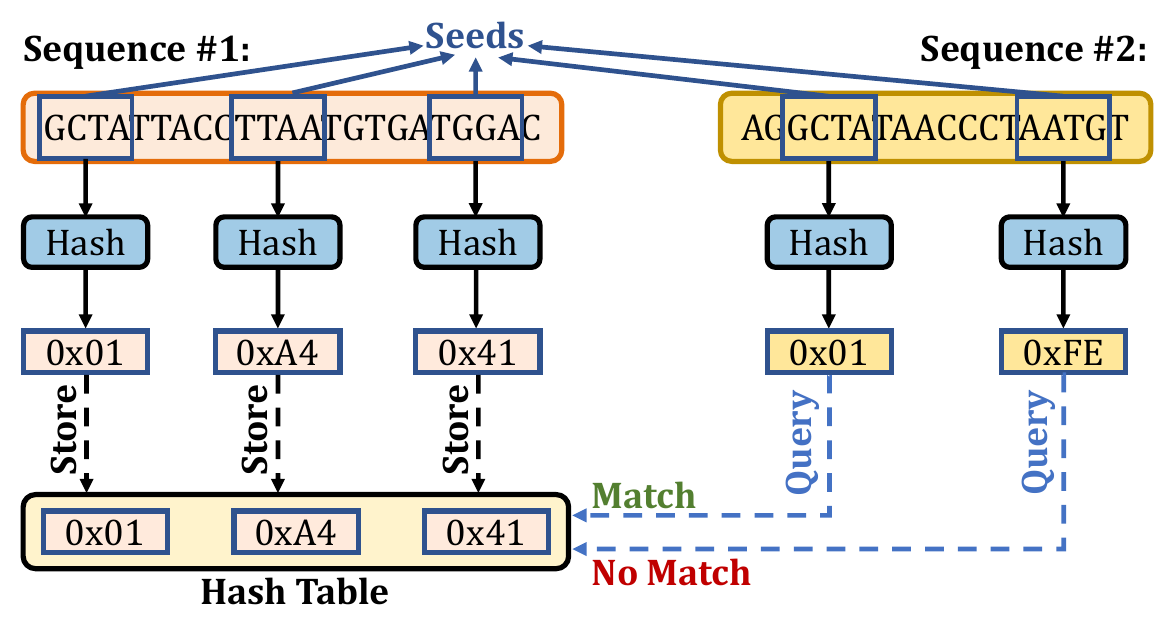}
  \caption{Finding seed matches with a single lookup of hash values.}
  \label{fig:hashing}
\end{figure}

At the core of these computational approaches, similarities between sequences must be identified to overcome the fundamental limitations of HTS technologies. However, identifying the similarities across \emph{all} pairs of sequences is not practical due to the costly algorithms used to calculate the distance between two sequences, such as sequence alignment algorithms using dynamic programming (DP) approaches~\cite{alser_technology_2021, alser_going_2022}. To practically identify similarities, it is essential to avoid calculating the distance between dissimilar sequence pairs.
A common heuristic is to find matching \emph{short} subsequences, called \emph{seeds}, between sequence pairs by using a hash table~\cite{altschul_basic_1990, altschul_gapped_1997, ning_ssaha_2001, kent_blatblast-like_2002, ma_patternhunter_2002, schwartz_humanmouse_2003, slater_automated_2005, wu_gmap_2005, ondov_efficient_2008, li_soap_2008, jiang_seqmap_2008, lin_zoom_2008, smith_using_2008, alkan_personalized_2009, homer_bfast_2009, schneeberger_simultaneous_2009, weese_razersfast_2009, rumble_shrimp_2009, li_soap2_2009, hach_mrsfast_2010, wu_fast_2010, rizk_gassst_2010, david_shrimp2_2011, egidi_better_2013, liu_rhat_2016, li_minimap_2016, li_minimap2_2018}. Sequences that have no or few seed matches are quickly filtered out from performing costly sequence alignment. There are several techniques that generate seeds from sequences, known as \emph{seeding techniques}.
To find the matching seeds efficiently, a common approach is to match the hash values of seeds with a \emph{single lookup} using a hash table that contains the hash values of all seeds of interest. Figure~\ref{fig:hashing} shows an overview of how hash tables are used to find seed matches between two sequences. Seeds in Figure~\ref{fig:hashing} are extracted from sequences based on a seeding technique. These seeds are used to find matches between sequences. To find seed matches, the hash values of seeds are used for filling and querying the hash table, as shown in Figure~\ref{fig:hashing}. Querying the hash table with hash values enables finding the positions where a seed from the second sequence appears in the first sequence with a single lookup.
The use of seeds drastically reduces the search space from all possible sequence pairs to similar sequence pairs to facilitate efficient distance calculations over many sequence pairs~\cite{baichoo_computational_2017, roberts_reducing_2004, schleimer2003winnowing}.

\enlargethispage{-30.1pt}

\begin{figure}[tbh]
  \centering
  \includegraphics[width=\linewidth]{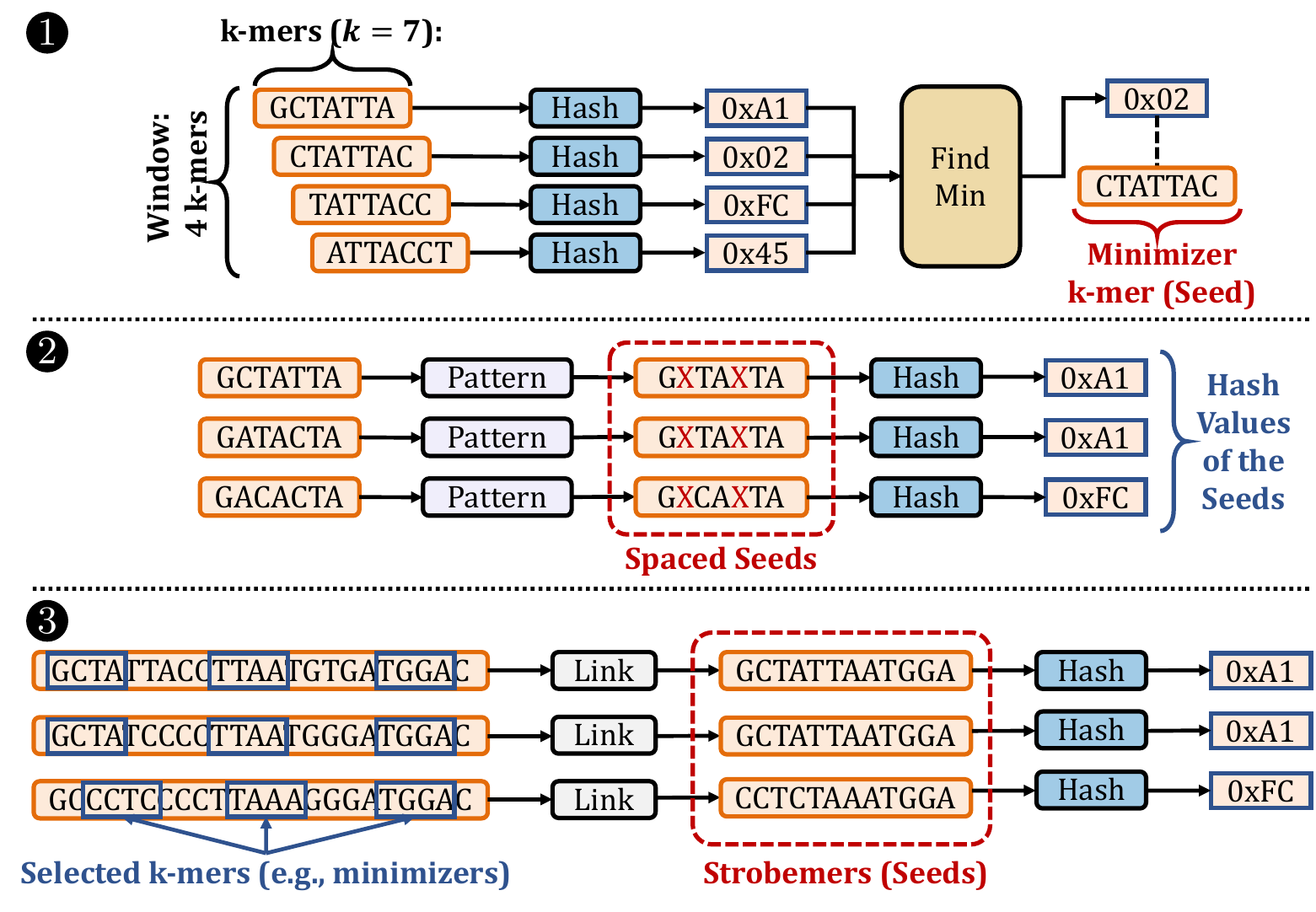}
  \caption{Examples of common seeding techniques.
  \circlednumber{$1$} Finding the minimizer k-mers.
  \circlednumber{$2$} A spaced seeding technique. Masked characters are highlighted by X in red.
  \circlednumber{$3$} A simple example of the strobemers technique. \vspace{-3pt}
  }
  \label{fig:seeding}
\end{figure}

Figure~\ref{fig:seeding} shows the three main directions that existing seeding techniques take. The first direction aims to minimize the computational overhead of using and storing seeds by selectively choosing fewer seeds from all fixed-length subsequences of reads, called \emph{k-mers}, where the fixed length is $k$.
The existing works such as minimap2~\cite{li_minimap2_2018}, MHAP~\cite{berlin_assembling_2015}, Winnowmap2~\cite{jain_long-read_2022, jain_weighted_2020}, re$M_{u}$val~\cite{deblasio_practical_2019}, and CAS~\cite{xin_context-aware_2020} use sampling techniques to choose a subset of k-mers from all k-mers of a read without significantly reducing their accuracy. For example, minimap2 uses only the k-mers with the \emph{minimum} hash value in a window of $w$ consecutive k-mers, known as the \emph{minimizer} k-mers~\cite{roberts_reducing_2004} (\circlednumber{$1$} in Figure~\ref{fig:seeding}). Such a sampling approach guarantees that one k-mer is sampled in each window to provide a fixed sampling ratio that can be tuned to increase the probability of matching k-mers between reads. Alternatively, MHAP uses the MinHash technique~\cite{broder_resemblance_1997} to generate many hash values from each k-mer of a read using many hash functions. For each hash function, only the k-mer with the minimum hash value is used as a seed with no windowing guarantees.
MHAP is mainly effective for matching sequences with similar lengths since the number of hash functions is fixed for all sequences, whereas it can generate too many seeds for shorter sequences when the sequence lengths vary greatly~\cite{li_minimap_2016}.
While these k-mer selection approaches reduce the number of seeds to use, 
all of these existing works find \emph{only} exact-matching k-mers with a single lookup, as they use hash functions with \emph{low-collision} rates to generate the hash values of these k-mers. The exact-matching requirement imposes challenges when determining the k-mer length. Longer k-mer lengths significantly decrease the probability of finding exact-matching k-mers between sequences due to genetic variations and sequencing errors. Short k-mer lengths (e.g., 8-21 bases) result in matching a large number of k-mers due to both the repetitive nature of most genomes and the high probability of finding the same short k-mer frequently in a long sequence of DNA letters~\cite{xin_accelerating_2013}. Although k-mers are commonly used as seeds, a seed is a more general concept that can allow substitutions, insertions and deletions (indels) when matching short subsequences between sequence pairs.

The second direction is to allow substitutions when matching k-mers by \emph{masking} (i.e., ignoring) certain characters of k-mers and using the masked k-mers as seeds\circlednumber{$2$}. Predefined \emph{patterns} determine the fixed masking positions for all k-mers. Seeds generated from masked k-mers are known as \emph{spaced seeds}~\cite{ma_patternhunter_2002}. The tools such as ZOOM!~\cite{lin_zoom_2008} and SHRiMP2~\cite{david_shrimp2_2011} use spaced seeds to improve the sensitivity when mapping short reads (i.e., Illumina paired-end reads). S-conLSH~\cite{chakraborty_conlsh_2020, chakraborty_s-conlsh_2021} generates many spaced seeds from each k-mer using different masking patterns to improve the sensitivity when matching spaced seeds with locality-sensitive hashing techniques.
There have been recent improvements in determining the masking patterns to improve the sensitivity of spaced seeds~\cite{petrucci_iterative_2020, mallik_ales_2021}.
Unfortunately, spaced seeds cannot find \emph{any arbitrary} fuzzy matches of k-mers with a single lookup due to 1)~fixed patterns that allow mismatches only at certain positions of k-mers and 2)~\emph{low-collision hashing} techniques that can be used for finding \emph{only} exact-matching spaced seeds, which are key limitations in improving the sensitivity of spaced seeds.

The third direction aims to allow both substitutions and indels when matching k-mers. A common approach is to link a few selected k-mers of a sequence to use these linked k-mers as seeds, such as paired-minimizers~\cite{chin_human_2019} and strobemers~\cite{sahlin_effective_2021, sahlin_flexible_2022}\circlednumber{$3$}. These approaches can ignore large gaps between the linked k-mers. For example, the strobemer technique concatenates a subset of selected k-mers of a sequence to generate a strobemer sequence, which is used as a seed. Strobealign~\cite{sahlin_flexible_2022} uses these strobemer seeds for mapping short reads with high accuracy and performance.
Strobemers enable masking some characters within sequences without requiring a fixed pattern, unlike spaced k-mers. This makes strobemers a more sensitive approach for detecting indels with varying lengths as well as substitutions. However, the nature of the hash function used in strobemers requires exact matches of \emph{all} concatenated k-mers in strobemer sequences when matching seeds. Such an exact match requirement introduces challenges for further improving the sensitivity of strobemers for detecting indels and substitutions between sequences.

To our knowledge, there is no work that can \emph{efficiently} find fuzzy matches of seeds \emph{without} requiring 1)~\emph{exact matches} of all k-mers (i.e., any k-mer can mismatch) and 2)~imposing high performance and memory space overheads. In this work, we observe that existing works have such a limitation mainly because they employ hash functions with low-collision rates when generating the hash values of seeds. Although it is important to reduce the collision rate for assigning different hash values for dissimilar seeds for accuracy and performance reasons, the choice of hash functions also makes it unlikely to assign the same hash value for similar seeds. Thus, seeds \emph{must} exactly match to find matches between sequences with a single lookup. Mitigating such a requirement so that similar seeds can have the same hash value has the potential to improve further the performance and sensitivity of the applications that use seeds with their ability to allow substitutions and indels at any arbitrary position when matching seeds.

A hashing technique, SimHash~\cite{charikar_similarity_2002, manku_detecting_2007}, provides useful properties for efficiently detecting highly similar seeds from their hash values. The SimHash technique can generate similar hash values for similar real-valued vectors or sets~\cite{charikar_similarity_2002}. Such a property enables estimating the cosine similarity between a pair of vectors~\cite{goemans_improved_1995} based on the Hamming distance of their hash values that SimHash generates (i.e., \emph{SimHash values})~\cite{charikar_similarity_2002, pratap_scaling_2020}. Although MinHash can provide better cosine similarity estimations than SimHash~\cite{shrivastava_defense_2014}, SimHash enables generating compact hash values that are practically useful for similarity estimations based on the Hamming distance. To efficiently find the pairs of SimHash values with a small Hamming distance, the number of matching most significant bits between different permutations of these SimHash values are computed~\cite{manku_detecting_2007}. This \emph{permutation-based} approach enables exploiting the Hamming distance similarity properties of the SimHash technique for various applications that find near-duplicate items~\cite{manku_detecting_2007, uddin_effectiveness_2011, sood_probabilistic_2011, feng_near-duplicate_2014, frobe_copycat_2021}.

In genomics, the properties of the SimHash and the permutation-based techniques are used for cell type classification~\cite{sun_reference-free_2021} and short sequence alignment~\cite{lederman_random-permutations-based_2013}. In read alignment, the permutation-based approach~\cite{manku_detecting_2007} is applied for detecting mismatches by permuting the sequences \emph{without} generating the hash values using the SimHash technique. This approach can find the longest prefix matches between a reference genome and a read since the mismatches between a pair of sequences \emph{may} move to the last positions of these sequences after applying different permutations while keeping the Hamming distance between sequences the same. This approach uses various versions of permutations to find the prefix matches. Apart from the permutation-based technique, a pigeonhole principle is also used for tolerating mismatches in read alignment~\cite{jiang_seqmap_2008, li_soap_2008, smith_using_2008, xin_shifted_2015, xin_context-aware_2020}. Unfortunately, none of these works can find highly similar seed matches that have the same hash value with a single lookup, which we call \emph{fuzzy seed matches}.

\textbf{Our goal} in this work is to enable finding \emph{fuzzy} matches of seeds as well as exact-matching seeds between sequences (e.g., reads) with a single lookup of hash values of these seeds.
To this end, we propose \emph{\proposal}, the \emph{first} efficient and accurate mechanism that can identify both exact-matching and highly similar seeds with a single lookup of their hash values.
The \textbf{key idea} in \proposal is to enable assigning the same hash value for highly similar seeds. To this end, \proposal~1)~exploits the SimHash technique~\cite{charikar_similarity_2002, manku_detecting_2007} and 2)~provides proper mechanisms for using any seeding technique with SimHash to find fuzzy seed matches with a single lookup of their hash values. This provides us with two key benefits.
First, \proposal can generate the same hash value for highly similar seeds \emph{without} imposing exact matches of seeds, unlike existing seeding mechanisms that use hash functions with low-collision rates.
Second, \proposal enables finding fuzzy seed matches with a single lookup of a hash value rather than 1)~using various permutations to find the longest prefix matches~\cite{lederman_random-permutations-based_2013} or 2)~matching many hash values for calculating costly similarity scores (e.g., Jaccard similarity~\cite{jaccard_nouvelles_1908}) that the conventional locality-sensitive hashing-based methods use, such as MHAP~\cite{berlin_assembling_2015} or S-conLSH~\cite{chakraborty_conlsh_2020, chakraborty_s-conlsh_2021}. These two ideas ensure that \proposal can efficiently find both 1)~all exact-matching seeds that a seeding technique finds using a conventional hash function with a low-collision rate and 2)~approximate seed matches that these conventional hashing mechanisms cannot find with a single lookup of a hash value.

\begin{figure}[tbh]
  \centering
  \includegraphics[width=\linewidth]{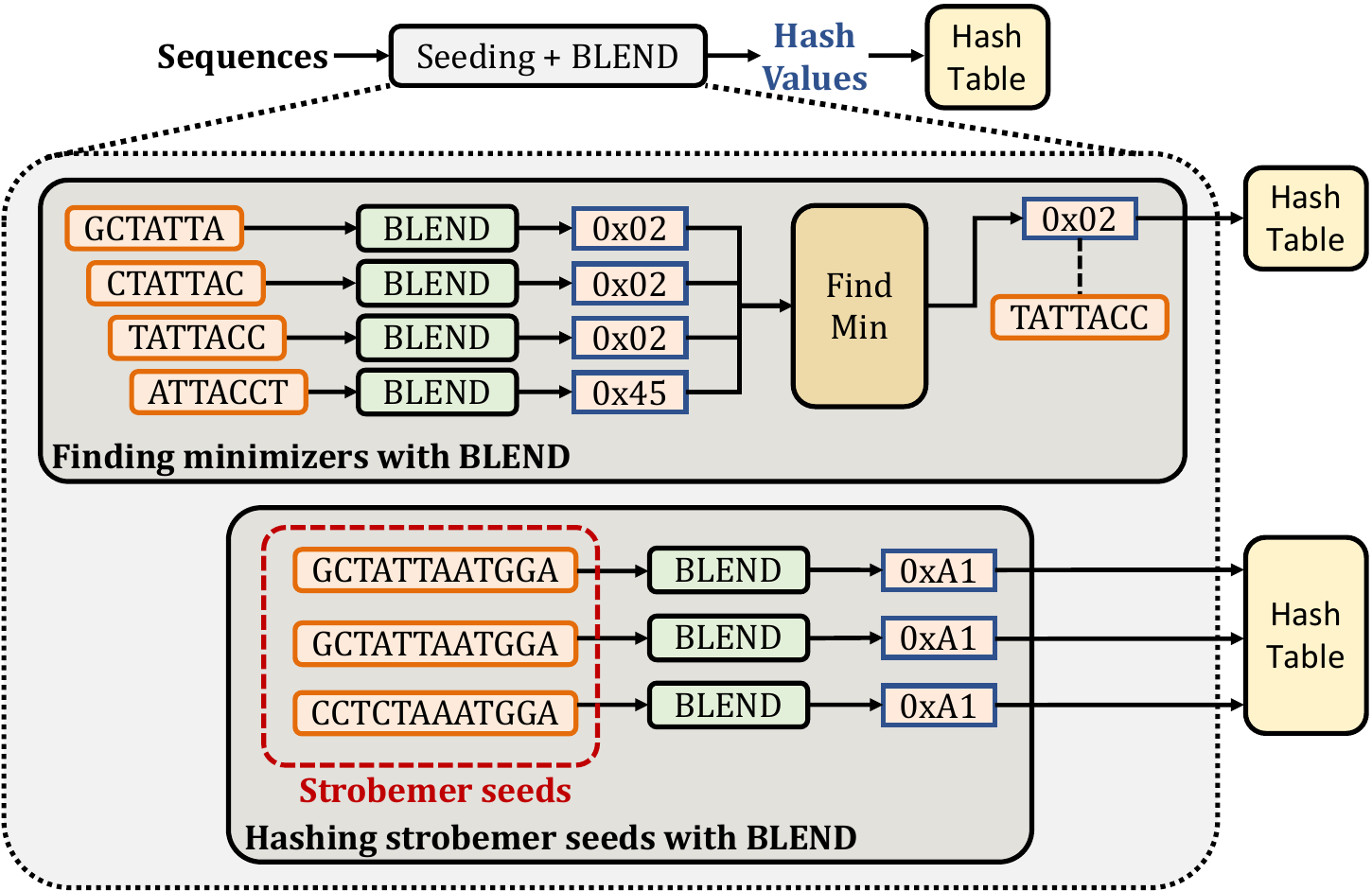}
  \caption{Replacing the hash functions in seeding techniques with \proposal.}
  \label{fig:seeding_and_blend}
\end{figure}

Figure~\ref{fig:seeding_and_blend} shows two examples of how \proposal can replace the conventional hash functions that the seeding techniques use in Figure~\ref{fig:seeding}. The \textbf{key challenge} is to accurately and efficiently define the items of sets from seeds that the SimHash technique requires. To achieve this, \proposal provides two mechanisms for converting seeds into sets of items: 1)~\texttt{\proposal-I} and 2)~\texttt{\proposal-S}. To perform a sensitive detection of substitutions, \texttt{\proposal-I} uses all overlapping smaller k-mers of a potential seed sequence as the items of a set for generating the hash value with SimHash. To allow mismatches between the linked k-mers that strobemers and similar seeding mechanisms use, \texttt{\proposal-S} uses only the linked k-mers as the set with SimHash. We envision that \proposal can be integrated with any seeding technique that uses hash values for matching seeds with a single lookup by replacing their hash function with \proposal and using the proper mechanism for converting seeds into a set of items.

Using erroneous (ONT and PacBio CLR), highly accurate (PacBio HiFi), and short (Illumina) reads, we experimentally show the benefits of \proposal on two important applications in genomics: 1)~read overlapping and 2)~read mapping. First, read overlapping aims to find overlaps between all pairs of reads based on seed matches. These overlapping reads are mainly useful for generating an assembly of the sequenced genome~\cite{pop_comparative_2004, li_minimap_2016}. We compare \proposal with minimap2 and MHAP by finding overlapping reads. We then generate the assemblies from the overlapping reads to compare the qualities of these assemblies. Second, read mapping uses seeds to find similar portions between a reference genome and a read before performing the read alignment. Aligning a read to a reference genome shows the edit operations (i.e., match, substitution, insertion, and deletions) to make the read identical to the portion of the reference genome, which is useful for downstream analysis (e.g., variant calling~\cite{mckenna_genome_2010}). We compare \proposal with minimap2, LRA~\cite{ren_lra_2021}, Winnowmap2, S-conLSH, and Strobealign by mapping long and paired-end short reads to their reference genomes. We evaluate the effect of the long read mapping results on downstream analysis by calling structural variants (SVs) and calculating the accuracy of SVs. This paper provides the following \textbf{key contributions} and  \textbf{major results}:

\begin{squishitem}
\item We introduce \proposal, the \emph{first} mechanism that can quickly and efficiently find \emph{fuzzy} seed matches between sequences with a single lookup.
\item We propose two mechanisms for converting seeds into a set of items that the SimHash technique requires: 1)~\texttt{\proposal-I} and 2)~\texttt{\proposal-S}. We show that \texttt{\proposal-S} provides better speedup and accuracy than \texttt{\proposal-I} when using PacBio HiFi reads for read overlapping and read mapping. When using ONT, PacBio CLR, and short reads, \texttt{\proposal-I} provides significantly better accuracy than \texttt{\proposal-S} with similar performance.
\item For read overlapping, we show that \proposal provides speedup compared to minimap2 and MHAP by \movpM-\ovpM (on average \avgovpM), \movpMH-\ovpMH (on average \avgovpMH) while reducing the memory overhead by \movmM-\ovmM (on average \avgovmM), \movmMH-\ovmMH (on average \avgovmMH), respectively.
\item We show that \proposal usually finds \emph{longer} overlaps between reads while using \emph{fewer} seed matches than other tools, which improves the performance and memory space efficiency for read overlapping.
\item We find that we can construct more accurate assemblies with similar contiguity by using the overlapping reads that \proposal finds compared to those that minimap2 finds.
\item For read mapping, we show that \proposal provides speedup compared to minimap2, LRA, Winnowmap2, and S-conLSH by \mrmpM-\rmpM (on average \avgrmpM), \mrmpL-\rmpL (on average \avgrmpL), \mrmpW-\rmpW (on average \avgrmpW), \mrmpS-\rmpS (on average \avgrmpS) while maintaining a similar memory overhead by \mrmmM-\rmmM (on average \avgrmmM), \mrmmL-\rmmL (on average \avgrmmL), \mrmmW-\rmmW (on average \avgrmmW), \mrmmS-\rmmS (on average \avgrmmS), respectively.
\item We show that \proposal provides a read mapping accuracy similar to minimap2, and Winnowmap2 usually provides the best read mapping accuracy.
\item We show that \proposal enables calling structural variants with the highest F1 score compared to minimap2, LRA, and Winnowmap2.
\item We open source our \proposal implementation as integrated into minimap2.
\item We provide the open-source SIMD implementation of the SimHash technique that \proposal employs.
\end{squishitem}

%% file: sections/2_methods.tex
\section{Material \& Methods} \label{sec:methods}
We propose \textbf{\emph{\proposal}}, a mechanism that can efficiently find fuzzy (i.e., approximate) seed matches with a single lookup of their hash values.
To find fuzzy seed matches, \proposal introduces a new mechanism that enables generating the same hash values for highly similar seeds. By combining this mechanism with any seeding approach (e.g., minimizer k-mers or strobemers), \proposal can find fuzzy seed matches between sequences with a single lookup of hash values.

\begin{figure}[tbh]
  \centering
  \includegraphics[width=\linewidth]{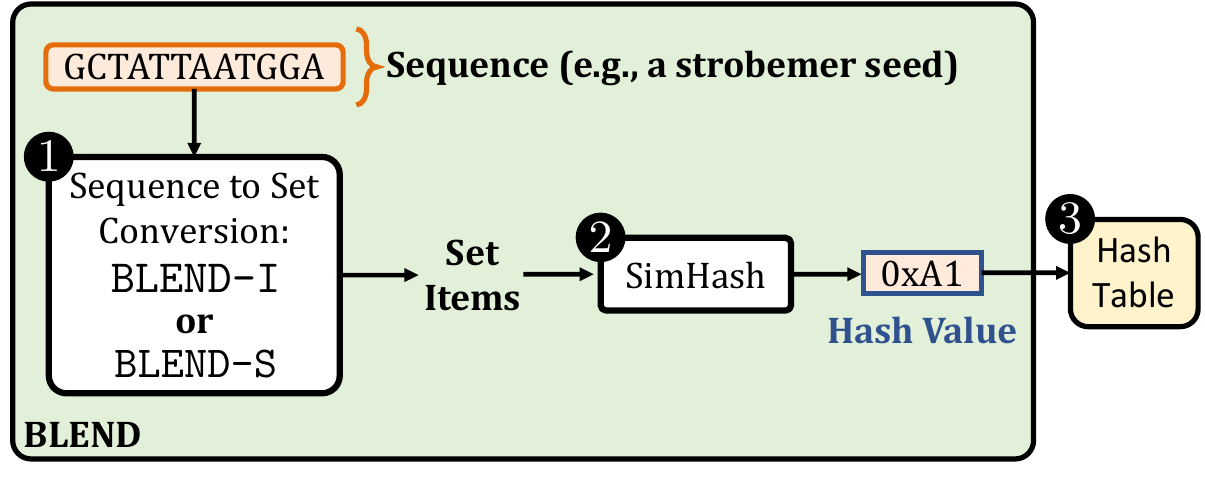}
  \caption{Overview of \proposal.\circlednumber{$1$}~\proposal uses \texttt{\proposal-I} or \texttt{\proposal-S} for converting a sequence into its set of items.\circlednumber{$2$}~\proposal generates the hash value of the input sequence using its set of items with the SimHash technique.\circlednumber{$3$}~\proposal uses hash tables for finding fuzzy seed matches with a single lookup of the hash values that \proposal generates.}
  \label{fig:blend-overview}
\end{figure}

Figure~\ref{fig:blend-overview} shows the overview of steps to find fuzzy seed matches with a single lookup in three steps.
First, \proposal starts with converting the input sequence it receives from a seeding technique (e.g., a strobemer sequence in Figure~\ref{fig:seeding_and_blend}) to its set representation as the SimHash technique generates the hash value of the set using its items\circlednumber{$1$}. To enable effective and efficient integration of seeds with the SimHash technique, \proposal proposes two mechanisms for identifying the items of the set of the input sequence: 1)~\texttt{\proposal-I} and 2)~\texttt{\proposal-S}.
Second, after identifying the items of the set, \proposal uses this set with the SimHash technique to generate the hash value for the input sequence\circlednumber{$2$}. \proposal uses the SimHash technique as it allows for generating the same hash value for highly similar sets. 
Third, \proposal uses the hash tables with the hash values it generates to enable finding fuzzy seed matches with a single lookup of their hash values\circlednumber{$3$}.

\subsection{Sequence to Set Conversion}\label{subsec:seed}
\begin{figure}[tbh]
  \centering
  \includegraphics[width=\linewidth]{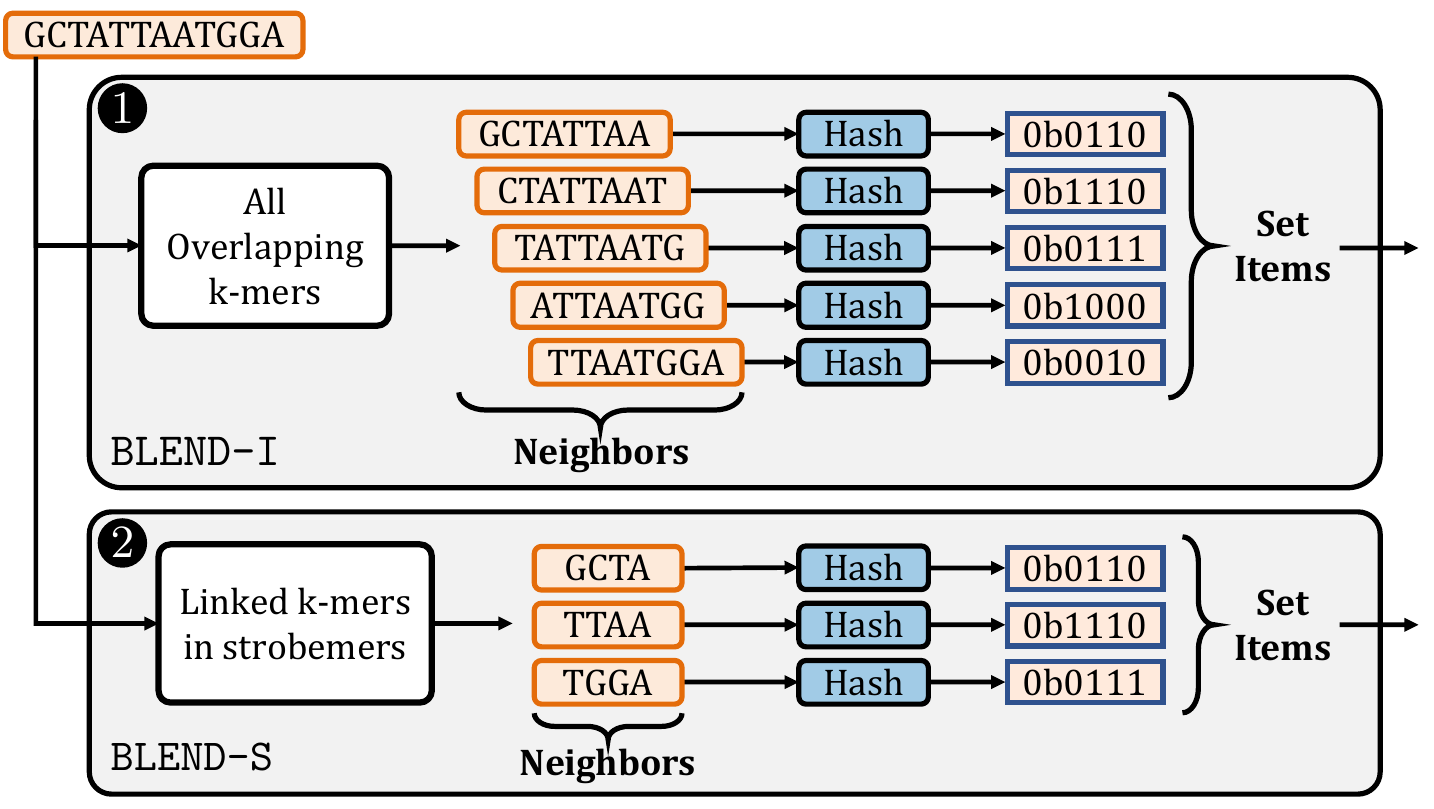}
  \caption{Overview of two mechanisms used for determining the set items of input sequences.\circlednumber{$1$}~\texttt{\proposal-I} uses the hash values of all the overlapping k-mers of an input sequence as the set items.\circlednumber{$2$}~\texttt{\proposal-S} uses the hash values of only the k-mers selected by the strobemer seeding mechanism.}
  \label{fig:blend-is}
\end{figure}

Our goal is to convert the input sequences that \proposal receives from any seeding technique (Figure~\ref{fig:seeding_and_blend}) to their proper set representations so that \proposal can use the items of sets for generating the hash values of input sequences with the SimHash technique.
To achieve effective and efficient conversion of sequences into their set representations in different scenarios, \proposal provides two mechanisms: 1)~\texttt{\proposal-I} and 2)~\texttt{\proposal-S}, as we show in Figure~\ref{fig:blend-is}.

The goal of the first mechanism, \texttt{\proposal-I}, is to provide high sensitivity for a single character change in the input sequences that seeding mechanisms provide when generating their hash values such that two sequences are likely to have the same hash value if they differ by a few characters. \texttt{\proposal-I} has three steps. First, \texttt{\proposal-I} extracts \emph{all} the overlapping k-mers of an input sequence, as shown in \circlednumber{$1$} of Figure~\ref{fig:blend-is}. For simplicity, we use the \emph{neighbors} term to refer to all the k-mers that \texttt{\proposal-I} extracts from an input sequence (Figure~\ref{fig:blend-is}). Second, \texttt{\proposal-I} generates the hash values of these k-mers using any hash function. Third, \texttt{\proposal-I} uses the hash values of the k-mers as the set items of the input sequence for SimHash. Although \texttt{\proposal-I} can be integrated with any seeding mechanism, we integrate it with the minimizer seeding mechanism, as shown in Figure~\ref{fig:seeding_and_blend} as proof of work.

The goal of the second mechanism, \texttt{\proposal-S}, is to allow indels and substitutions when matching the sequences such that two sequences are likely to have the same hash value if these sequences differ by a few k-mers. \texttt{\proposal-S} has three steps. First, \texttt{\proposal-S} uses \emph{only} the selected k-mers that the strobemer-like seeding mechanisms find and link~\cite{sahlin_effective_2021} as neighbors, as shown in \circlednumber{$2$} of Figure~\ref{fig:blend-is}. \texttt{\proposal-S} can enable a few of these linked k-mers to mismatch between strobemer sequences because a single character difference does not propagate to the other linked k-mers as opposed to the effect of a single character difference propagating to several overlapping k-mers in \texttt{\proposal-I}. To ensure the correctness of strobemer seeds when matching them based on their hash values, \texttt{\proposal-S} uses \emph{only} the selected k-mers from the same strand. Second, \texttt{\proposal-S} generates the hash values of these linked k-mers using any hash function. Third, \texttt{\proposal-S} uses the hash values of all such selected k-mers as the set items of the input sequence for SimHash.

\subsection{Integrating the SimHash Technique}\label{subsec:seedhash}

Our goal is to enable efficient comparisons of equivalence or high similarity between seeds with a single lookup by generating the same hash value for highly similar or equivalent seeds.
To enable generating the same hash value for these seeds, \proposal uses the SimHash technique~\cite{charikar_similarity_2002}. The SimHash technique takes a set of items and generates a hash value for the set using its items. The key benefit of the SimHash technique is that it allows generating the same hash value for highly similar sets while enabling any \emph{arbitrary} items to mismatch between sets. To exploit the key benefit of the SimHash technique, \proposal efficiently and effectively integrates the SimHash technique with the set items that \texttt{\proposal-I} or \texttt{\proposal-S} determine. \proposal uses these set items for generating the hash values of seeds such that highly similar seeds can have the same hash value to enable finding fuzzy seed matches with a single lookup of their hash values.

\begin{figure}[tbh]
  \centering
  \includegraphics[width=\linewidth]{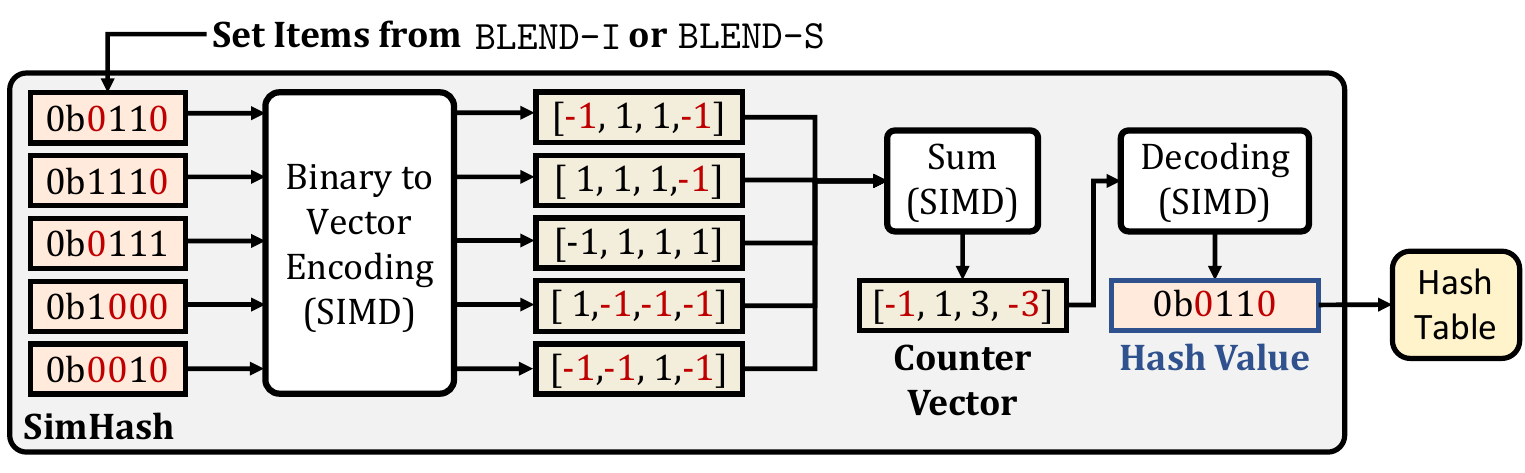}
  \caption{The overview of the steps in the SimHash technique for calculating the hash value of a given set of items. The set items are the hash values represented in their binary form. Binary to Vector Encoding converts these set items to their corresponding vector representations. Sum performs the vector additions and stores the result in a separate vector that we call the \emph{counter vector}. Decoding generates the hash value of the set based on the values in the counter vector. \proposal uses SIMD operations for these three steps, as indicated by SIMD. We highlight in red how $0$ bits are converted and propagated in the SimHash technique.}
  \label{fig:simhash}
\end{figure}

\proposal employs the SimHash technique in three steps: 1)~encoding the set items as vectors, 2)~performing vector additions, and 3)~decoding the vector to generate the hash value for the set that \texttt{\proposal-I} or \texttt{\proposal-S} determine, as we show in Figure~\ref{fig:simhash}. To enable efficient computations between vectors, \proposal uses SIMD operations when performing all these three steps. We provide the details of our SIMD implementation in Supplementary Section~\ref{suppsec:simd_implementation} and Supplementary Figures~\ref{suppfig:simd1} and~\ref{suppfig:simd2}.

First, the goal of the \emph{binary to vector encoding} step is to transform all the hash values of set items from the binary form into their corresponding vector representations so that \proposal can efficiently perform the bitwise arithmetic operations that the SimHash technique uses in the vector space.
For each hash value in the set item, the encoding can be done in two steps. The first step creates a vector of $n$ elements for an $n$-bit hash value. We assume that all the elements in the vector are initially set to $1$. For each bit position $t$ of the hash value, the second step assigns $-1$ to the $t^{th}$ element in the vector if the bit at position $t$ is $0$, as we highlight in Figure~\ref{fig:simhash} with red colors of $0$ bits and their corresponding $-1$ values in the vector space. For each hash value in set items, the resulting vector includes $1$ for the positions where the corresponding bit of a hash value is $1$ and $-1$ for the positions where the bit is $0$.

Second, the goal of the vector addition operation is to determine the bit positions where the number of $1$ bits is greater than the number of $0$ bits among the set items, which we call determining the \emph{majority} bits. The key insight in determining these majority bits is that highly similar sets are likely to result in \emph{similar} majority results because a few differences between two similar sets are unlikely to change the majority bits at each position, given that there is a sufficiently large number of items involved in this majority calculation. To efficiently determine the majority of bits at each position, \proposal counts the number of $1$ and $0$ bits at a position by using the vectors it generates in the vector encoding step, as shown with the addition step (Sum) in Figure~\ref{fig:simhash}. The vector addition performs simple additions of $+1$ or $-1$ values between the vector elements and stores the result in a separate \emph{counter} vector. The values in this counter vector show the majority of bits at each position of the set items. Since \proposal assigns $-1$ for $0$ bits and $1$ for $1$ bits, the majority of bits at a position is either 1)~$1$ if the corresponding value in the counter vector is greater than $0$ or 2)~$0$ if the values are less than or equal to $0$.

Third, to generate the hash value of a set, \proposal uses the majority of bits that it determines by calculating the counter vector. To this end, \proposal decodes the counter vector into a hash value in its binary form, as shown in Figure~\ref{fig:simhash} with the decoding step. The decoding operation is a simple conditional operation where each bit of the final hash value is determined based on its corresponding value at the same position in the counter vector. \proposal assigns the bit either 1) $1$ if the value at the corresponding position of the counter vector is greater than $0$ or 2) $0$ if otherwise. Thus, each bit of the final hash value of the set shows the majority voting result of set items of a seed. We use this final hash value for the input sequence that the seeding techniques provide because highly similar sequences are likely to have many characters or k-mers in common, which essentially leads to generating \emph{similar} set items by using \texttt{\proposal-I} or \texttt{\proposal-S}. Properly identifying the set items of similar sequences enables \proposal to find similar majority voting results with the SimHash technique, which can lead to generating the same final hash value for similar sequences. This enables \proposal to find fuzzy seed matches with a single lookup using these hash values. We provide a step-by-step example of generating the hash values for two different seeds in Supplementary Section~\ref{suppsec:real} and Supplementary Tables~\ref{supptab:kmers7-15-1-32}-~\ref{supptab:kmers15-7-2-16}.

\subsection{Using the Hash Tables}\label{subsec:storeseed}

Our goal is to enable an efficient lookup of the hash values of seeds to find fuzzy seed matches with a single lookup. To this end, \proposal uses hash tables in two steps. First, \proposal stores the hash values of all the seeds of target sequences (e.g., a reference genome) in a hash table, usually known as the \emph{indexing} step. Keys of the hash table are hash values of seeds, and the value that a key returns is a \emph{list} of metadata information (i.e., seed length, position in the target sequence, and the unique name of the target sequence). \proposal keeps minimal metadata information for each seed sufficient to locate seeds in target sequences. Since similar or equivalent seeds can share the same hash value, \proposal stores these seeds using the same hash value in the hash table. Thus, a query to the hash table returns all fuzzy seed matches with the same hash value.

Second, \proposal iterates over all query sequences (e.g., reads) and uses the hash table from the indexing step to find fuzzy seed matches between query and target sequences. The query to the hash table returns the list of seeds of the target sequences that have the same hash value as the seed of a query sequence. Thus, the list of seeds that the hash table returns is the list of fuzzy seed matches for a seed of a query sequence as they share the same hash value. \proposal can find fuzzy seed matches with a single lookup using the hash values it generates for the seeds from both query and target sequences.

\proposal finds fuzzy seed matches mainly for two important genomics applications: read overlapping and read mapping. For these applications, \proposal stores all the list of fuzzy seed matches between query and target sequences to perform \emph{chaining} among fuzzy seed matches that fall in the same target sequence (overlapping reads) optionally, followed by alignment (read mapping) as described in minimap2~\cite{li_minimap2_2018}.

%% file: sections/3_evaluation.tex
\section{Results} \label{sec:evaluation}
\subsection{Evaluation Methodology}\label{subsec:evaluation-methodology}
We replace the mechanism in minimap2 that generates hash values for seeds with \proposal to find fuzzy seed matches when performing end-to-end read overlapping and read mapping. We also incorporate the \texttt{\proposal-I} and \texttt{\proposal-S} mechanisms in the implementation and provide the user to choose either of these mechanisms when using \proposal. We provide a set of default parameters we optimize based on sequencing technology and the application to perform (e.g., read overlapping). We explain the details of the \proposal parameters in Supplementary Table~\ref{supptab:pardef} and the parameter configurations we use for each tool and dataset in Supplementary Tables~\ref{supptab:ovpars} and~\ref{supptab:mappars}. We determine these default parameters empirically by testing the performance and accuracy of \proposal with different values for some parameters (i.e., k-mer length, number of k-mers to include in a seed, and the window length) as shown in Supplementary Table~\ref{supptab:parameter_exploration}. We show the trade-offs between the seeding mechanisms \texttt{\proposal-I} and \texttt{\proposal-S} in Supplementary Figures~\ref{suppfig:overlap_perf-blend} and ~\ref{suppfig:read_mapping_perf-blend} and Supplementary Tables~\ref{supptab:overlap_assembly-blend} -~\ref{supptab:mapping_accuracy-blend} regarding their performance and accuracy.

For our evaluation, we use real and simulated read datasets as well as their corresponding reference genomes. We list the details of these datasets in Table~\ref{tab:dataset}. To evaluate \proposal in several common scenarios in read overlapping and read mapping, we classify our datasets into three categories: 1)~highly accurate long reads (i.e., PacBio HiFi), 2)~erroneous long reads (i.e., PacBio CLR and Oxford Nanopore Technologies), and 3)~short reads (i.e., Illumina). We use PBSIM2~\cite{ono_pbsim2_2021} to simulate the erroneous PacBio and Oxford Nanopore Technologies (ONT) reads from the Yeast genome.
To use realistic depth of coverage, we use SeqKit~\cite{shen_seqkit_2016} to down-sample the original \emph{E. coli}, and \emph{D. ananassae} reads to $100\times$ and $50\times$ sequencing depth of coverage, respectively.

\begin{table}[tbh]
\centering
\input{tables/dataset}
\label{tab:dataset}
\end{table}

We evaluate \proposal based on two use cases: 1)~read overlapping and 2)~read mapping to a reference genome. For read overlapping, we perform \emph{all-vs-all overlapping} to find all pairs of overlapping reads within the same dataset (i.e., the target and query sequences are the same set of sequences). To calculate the overlap statistics, we report the overall number of overlaps, the average length of overlaps, and the number of seed matches per overlap. To evaluate the quality of overlapping reads based on the accuracy of the assemblies we generate from overlaps, we use miniasm~\cite{li_minimap_2016}. We use miniasm because it does not perform error correction when generating \emph{de novo} assemblies, which allows us to directly assess the quality of overlaps without using additional approaches that externally improve the accuracy of assemblies. We use \texttt{mhap2paf.pl} package as provided by miniasm to convert the output of MHAP to the format miniasm requires (i.e., PAF). We use QUAST~\cite{gurevich_quast_2013} to measure statistics related to the contiguity, length, and accuracy of \emph{de novo} assemblies, such as the overall assembly length, largest contig, NG50, and NGA50 statistics (i.e., statistics related to the length of the shortest contig at the half of the overall reference genome length), k-mer completeness (i.e., amount of shared k-mers between the reference genome and an assembly), number of mismatches per 100Kb, and GC content (i.e., the ratio of G and C bases in an assembly). We use dnadiff~\cite{marcais_mummer4_2018} to measure the accuracy of \emph{de novo} assemblies based on 1)~the average identity of an assembly when compared to its reference genome and 2)~the fraction of overall bases in a reference genome that align to a given assembly (i.e., genome fraction). We compare \proposal with minimap2~\cite{li_minimap2_2018} and MHAP~\cite{berlin_assembling_2015} for read overlapping. For the human genomes, MHAP either 1)~requires a memory space larger than what we have in our system (i.e., 1TB) or 2)~generates a large output such that we cannot generate the assembly as miniasm exceeds the memory space we have.

For read mapping, we map all reads in a dataset (i.e., query sequences) to their corresponding reference genome (i.e., target sequence). We evaluate read mapping in terms of accuracy, quality, and the effect of read mapping on downstream analysis by calling structural variants. We compare \proposal with minimap2, LRA~\cite{ren_lra_2021}, Winnowmap2~\cite{jain_long-read_2022, jain_weighted_2020}, S-conLSH~\cite{chakraborty_conlsh_2020, chakraborty_s-conlsh_2021}, and Strobealign~\cite{sahlin_flexible_2022}. We do not evaluate 1)~LRA, Winnowmap2, and S-conLSH for short reads as these tools do not support mapping paired-end short reads, 2)~Strobealign for long reads as it is a short read aligner, 3)~S-conLSH for the \emph{D. ananassae} as S-conLSH crashes due to a segmentation fault when mapping reads to the \emph{D. ananassae} reference genome, and 4)~S-conLSH for mapping HG002 reads as its output cannot be converted into a sorted BAM file, which is required for variant calling. We do not evaluate the read mapping accuracy of LRA and S-conLSH because 1)~LRA generates a CIGAR string with characters that the \texttt{paftools mapeval} tool cannot parse to calculate alignment positions, and 2)~S-conLSH due to its poor accuracy results we observe in our preliminary analysis.

\textbf{Read mapping accuracy.} We measure 1)~the overall read mapping error rate and 2)~the distribution of the read mapping error rate with respect to the fraction of mapped reads. To generate these results, we use the tools in \texttt{paftools} provided by minimap2 in two steps. First, the \texttt{paftools pbsim2fq} tool annotates the read IDs with their true mapping information that PBSIM2 generates. The \texttt{paftools mapeval} tool calculates the error rate of read mapping tools by comparing the mapping regions that the read mapping tools find with their true mapping regions annotated in read IDs. The error rate shows the ratio of reads mapped to incorrect regions over the entire mapped reads.

\textbf{Read mapping quality.} We measure 1)~the breadth of coverage (i.e., percentage of bases in a reference genome covered by at least one read), 2)~the average depth of coverage (i.e., the average number of read alignments per base in a reference genome), 3)~mapping rate (i.e., number of aligned reads) and 4)~rate of properly paired reads for paired-end mapping. To measure the breadth and depth of coverage of read mapping, we use BEDTools~\cite{quinlan_bedtools_2010} and Mosdepth~\cite{pedersen_mosdepth_2018}, respectively. To measure the mapping rate and properly paired reads, we use BAMUtil~\cite{jun_efficient_2015}.

\textbf{Downstream analysis.} We use sniffles2~\cite{sedlazeck_accurate_2018, smolka_comprehensive_2022} to call structural variants (SVs) from the HG002 long read mappings. We use Truvari~\cite{english_truvari_2022} to compare the resulting SVs with the benchmarking SV set (i.e., the \emph{Tier 1} set) released by the Genome in a Bottle (GIAB) consortium~\cite{zook_robust_2020} in terms of their true positives ($TP$), false positives ($FP$), false negatives ($FN$), precision ($P = TP/(TP+FP)$), recall ($R = TP/(TP + FN)$) and the $F_1$ scores ($F_{1} = 2 \times (P \times R)/(P+R)$). False positives show the number of the called SVs missing in the benchmarking set. False negatives show the number of SVs in the benchmarking set missing from the called SV set. The Tier 1 set includes 12,745 sequence-resolved SVs that include the \texttt{PASS} filter tag. GIAB provides the high-confidence regions of these SVs with low errors. We follow the benchmarking strategy that GIAB suggests~\cite{zook_robust_2020}, where we compare the SVs with the \texttt{PASS} filter tag within the high-confidence regions.

For both use cases, we use the \texttt{time} command in Linux to evaluate the performance and peak memory footprints. We provide the average speedups and memory overhead of \proposal compared to each tool, while dataset-specific results are shown in our corresponding figures. When applicable, we use the default parameters of all the tools suggested for certain use cases and sequencing technologies (e.g., mapping HiFi reads in minimap2). Since minimap2 and MHAP do not provide default parameters for read overlapping using HiFi reads, we use the parameters that HiCanu~\cite{nurk_hicanu_2020} uses for overlapping HiFi reads with minimap2 and MHAP. We provide the details regarding the parameters and versions we use for each tool in Supplementary Tables~\ref{supptab:ovpars}, \ref{supptab:mappars}, and \ref{supptab:version}. When applicable in read overlapping, we use the same window and the seed length parameters that \proposal uses in minimap2 and show the performance and accuracy results in Supplementary Figure~\ref{suppfig:overlap_perf-eq} and Supplementary Table~\ref{supptab:overlap_assembly-eq}. For read mapping, the comparable default parameters in \proposal are already the same as in minimap2.

\subsection{Empirical Analysis of Fuzzy Seed Matching}\label{subsec:fuzzy_matching}
We evaluate the effectiveness of fuzzy seed matching by finding non-identical seeds with the same hash value (i.e., collisions) when using a low-collision hash function that minimap2 uses (\texttt{hash64}) and \proposal in two ways.

\subsubsection{Finding minimizer collisions.}\label{subsubsec:mincollisions}
\begin{figure*}[htb]
\centering
\includegraphics[width=\linewidth]{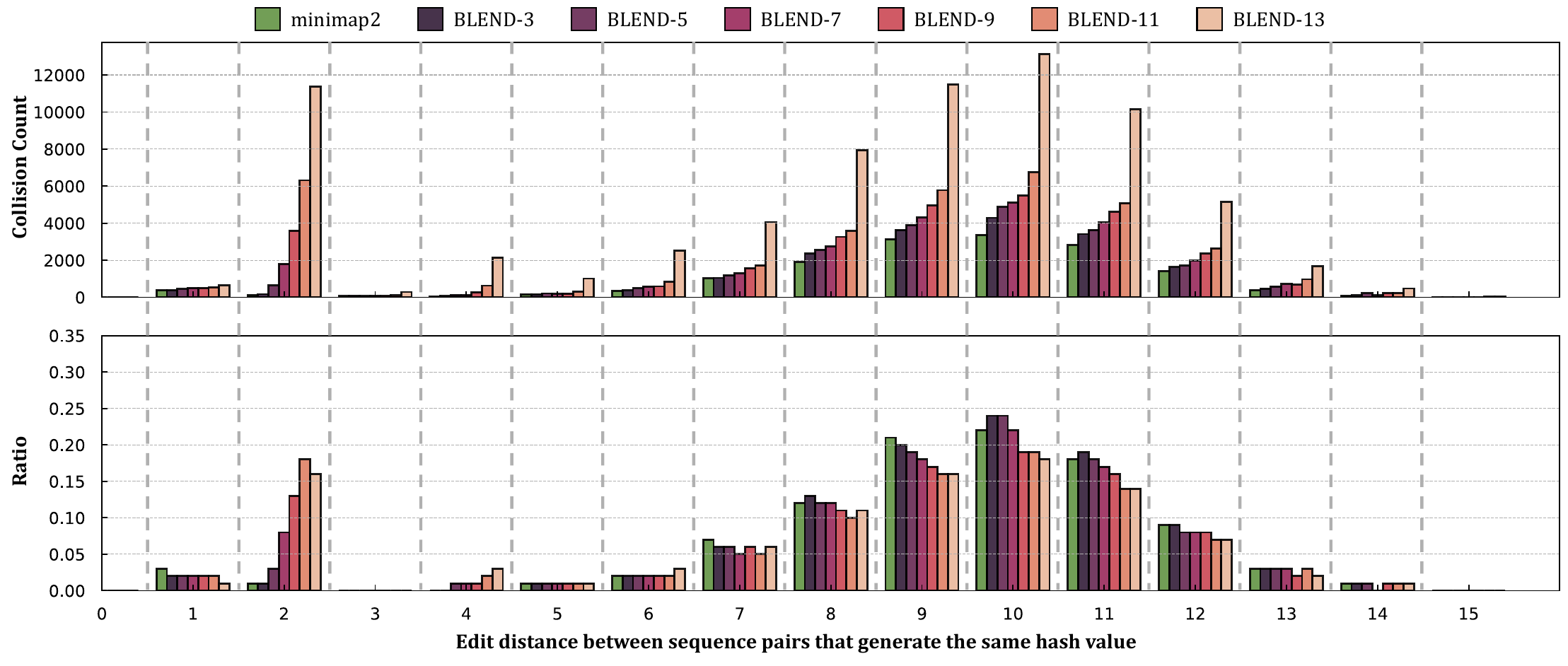}
\caption{Fuzzy seed matching statistics. Collision count shows the number of non-identical seeds that generate the same hash value and the edit distance between these sequences. Ratio is the proportion of collisions between non-identical sequences at a certain edit distance over all collisions. \proposal-$n$ shows the number of neighbors ($n$) that \proposal uses.}
\label{fig:fuzzy_seed_matching}
\end{figure*}

Our goal is to evaluate the effects of using a low-collision hash function and the \proposal mechanism on the hash value collisions between non-identical minimizers. We use minimap2 and \proposal to find all the minimizer seeds in the \emph{E. coli} reference genome~\cite{tvedte_comparison_2021}, as explained in Supplementary Section~\ref{suppsec:min_fuzzy_statistics}. Figure~\ref{fig:fuzzy_seed_matching} shows the edit distance between non-identical seeds with hash collision when using minimap2 and \proposal. We evaluate \proposal for various numbers of \emph{neighbors} (n) as explained in the~\emph{\nameref{subsec:seed}} section, which we show as \proposal-n in Figure~\ref{fig:fuzzy_seed_matching}, Supplementary Tables~\ref{supptab:min_fuzzy_seed_matching} and~\ref{supptab:seq_fuzzy_seed_matching}. We make three key observations. 
First, \proposal significantly increases the ratio of hash collisions between highly similar minimizer pairs (e.g., edit distance less than 3) compared to using a low-collision hash function in minimap2. This result shows that \proposal favors increasing the collisions for highly similar seeds (i.e., fuzzy seed matching) than uniformly increasing the number of collisions by keeping the same ratio across all edit distance values.
Second, the number of collisions that minimap2 and \proposal find are similar to each other for the minimizer pairs that have a large edit distance between them (e.g., larger than 6). The only exception to this observation is \proposal-13, which substantially increases all collisions for any edit distance due to using many small k-mers (i.e., thirteen 4-mers) when generating the hash values of 16-character long seeds. \Copy{R3/2A}{\rev{We note that the number of collisions is significantly higher when the edit distance between minimizers is $2$ compared to the collisions with edit distance $1$. We argue that this may be due to the distribution of the edit distances between minimizer pairs where there may be significantly a large number of minimizer pairs with edit distance $2$ than $1$.}}
Third, increasing the number of neighbors can effectively reduce the average edit distance between fuzzy seed matches with the cost of increasing the overall number of minimizer seeds, as shown in Supplementary Table~\ref{supptab:min_fuzzy_seed_matching}. We conclude that \proposal can effectively find highly similar seeds with the same hash value as it increases the ratio of collisions between similar seeds while providing a collision ratio similar to minimap2 for dissimilar seeds.

\subsubsection{Identifying similar sequences.}
Our goal is to find non-identical k-mer matches with the same hash value (i.e., fuzzy k-mer matches) between highly similar sequence pairs, as explained in Supplementary Section~\ref{suppsec:seq_fuzzy_statistics}.
Supplementary Table~\ref{supptab:seq_fuzzy_seed_matching} shows the number and portion of similar sequence pairs that we can find using \emph{only} fuzzy k-mer matches. We make two key observations.
First, \proposal is the only mechanism that can identify similar sequences from their fuzzy k-mer matches since low-collision hash functions cannot increase the collision rates for high similarity matches. Second, \proposal can identify a larger number of similar sequence pairs with an increasing number of neighbors. For the number of neighbors larger than 5, the percentage of these similar sequence pairs that \proposal can identify ranges from $1.2\%$ to $7.9\%$ of the overall number of sequences we use in our dataset. We conclude that \proposal enables finding similar sequence pairs from fuzzy k-mer matches that low-collision hash functions cannot find.

\subsection{Use Case 1: Read Overlapping}\label{subsec:overlap}
\subsubsection{Performance.}\label{subsubsec:overlap-performance}

\begin{figure*}[hb!]
\centering
\includegraphics[width=0.9\linewidth]{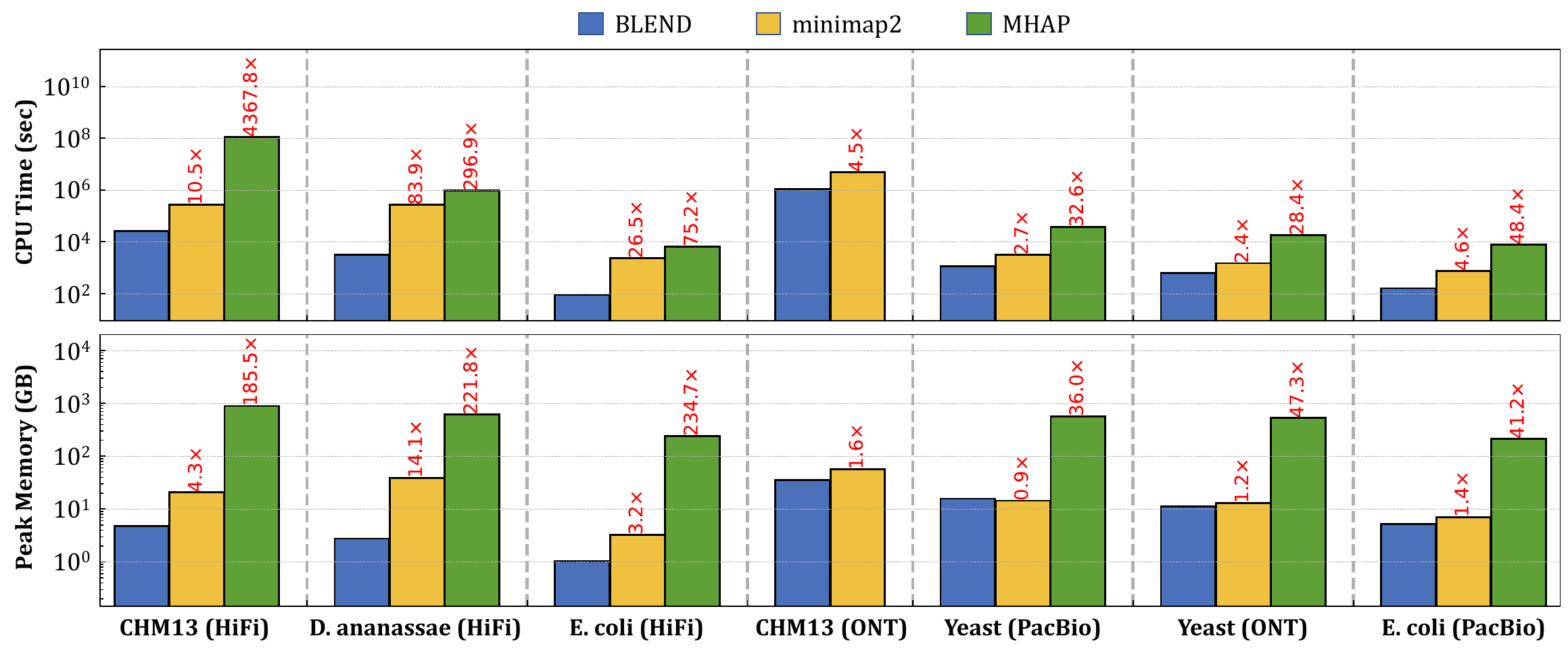}
\caption{CPU time and peak memory footprint comparisons of read overlapping.}
\label{fig:overlap_perf}
\end{figure*}

Figure~\ref{fig:overlap_perf} shows the CPU time and peak memory footprint comparisons for read overlapping. We make the following five observations.
First, \proposal provides an average speedup of \avgovpM and \avgovpMH while reducing the memory footprint by \avgovmM and \avgovmMH compared to minimap2 and MHAP, respectively. \proposal is significantly more performant and provides less memory overheads than MHAP because MHAP generates many hash values for seeds regardless of the length of the sequences, while \proposal allows sampling the number of seeds based on the sequence length with the windowing guarantees of minimizers and strobemer seeds.
Second, when considering only HiFi reads, \proposal provides significant speedups by \avgovpHM and \avgovpHMH while reducing the memory footprint by \avgovmHM and \avgovmHMH compared to minimap2 and MHAP, respectively.
HiFi reads allow \proposal to increase the window length (i.e., $w=200$) when finding the minimizer k-mer of a seed, which improves the performance and reduces the memory overhead without reducing the accuracy. This is possible mainly because \proposal can find \emph{both} fuzzy and exact seed matches, which enables \proposal to find \emph{unique} fuzzy seed matches that minimap2 \emph{cannot} find due to its exact-matching seed requirement.
Third, we find that \proposal requires less than 16GB of memory space for almost all the datasets, making it largely possible to find overlapping reads even with a personal computer with relatively small memory space.
\proposal has a lower memory footprint because 1)~\proposal uses as many seeds as the number of minimizer k-mers per sequence to benefit from the reduced storage requirements that minimizer k-mers provide, and 2)~the window length is larger than minimap2 as \proposal can tolerate increasing this window length with the fuzzy seed matches without reducing the accuracy.
Fourth, when using erroneous reads (i.e., PacBio CLR and ONT), \proposal performs better than other tools with memory overheads similar to minimap2. The set of parameters we use for erroneous reads prevents \proposal from using large windows (i.e., $w=10$ instead of $w=200$) without reducing the accuracy of read overlapping. Smaller window lengths generate more seeds, which increases the memory space requirements.
Fifth, we use the same parameters (i.e., the seed length and the window length) with minimap2 that \proposal uses to observe the benefits that \proposal provides with PacBio CLR and ONT datasets. We cannot perform the same experiment for the HiFi datasets because \proposal uses strobemer seeds of length $31$, which minimap2 cannot support due to its minimizer seeds and the maximum seed length limitation in its implementation (i.e., max. $28$). We use \emph{minimap2-Eq} to refer to the version of minimap2, where it uses the parameters equivalent to the \proposal parameters for a given dataset in terms of the seed and window lengths.
We show in Supplementary Figure~\ref{suppfig:overlap_perf-eq} that minimap2-Eq performs, on average, $\sim5\%$ better than \proposal with similar memory space requirements when using the same set of parameters with the \texttt{\proposal-I} technique. Minimap2-Eq provides worse accuracy than \proposal when generating the ONT assemblies, as shown in Supplementary Table~\ref{supptab:overlap_assembly-eq}, while the erroneous PacBio assemblies are more accurate with minimap2-Eq. The main benefit of \proposal is to provide overall higher accuracy than both the baseline minimap2 and minimap-Eq, which we can achieve by finding unique fuzzy seed matches that minimap2 cannot find.
We conclude that \proposal is significantly more memory-efficient and faster than other tools to find overlaps, especially when using HiFi reads with its ability to sample many seeds using large values of $w$ without reducing the accuracy.

\subsubsection{Overlap Statistics.}\label{subsubsec:overlap-statistics}
\begin{figure}[t]
\centering
  \includegraphics[width=\linewidth]{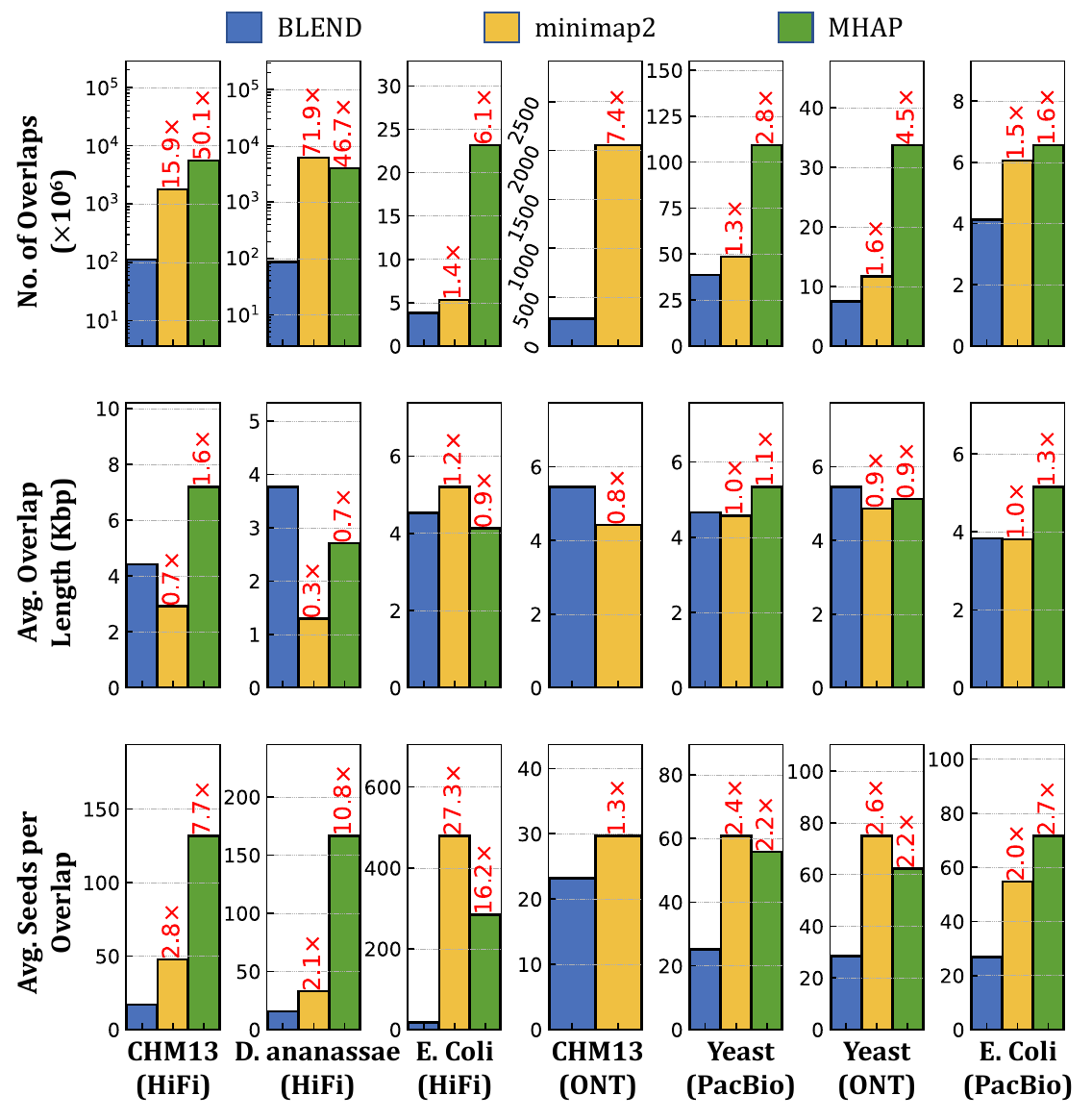}
    \caption{\Copy{R3/5A}{\rev{Average number and length of overlaps}, and average number of seeds used to find a single overlap.}}
    \label{fig:overlap_stats}
\end{figure}

Figure~\ref{fig:overlap_stats} shows the overall number of overlaps, the average length of overlaps, and the average number of seed matches that each tool finds to identify the overlaps between reads. The combination of the overall number of overlaps and the average number of seed matches provides the overall number of seeds found by each method. We make the following four key observations.
First, we observe that \proposal finds overlaps longer than minimap2 and MHAP can find in most cases. \proposal can 1)~uniquely find the fuzzy seed matches that the exact-matching-based tools cannot find and 2)~perform chaining on these fuzzy seed matches to increase the length of overlap using many fuzzy seed matches that are relatively close to each other. \Copy{R3/3A}{Finding more distinct seeds and chaining these seeds enable \proposal to find longer overlaps than other tools. \rev{Although these unique features of \proposal can lead to chaining longer overlaps, we also note that \proposal may not be able to find very short overlaps due to the larger window lengths it uses, which can also contribute to increasing the average length of overlaps.}}
Second, \proposal uses significantly fewer seed matches per overlap than other tools, up to \ovmaxseed, to find these longer overlaps. This is mainly because \proposal needs much fewer seeds per overlap as it uses 1)~larger window lengths than minimap2 and 2)~provides windowing guarantees, unlike MHAP.
Third, finding fewer seed matches per overlap leads to 1)~finding fewer overlaps than minimap2 and MHAP find and 2)~reporting fewer seed matches overall. These overlaps that \proposal cannot find are mainly because of the strict parameters that minimap2 and MHAP use due to their exact seed matching limitation (e.g., smaller window lengths). \proposal can increase the window length while producing more accurate and complete assemblies than minimap2 and MHAP (Table~\ref{tab:overlap_assembly}). This suggests that minimap2 and MHAP find redundant overlaps and seed matches that have no significant benefits in generating accurate and complete assemblies from these overlaps. 
Fourth, the sequencing depth of coverage has a larger impact on the number of overlaps that \proposal can find compared to the impact on minimap2 and MHAP. We observe this trend when comparing the number of overlaps found using the PacBio ($200\times$ coverage) and ONT ($100\times$ coverage) reads of the Yeast genome. The gap between the number of overlaps found by \proposal and other tools increases as the sequencing coverage decreases. This suggests that \proposal can be less robust to the sequencing depth of coverage. \Copy{R3/4A}{Such a trend does not impact the accuracy of the assemblies that we generate using the \proposal overlaps, \rev{while it provides lower NGA50 and NG50 values as shown in Table~\ref{tab:overlap_assembly}.}}
We conclude that the performance and memory-efficiency improvements in read overlapping are proportional to the reduction in the seed matches that \proposal uses to find overlapping reads. Thus, finding fewer non-redundant seed matches can dramatically improve the performance and memory space usage without reducing the accuracy.

\subsubsection{Assembly Quality Assessment.}\label{subsubsec:overlap-assembly}

Our goal is to assess the quality of assemblies generated using the overlapping reads found by \proposal, minimap2, and MHAP. Table~\ref{tab:overlap_assembly} shows the statistics related to the accuracy of assemblies (i.e., the six statistics on the leftmost part of the table) and the statistics related to assembly length and contiguity (i.e., the four statistics on the rightmost part of the table) when compared to their respective reference genomes. We make the following five key observations based on the accuracy results of assemblies.
\input{tables/assembly_quality}
First, we observe that we can construct more accurate assemblies in terms of average identity and k-mer completeness when we use the overlapping reads that \proposal finds than those minimap2 and MHAP find. These results show that the assemblies we generate using the \proposal overlaps are more similar to their corresponding reference genome. \proposal can find unique and accurate overlaps using fuzzy seed matches that lead to more accurate \emph{de novo} assemblies than the minimap2 and MHAP overlaps due to their lack of support for fuzzy seed matching.
Second, we observe that assemblies generated using \proposal overlaps usually cover a larger fraction of the reference genome than minimap2 and MHAP overlaps.
Third, although the average identity and genome fraction results seem mixed for the PacBio CLR and ONT reads such that \proposal is best in terms of either average identity or genome fraction, we believe these two statistics should be considered together (e.g., by multiplying both results). This is because a highly accurate but much smaller fraction of the assembly can align to a reference genome, giving the best results for the average identity. We observe that this is the case for the \emph{D. ananassae} and \emph{Yeast} (PacBio CLR) genomes such that MHAP provides a very high average identity only for the much smaller fraction of the assemblies than the assemblies generated using \proposal and minimap2 overlaps. Thus, when we combine average identity and genome fraction results, we observe that \proposal consistently provides the best results for all the datasets.
Fourth, \proposal usually provides the best results in terms of the aligned length and the number of mismatches per 100Kb. In some cases, QUAST cannot generate these statistics for the MHAP results as a small portion of the assemblies aligns the reference genome when the MHAP overlaps are used.
Fifth, we find that assemblies generated from \proposal overlaps are less biased than minimap2 and MHAP overlaps, based on the average GC content results that are mostly closer to their corresponding reference genomes. We conclude that \proposal overlaps yield assemblies with higher accuracy and less bias than the assemblies that the minimap2 and MHAP overlaps generate in most cases.

Table~\ref{tab:overlap_assembly} shows the results related to assembly length and contiguity on its rightmost part. We make the following three observations.
First, we show that \proposal yields assemblies with better contiguity when using HiFi reads based on the largest NG50, NGA50, and contig length results compared to minimap2 with the exception of the human genome.
Second, minimap2 provides better contiguity for the human genomes and erroneous reads.
Third, the overall length of all assemblies is mostly closer to the reference genome assembly.
We conclude that minimap2 provides better contiguity for the assemblies from erroneous and human reads while \proposal is usually better suited for using the HiFi reads.

\subsection{Use Case 2: Read Mapping}\label{subsec:mapping}
\subsubsection{Performance.}\label{subsubsec:mapping-perf}

\begin{figure*}[htb]
\centering
  \includegraphics[width=0.9\linewidth]{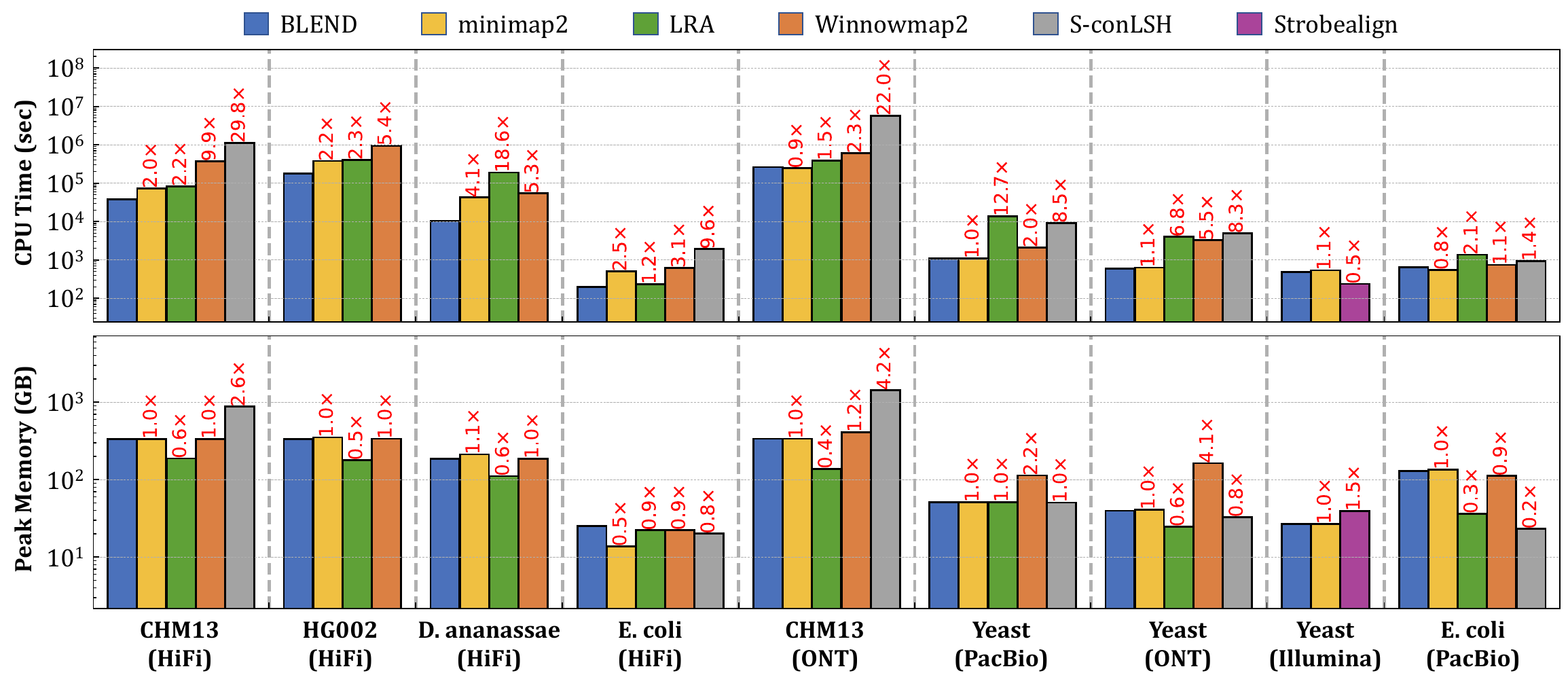}
    \caption{CPU time and peak memory footprint comparisons of read mapping.}
    \label{fig:mapping_perf}
\end{figure*}

Figure~\ref{fig:mapping_perf} shows the CPU time and the peak memory footprint comparisons when performing read mapping to the corresponding reference genomes. We make the following four key observations.
First, we observe that \proposal provides an average speedup of \avgrmpM, \avgrmpL, \avgrmpW, and \avgrmpS over minimap2, LRA, Winnowmap2, and S-conLSH, respectively. Although \proposal performs better than most of these tools, the speedups we see are usually lower than those we observe in read overlapping. Read mapping includes an additional computationally costly step that read overlapping skips, which is the read alignment. The extra overhead of read alignment slightly hinders the benefits that \proposal provides that we observe in read overlapping.
Second, we find that LRA and minimap2 require \avgrmmL and \avgrmmM of the memory space that \proposal uses, while Winnowmap2 and S-conLSH have a larger memory footprint by \avgrmmW and \avgrmmS, respectively. \proposal cannot provide similar reductions in the memory overhead that we observe in read overlapping due to the narrower window length ($w=50$ instead of $w=200$) it uses to find the minimizer k-mers for HiFi reads. Using a narrow window length generates more seeds to store in a hash table, which proportionally increases the peak memory space requirements.
Third, \proposal provides performance and memory usage similar to minimap2 when mapping the erroneous ONT and PacBio reads because \proposal uses the same parameters as minimap2 for these reads (i.e., same $w$ and seed length).
Fourth, Strobealign is the best-performing tool for mapping short reads with the cost of larger memory overhead.
We conclude that \proposal, on average, 1)~performs better than all tools for mapping long reads and 2)~provides a memory footprint similar to or better than minimap2, Winnowmap2, S-conLSH, and Strobealign, while LRA is the most memory-efficient tool.

\subsubsection{Read Mapping Accuracy.}\label{subsubsec:mapping-accuracy}

\begin{figure*}[htb]
\centering
  \includegraphics[width=0.9\linewidth]{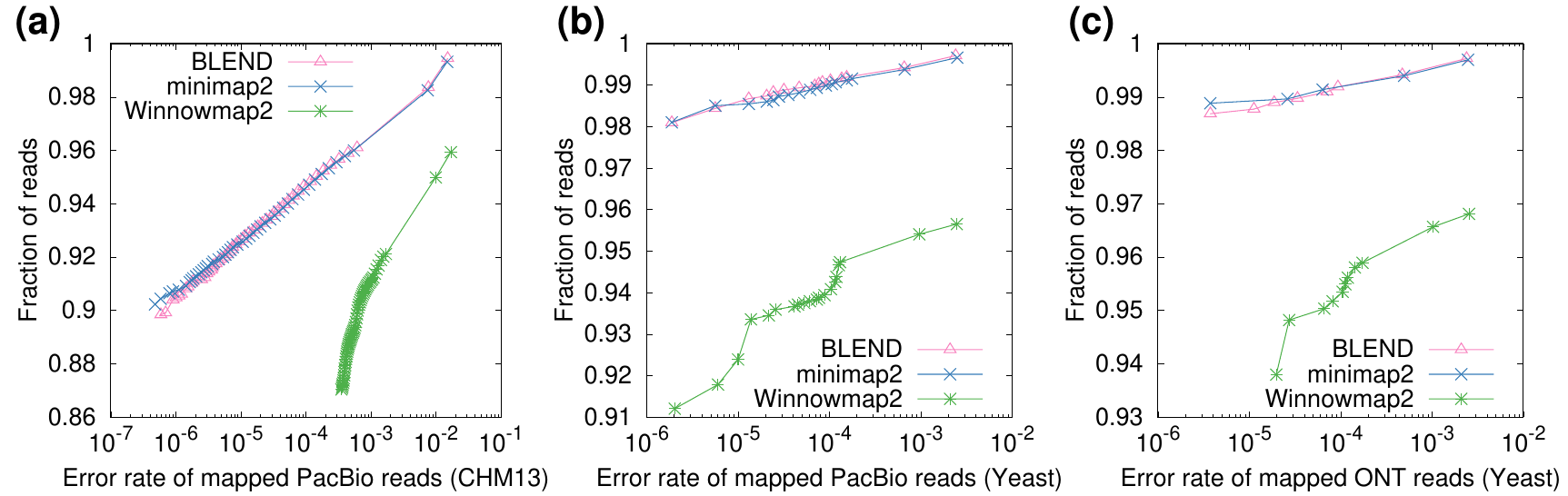}
    \caption{\Copy{R3/6A}{\rev{Fraction of simulated reads with an average mapping error rate. Reads are binned by their mapping quality scores. There is a bin for each mapping quality score as reported by the read mapper, and bins are sorted based on their mapping quality scores in descending order. For each tool, the $n^{th}$ data point from the left side of the x-axis shows the rate of incorrectly mapped reads among the reads in the first $n$ bins. We show the number of reads in these bins in terms of the fraction of the overall number of reads in the dataset. The data point with the largest fraction shows the average mapping error rate of all mapped reads.}}}
    \label{fig:mapping_accuracy}
\end{figure*}

Table~\ref{tab:mapping_accuracy} and Figure~\ref{fig:mapping_accuracy} show the overall read mapping accuracy and fraction of mapped reads with their average mapping accuracy, respectively. We make two observations. First, we observe that \proposal generates the most accurate read mapping in most cases, while minimap2 provides the most accurate read mapping for the human genome. These two tools are on par in terms of their read mapping accuracy and the fraction of mapped reads. Second, although Winnowmap2 provides more accurate read mapping than minimap2 for the PacBio reads from the Yeast genome, Winnowmap2 always maps a smaller fraction of reads than those \proposal and minimap2 map. We conclude that although the results are mixed, \proposal is the only tool that generates either the most or the second-most accurate read mapping in all datasets, providing the overall best accuracy results.

\input{tables/mapping_accuracy}

\subsubsection{Read Mapping Quality.}\label{subsubsec:mapping-quality}

Our goal is to assess the quality of read mappings in terms of four metrics: average depth of coverage, breadth of coverage, number of aligned reads, and the ratio of the paired-end reads that are properly paired in mapping. Table~\ref{tab:mapping_quality} shows the quality of read mappings based on these metrics when using \proposal, minimap2, LRA, Winnowmap2, and Strobealign. We exclude S-conLSH from the read mapping quality comparisons as we cannot convert its SAM output to BAM format to properly index the BAM file due to issues with its SAM output format. We make five observations.

\input{tables/mapping_quality}

First, all tools cover a large portion of the reference genomes based on the breadth of coverage of the reference genomes. Although LRA provides the lowest breadth of coverage in most cases compared to the other tools, it also provides the best breadth of coverage after mapping the human HG002 reads. This result shows that these tools are less biased in mapping reads to particular regions with their high breadth of coverage, and the best tool for covering the largest portion of the genome depends on the dataset.

Second, both \proposal and minimap2 map an almost complete set of reads to the reference genome for all the datasets, while Winnowmap2 suffers from a slightly lower number of aligned reads when mapping erroneous PacBio CLR and ONT reads. The only exception to this observation is the HG002 dataset, where \proposal provides a smaller number of aligned reads compared to other tools, while \proposal provides the same breadth of coverage as minimap2. We investigate if such a smaller number of aligned reads leads to a coverage bias genome-wide in Supplementary Figures~\ref{suppfig:genomedepth-1}, \ref{suppfig:genomedepth-13}, and \ref{suppfig:genomedepth_diff}. We find that the distribution of the depth of coverage of \proposal is mostly similar to minimap2. There are a few regions in the reference genome where minimap2 provides substantially higher coverage than \proposal provides, as we show in Supplementary Figure~\ref{suppfig:genomedepth_diff}, which causes \proposal to align a smaller number of reads than minimap2 aligns. Since these regions are still covered by both \proposal and minimap2 with different depths of coverage, these two tools generate the same breadth of coverage without leading to no significant coverage bias genome-wide.

Third, we find that all the tools generate read mappings with a depth of coverage significantly close to their sequencing depth of coverage. This shows that almost all reads map to the reference genome evenly. 
Fourth, Strobealign generates the largest number of 1)~short reads mappings to the reference genome and 2)~properly paired reads compared to \proposal and minimap2. Strobealign can map more reads using less time (Figure~\ref{fig:mapping_perf}, which makes its throughput much higher than \proposal and minimap2.
Fifth, although Strobealign can map more reads, it covers the smallest portion of the reference genome based on the breadth of coverage compared to \proposal and minimap2. This suggests that Strobealign provides a higher depth of coverage at certain regions of the reference genome than \proposal and minimap2 while leaving larger gaps in the reference genome.
We conclude that the read mapping qualities of \proposal, minimap2, and Winnowmap2 are highly similar, while LRA provides slightly worse results. It is worth noting that \proposal provides a better breadth of coverage than minimap2 provides in most cases while using the same parameters in read mapping. \proposal does this by finding unique fuzzy seed matches that the other tools cannot find due to their exact-matching seed requirements.

\subsubsection{Downstream Analysis.}\label{subsubsec:sv_calling}
To evaluate the effect of read mapping on downstream analysis, we call SVs from the HG002 long read mappings that \proposal, minimap2, LRA, and Winnowmap2 generate. Table~\ref{tab:sv_calling} shows the benchmarking results. We make two key observations. First, we find that \proposal provides the best overall accuracy in downstream analysis based on the best $F_1$ score compared to other tools. This is because \proposal provides the best true positive and false negative numbers while providing the second-best false positive numbers after LRA. These two best values overall contribute to achieving the best recall and second-best precision that is on par with the precision LRA provides. Second, although LRA generates the second-best $F_1$ score, it provides the worst recall results due to the largest number of false negatives. We conclude that \proposal is consistently either the best or second-best in terms of the metrics we show in Table~\ref{tab:sv_calling}, which leads to providing the best overall $F_1$ accuracy in structural variant calling.

\begin{table}[h]
\centering
\input{tables/sv_calling}
\label{tab:sv_calling}
\end{table}

%% file: tables/dataset.tex
\caption{Details of datasets used in evaluation.}
\resizebox{\linewidth}{!}{
\begin{tabular}{@{}llrrll@{}}\toprule
\textbf{Organism} & \textbf{Library} & \textbf{Reads (\#)} & \textbf{Seq.} & \textbf{SRA} & \textbf{Reference}\\
\textbf{} & \textbf{} & \textbf{} & \textbf{Depth} & \textbf{Accession} & \textbf{Genome}\\\midrule
\emph{Human CHM13} & PacBio HiFi & 3,167,477 & 16 & SRR1129212{2-3} & T2T-CHM13 (v1.1) \\
                   & ONT$^{*}$ & 10,380,693 & 30 & Simulated R9.5 & T2T-CHM13 (v2.0) \\\midrule
\emph{Human HG002} & PacBio HiFi & 11,714,594 & 52 & SRR1038224{4-9} & GRCh37 \\\midrule
\emph{D. ananassae} & PacBio HiFi & 1,195,370 & 50 & SRR11442117 & \cite{tvedte_comparison_2021} \\\midrule
\emph{Yeast} & PacBio CLR$^{*}$ & 270,849 & 200 & Simulated P6-C4 & GCA\_000146045.2\\
             & ONT$^{*}$ & 135,296 & 100 & Simulated R9.5 & GCA\_000146045.2\\
             & Illumina MiSeq & 3,318,467 & 80 & ERR1938683 & GCA\_000146045.2\\\midrule
\emph{E. coli} & PacBio HiFi & 38,703 & 100 & SRR11434954 & \cite{tvedte_comparison_2021}\\
               & PacBio CLR & 76,279 & 112 & SRR1509640 & GCA\_000732965.1\\\bottomrule
\multicolumn{6}{l}{\footnotesize $^{*}$ We use PBSIM2 to generate the simulated PacBio and ONT reads.}\\
\multicolumn{6}{l}{We show the simulated chemistry under the SRA Accession column.}\\
\end{tabular}}

%% file: tables/assembly_quality.tex
\begin{table*}[htb]
\centering
\caption{Assembly quality comparisons.}
\resizebox{0.95\linewidth}{!}{
\begin{tabular}{@{}clrrrrrrrrrr@{}}\toprule
\textbf{Dataset} 	  & \textbf{Tool} & \textbf{Average}	   & \textbf{Genome}		& \textbf{K-mer}	   & \textbf{Aligned}	   & \textbf{Mismatch per} & \textbf{Average} & \textbf{Assembly}	  & \textbf{Largest} 	  & \textbf{NGA50}    & \textbf{NG50}	  \\
					  & 			  & \textbf{Identity (\%)} & \textbf{Fraction (\%)} & \textbf{Compl. (\%)} & \textbf{Length (Mbp)} & \textbf{100Kbp (\#)}  & \textbf{GC (\%)} & \textbf{Length (Mbp)} & \textbf{Contig (Mbp)} & \textbf{(Kbp)}    & \textbf{(Kbp)}    \\\midrule
\emph{CHM13} 	  	  & \proposal 	  & \textbf{99.8526}	   & \textbf{98.4847}		& \textbf{90.15}	   & 3,092.54 			   & \textbf{22.02}	  	   & \textbf{40.78}	  & \textbf{3,095.21} 	  & 22.8397 			  & 5,442.25 		  & 5,442.31 		  \\
(HiFi)				  & minimap2 	  & 99.7421				   & 97.1493 				& 83.05 			   & \textbf{3,094.79} 	   & 55.96 			  	   & 40.71 			  & 3,100.97 			  & \textbf{47.1387} 	  & \textbf{7,133.43} & \textbf{7,134.31} \\
					  & MHAP 		  & N/A					   & N/A 					& N/A 				   & N/A 				   & N/A 				   & N/A 			  & N/A 				  & N/A 				  & N/A 		      & N/A 		      \\
					  & Reference 	  & 100 				   & 100 					& 100 				   & 3,054.83 			   & 0.00 				   & 40.85 			  & 3,054.83 			  & 248.387 			  & 154,260 	      & 154,260 	   	  \\\midrule
\emph{D. ananassae}   & \proposal 	  & \textbf{99.7856}	   & \textbf{97.2308}		& \textbf{86.43}	   & 240.391 			   & \textbf{143.13} 	   & \textbf{41.75}	  & \textbf{247.153} 	  & \textbf{6.23256} 	  & \textbf{792.407}  & \textbf{798.913}  \\
(HiFi)				  & minimap2 	  & 99.7044 			   & 96.3190 				& 72.33 			   & \textbf{289.453} 	   & 191.53	  			   & 41.68 			  & 298.28 			  	  & 4.43396 			  & 273.398 		  & 278.775 		  \\
					  & MHAP 		  & 99.5551 			   & 0.7276 				& 0.21 				   & 2.29 				   & 239.76 			   & 42.07 			  & 2.34951 			  & 0.028586 			  & N/A 		   	  & N/A 		   	  \\
					  & Reference 	  & 100 				   & 100 					& 100 				   & 213.805 			   & 0.00 				   & 41.81 			  & 213.818 			  & 30.6728 			  & 26,427.4 		  & 26,427.4 		  \\\midrule
\emph{E. coli} 		  & \proposal 	  & \textbf{99.8320} 	   & \textbf{99.8801} 		& \textbf{87.91}	   & \textbf{5.12155} 	   & \textbf{3.77}	  	   & \textbf{50.53}	  & 5.12155 			  & \textbf{3.41699}	  & \textbf{3,416.99} & \textbf{3,416.99} \\
(HiFi)				  & minimap2 	  & 99.7064 			   & 99.8748 				& 79.27 			   & 5.09249 			   & 19.71 			  	   & 50.47 			  & \textbf{5.09436} 	  & 3.08849 			  & 3,087.05 		  & 3,087.05 		  \\
					  & MHAP 		  & N/A 				   & N/A 					& N/A 				   & N/A 				   & N/A 				   & N/A 			  & N/A 				  & N/A 				  & N/A 		   	  & N/A 		   	  \\
					  & Reference 	  & 100 				   & 100 					& 100 				   & 5.04628 			   & 0.00 				   & 50.52 			  & 5.04628 			  & 4.94446 			  & 4,944.46 		  & 4,944.46 		  \\\midrule
\emph{CHM13} 	  	  & \proposal 	  & N/A	   				   & N/A					& \textbf{29.26}	   & \textbf{2,891.28} 	   & \textbf{4,077.53}	   & \textbf{41.32}	  & 2,897.87 	  		  & 25.2071 			  & 5,061.52 		  & 5,178.59 		  \\
(ONT)				  & minimap2 	  & N/A	   				   & N/A					& 28.32 			   & 2,860.26 	   		   & 4,660.73 			   & 41.36 			  & \textbf{2,908.55} 	  & \textbf{66.7564}	  & \textbf{13,189.2} & \textbf{13,820.3} \\
					  & Reference 	  & 100 				   & 100 					& 100 			   	   & 3,117.29 			   & 0.00 				   & 40.75 			  & 3,117.29 			  & 248.387 			  & 150,617 	   	  & 150,617 	      \\\midrule
\emph{Yeast} 		  & \proposal 	  & 89.1677 	   		   & \textbf{97.0854} 		& \textbf{33.81}	   & \textbf{12.3938} 	   & 2,672.37 	   		   & 38.84	  		  & 12.4176 			  & 1.54807 			  & 635.966 		  & 636.669 		  \\
(PacBio)			  & minimap2 	  & 88.9002 			   & 96.9709 				& 33.38 			   & 12.0128 			   & 2,684.38 			   & 38.85 			  & \textbf{12.3325} 	  & \textbf{1.56078}	  & \textbf{810.046}  & \textbf{828.212}  \\
					  & MHAP 		  & \textbf{89.2182} 	   & 88.5928 				& 32.39 			   & 10.9039 			   & \textbf{2,552.05} 	   & \textbf{38.81}   & 10.9896 			  & 1.02375 			  & 85.081 		   	  & 436.285 		  \\
					  & Reference 	  & 100 				   & 100 					& 100 				   & 12.1571 			   & 0.00 				   & 38.15 			  & 12.1571 			  & 1.53193 			  & 924.431 		  & 924.431 		  \\\midrule
\emph{Yeast} 	      & \proposal 	  & \textbf{89.6889} 	   & 99.2974 				& \textbf{35.95}	   & \textbf{12.3222} 	   & \textbf{2,529.47} 	   & \textbf{38.64}	  & \textbf{12.3225} 	  & 1.10582 			  & 793.046 		  & 793.046 		  \\
(ONT)			      & minimap2 	  & 88.9393 			   & \textbf{99.6878}		& 34.84 			   & 12.304 			   & 2,782.59 			   & 38.74 			  & 12.3725 			  & \textbf{1.56005}	  & \textbf{796.718}  & \textbf{941.588}  \\
					  & MHAP 		  & 89.1970 			   & 89.2785 				& 33.58 			   & 10.8302 			   & 2,647.19 			   & 38.84 			  & 10.9201 			  & 1.44328 			  & 118.886 		  & 618.908 		  \\
					  & Reference 	  & 100 				   & 100 					& 100 				   & 12.1571 			   & 0.00 				   & 38.15 			  & 12.1571 			  & 1.53193 			  & 924.431 		  & 924.431 		  \\\midrule
\emph{E. coli} 		  & \proposal 	  & \textbf{88.5806} 	   & \textbf{96.5238} 		& \textbf{32.32}	   & \textbf{5.90024} 	   & \textbf{1,857.56}	   & \textbf{49.81}	  & 6.21598 			  & 2.40671	  			  & \textbf{769.981}  & 2,060.4 	 	  \\
(PacBio)			  & minimap2 	  & 88.1365 			   & 92.7603 				& 30.74 			   & 5.37728 			   & 2,005.72 			   & 49.66 			  & \textbf{6.02707} 	  & \textbf{3.77098} 	  & 367.442    	   	  & \textbf{3,770.98} \\
					  & MHAP 		  & 88.4883 			   & 90.5533 				& 31.32 			   & 5.75159 			   & 1,999.48 			   & 49.69 			  & 6.26216 			  & 1.04286 			  & 110.535    	   	  & 456.01 		   	  \\
					  & Reference 	  & 100 				   & 100 					& 100 				   & 5.6394 			   & 0.00 				   & 50.43 			  & 5.6394 				  & 5.54732 			  & 5,547.32 		  & 5,547.32 		  \\\bottomrule
\multicolumn{12}{l}{\footnotesize Best results are highlighted with \textbf{bold} text. For most metrics, the best results are the ones closest to the corresponding value of the reference genome.}\\
\multicolumn{12}{l}{\footnotesize The best results for \emph{Aligned Length} are determined by the highest number within each dataset. We do not highlight the reference results as the best results.}\\
\multicolumn{12}{l}{\footnotesize N/A indicates that we could not generate the corresponding result because tool, QUAST, or dnadiff failed to generate the statistic.} \\
\end{tabular}}
\label{tab:overlap_assembly}
\end{table*}

%% file: tables/mapping_accuracy.tex
\begin{table}[htb]
\centering
\caption{Read mapping accuracy comparisons.}
\resizebox{\linewidth}{!}{
\begin{tabular}{@{}lrrr@{}}\toprule
\textbf{Dataset} 		& \multicolumn{3}{c}{\textbf{Overall Error Rate (\%)}} \\\cmidrule{2-4}
				 		& \proposal          & minimap2           & Winnowmap2 \\\midrule
\emph{CHM13} (ONT) 	    & 1.5168427 	     & \textbf{1.4914009} & 1.7001222  \\\midrule
\emph{Yeast} (PacBio) 	& \textbf{0.2403134} & 0.2504307          & 0.2474206  \\\midrule
\emph{Yeast} (ONT) 		& \textbf{0.2386617} & 0.2468770          & 0.2534777  \\\bottomrule
\multicolumn{4}{l}{\footnotesize Best results are highlighted with \textbf{bold} text.} \\
\end{tabular}}
\label{tab:mapping_accuracy}
\end{table}

%% file: tables/mapping_quality.tex
\begin{table}[b]
\centering
\caption{Read mapping quality comparisons.}
\resizebox{\linewidth}{!}{
\begin{tabular}{@{}clrrrr@{}}\toprule
\textbf{Dataset} 		& \textbf{Tool} & \textbf{Average} 			 & \textbf{Breadth of} & \textbf{Aligned} 	 & \textbf{Properly}\\
				 		& 				& \textbf{Depth of} 		 & \textbf{Coverage}   & \textbf{Reads} 	 & \textbf{Paired}\\
				 		& 				& \textbf{Cov. (${\times}$)} & \textbf{(\%)} 	   & \textbf{(\#)} 		 & \textbf{(\%)}\\\midrule
\emph{CHM13} 	        & \proposal 	& \textbf{16.58} 			 & \textbf{99.991} 	   & 3,171,916           & NA \\
(HiFi)					& minimap2 		& \textbf{16.58} 			 & \textbf{99.991} 	   & \textbf{3,172,261}  & NA \\
						& LRA 			& 16.37 					 & 99.064 			   & 3,137,631 			 & NA \\
						& Winnowmap2 	& \textbf{16.58} 			 & 99.990 	           & 3,171,313 			 & NA \\\midrule
\emph{HG002} 	        & \proposal 	& 51.25 			         & 92.245 	           & 11,424,762          & NA \\
(HiFi)					& minimap2 		& 53.08 			         & 92.242 	           & 12,407,589          & NA \\
						& LRA 			& 52.48 					 & \textbf{92.275} 	   & \textbf{13,015,195} & NA \\
						& Winnowmap2 	& \textbf{53.81} 			 & 92.248 	           & 12,547,868 	     & NA \\\midrule
\emph{D. ananassae} 	& \proposal 	& 57.37 					 & 99.662 			   & 1,223,388 			 & NA \\
(HiFi)					& minimap2 		& \textbf{57.57} 			 & \textbf{99.665} 	   & 1,245,931 			 & NA \\
						& LRA 			& 57.06 					 & 99.599 			   & 1,235,098 			 & NA \\
						& Winnowmap2 	& 57.40 			 		 & 99.663 	   		   & \textbf{1,249,575}  & NA \\\midrule
\emph{E. coli} 			& \proposal 	& \textbf{99.14} 			 & 99.897 			   & 39,048 			 & NA\\
(HiFi)					& minimap2 		& \textbf{99.14} 			 & 99.897 			   & \textbf{39,065} 	 & NA\\
						& LRA 			& 99.10 			 		 & 99.897 			   & 39,063 			 & NA\\
						& Winnowmap2 	& \textbf{99.14} 			 & 99.897 			   & 39,036 			 & NA\\\midrule
\emph{CHM13} 	        & \proposal 	& \textbf{29.34} 			 & \textbf{99.999} 	   & \textbf{10,322,767} & NA \\
(ONT)			        & minimap2 		& 29.33 			         & \textbf{99.999} 	   & 10,310,182          & NA \\
						& LRA 			& 28.84 					 & 99.948 			   & 9,999,432 			 & NA \\
						& Winnowmap2 	& 28.98 			         & 99.936 	           & 9,958,402 			 & NA \\\midrule
\emph{Yeast} 	        & \proposal 	& \textbf{195.87} 			 & \textbf{99.980} 	   & \textbf{270,064}    & NA\\
(PacBio)		        & minimap2 		& 195.86 			         & \textbf{99.980} 	   & 269,935 	         & NA\\
						& LRA 			& 194.65 					 & 99.967 			   & 267,399 			 & NA\\
						& Winnowmap2 	& 192.35 			 		 & 99.977	           & 259,073 			 & NA\\\midrule
\emph{Yeast} 		    & \proposal 	& \textbf{97.88} 			 & \textbf{99.964} 	   & \textbf{134,919} 	 & NA\\
(ONT)			        & minimap2 		& \textbf{97.88} 			 & \textbf{99.964} 	   & 134,885 	         & NA\\
						& LRA 			& 97.25 					 & 99.952 			   & 132,862 			 & NA\\
						& Winnowmap2 	& 97.04 			 		 & 99.963 	   		   & 130,978 			 & NA\\\midrule
\emph{Yeast}            & \proposal 	& \textbf{79.92} 			 & \textbf{99.975} 	   & 6,493,730           & 95.88 \\
(Illumina)				& minimap2 		& 79.91 					 & 99.974 			   & 6,492,994 			 & 95.89 \\
				        & Strobealign   & \textbf{79.92} 			 & 99.970 			   & \textbf{6,498,380}  & \textbf{97.59} \\\midrule
\emph{E. coli} 			& \proposal 	& \textbf{97.51} 			 & 100 			       & 83,924 			 & NA\\
(PacBio)			    & minimap2 		& 97.29 			         & 100 			       & \textbf{85,326} 	 & NA\\
						& LRA 			& 93.61 			 		 & 100 			       & 80,802 			 & NA\\
						& Winnowmap2 	& 89.78 			         & 100 			       & 69,884 			 & NA\\\bottomrule
\multicolumn{6}{l}{\footnotesize Best results are highlighted with \textbf{bold} text.} \\
\multicolumn{6}{l}{\footnotesize Properly paired rate is only available for paired-end Illumina reads.} \\
\end{tabular}}
\label{tab:mapping_quality}
\end{table}

%% file: tables/sv_calling.tex
\caption{Benchmarking the structural variant (SV) calling results.}
\resizebox{\columnwidth}{!}{
\begin{tabular}{@{}lrrrrrr@{}}\toprule
              & \multicolumn{6}{c}{HG002 SVs (High-confidence Tier 1 SV Set)} \\\cmidrule{2-7}
\textbf{Tool} & \textbf{TP (\#)} &\textbf{FP (\#)} & \textbf{FN (\#)} & \textbf{Precision} & \textbf{Recall} & \textbf{$F_1$}  \\\midrule
\proposal	  & \textbf{9,229}   & 855             & \textbf{412}     & 0.9152 		       & \textbf{0.9573} & \textbf{0.9358} \\
minimap2	  & 9,222            & 915             & 419              & 0.9097 			   & 0.9565		     & 0.9326          \\
LRA			  & 9,155            & \textbf{830}    & 486              & \textbf{0.9169}    & 0.9496			 & 0.9329          \\
Winnowmap2	  & 9,170            & 1029            & 471              & 0.8991 			   & 0.9511		     & 0.9244          \\\bottomrule
\multicolumn{6}{l}{\footnotesize Best results are highlighted with \textbf{bold} text.} \\
\end{tabular}}

%% file: sections/4_discussion.tex
\section{Discussion} \label{sec:discussion}
We demonstrate that there are usually too many redundant \emph{short} and \emph{exact-matching} seeds used to find overlaps between sequences, as shown in Figure~\ref{fig:overlap_stats}. These redundant seeds usually exacerbate the performance and peak memory space requirement problems that read overlapping and read mapping suffer from as the number of chaining and alignment operations proportionally increases with the number of seed matches between sequences~\cite{li_minimap2_2018}. Such redundant computations have been one of the main limitations against developing population-scale genomics analysis due to the high runtime of a single high-coverage genome analysis.

There has been a clear interest in using long or fuzzy seed matches because of their potential to find similarities between target and query sequences efficiently and accurately~\cite{alser_technology_2021}. To achieve this, earlier works mainly focus on either 1)~chaining the exact k-mer matches by tolerating the gaps between them to increase the seed region or 2)~linking multiple consecutive minimizer k-mers such as strobemer seeds. Chaining algorithms are becoming a bottleneck in read mappers as the complexity of chaining is determined by the number of seed matches~\cite{l_guo_hardware_2019}. Linking multiple minimizer k-mers enables tolerating indels when finding the matches of short subsequences between genomic sequence pairs, but these seeds (e.g., strobemer seeds) should still exactly match due to the nature of the hash functions used to generate the hash values of seeds. This requires the seeding techniques to generate exactly the same seed to find either exact-matching or approximate matches of short subsequences.
We state that any arbitrary k-mer in the seeds should be tolerated to mismatch to improve the sensitivity of any seeding technique, which has the potential for finding more matching regions while using fewer seeds. Thus, we believe \proposal solves the main limitation of earlier works such that it can generate the same hash value for similar seeds to find fuzzy seed matches with a single lookup while improving the performance, memory overhead, and accuracy of the applications that use seeds.

We hope that \proposal advances the field and inspires future work in several ways, some of which we list next. 
First, we observe that \proposal is \emph{most effective} when using high coverage and highly accurate long reads. Thus, \proposal is already ready to scale for longer and more accurate sequencing reads. 
Second, the vector operations are suitable for hardware acceleration to improve the performance of \proposal further. Such an acceleration is mainly useful when a massive amount of k-mers in a seed are used to generate the hash value for a seed, as these calculations can be done in parallel. We already provide the SIMD implementation to calculate the hash values \proposal. We encourage implementing our mechanism for the applications that use seeds to find sequence similarity using processing-in-memory and near-data processing~\cite{cali_segram_2022, mansouri_ghiasi_genstore_2022, shahroodi_demeter_2022, diab2022high, khalifa_filtpim_2021, khatamifard2021genvom, cali_genasm_2020, chen2020parc, kaplan2020bioseal, laguna2020seed, angizi_pim-aligner_2020, nag_gencache_2019, kim_grim-filter_2018}, GPUs~\cite{sadasivan_accelerating_2022, a_zeni_logan_2020, goenka_segalign_2020}, and FPGAs and ASICs~\cite{singh_fpga-based_2021, chen2021high, yan_accel-align_2021, fujiki2020seedex, alser_sneakysnake_2020, turakhia_darwin_2018} to exploit the massive amount of embarrassingly parallel bitwise operations in \proposal to find fuzzy seed matches.
Third, we believe it is possible to apply the hashing technique we use in \proposal for many seeding techniques with a proper design. We already show we can apply SimHash in regular minimizer k-mers or strobemers. Strobemers can be generated using k-mer sampling strategies other than minimizer k-mers, which are based on syncmers and random selection of k-mers (i.e., randstrobes)~\cite{sahlin_flexible_2022}. It is worth exploring and rethinking the hash functions used in these seeding techniques.
Fourth, potential machine learning applications can be used to generate more sensitive hash values for fuzzy seed matching based on learning-to-hash approaches~\cite{j_wang_survey_2018} and recent improvements on SimHash for identifying nearest neighbors in machine learning and bioinformatics~\cite{sharma_improving_2018, chen_using_2020, sinha_fruit-fly_2021}.

\subsubsection{Limitations.}\label{subsubsec:limitations}
\Copy{R3/2B}{\rev{We identify two main limitations of our work that requires further improvements. First, \proposal may generate the same hash values for $1\%-8\%$ of all the similar sequence pairs in a dataset, as we show in Supplementary Table~\ref{supptab:seq_fuzzy_seed_matching}. These $1\%-8\%$ of similar sequence pairs that cannot be found using low-collision hash functions can be significant in improving the accuracy and performance of some genomics applications. However, such a percentage may also be considered low for other use cases. We observe that increasing the number of neighbors ($n$) can increase the percentage of similar sequence pairs that \proposal can find with the cost of causing more collisions for dissimilar sequence pairs. A newer generation of the SimHash-like hash functions such as DenseFly~\cite{sharma_improving_2018} or FlyHash~\cite{dasgupta_neural_2017} has the potential to improve the rate of similar sequence pairs with the same hash value.}} \Copy{R3/1A}{\rev{Second, the advantage of \proposal is mainly observed when using highly accurate and long reads with high sequencing depth of coverage in read overlapping and downstream analysis, while the improvements are lower in other datasets. Although \proposal scales better as the sequencing technologies become cheaper and generate longer and highly accurate reads, it is also essential to further improve its accuracy and performance for existing read datasets with erroneous long reads and short reads. This requires further optimizations in the parameter settings for erroneous long reads and short reads. We leave these two limitations as future work along with the other potential future works that we discuss earlier.}}

%% file: sections/5_conclusion.tex
\subsubsection{Conclusion.}\label{subsubsec:conclusion} 
We propose \proposal, a mechanism that can efficiently find fuzzy seed matches between sequences to improve the performance, memory space efficiency, and accuracy of two important applications significantly: 1)~read overlapping and 2)~read mapping. Based on the experiments we perform using real and simulated datasets, we make six key observations.
First, for read mapping, \proposal provides an average speedup of \avgovpM and \avgovpMH while reducing the peak memory footprint by \avgovmM and \avgovmMH compared to minimap2 and MHAP.
Second, we observe that \proposal finds longer overlaps, in general, while using significantly fewer seed matches by up to \ovmaxseed to find these overlaps. 
Third, we find that we can usually generate more \emph{accurate} assemblies when using the overlaps that \proposal finds than those found by minimap2 and MHAP.
Fourth, for read mapping, we find that \proposal, on average, provides speedup by 1) \avgrmpM, \avgrmpL, \avgrmpW, and \avgrmpS compared to minimap2, LRA, Winnowmap2, and S-conLSH, respectively. Fifth, Strobealign performs best for short read mapping, while \proposal provides better memory space usage than Strobealign.
Sixth, we observe that \proposal, minimap2, and Winnowmap2 provide both high quality and better accuracy in read mapping in all datasets, while \proposal and LRA provide the best SV calling results in terms of downstream analysis accuracy. We conclude that \proposal can use fewer fuzzy seed matches to significantly improve the performance and reduce the memory overhead of read overlapping without losing accuracy, while \proposal, on average, provides better performance and a similar memory footprint in read mapping without reducing the read mapping quality and accuracy.

%% file: sections/supp.tex
\onecolumn
\setcounter{secnumdepth}{3}
\clearpage
\begin{center}
\textbf{\LARGE Supplementary Material for\\ \ltitle}
\end{center}

\setcounter{section}{0}
\setcounter{equation}{0}
\setcounter{figure}{0}
\setcounter{table}{0}
\setcounter{page}{1}
\makeatletter
\renewcommand{\theequation}{S\arabic{equation}}
\renewcommand{\thetable}{S\arabic{table}}
\renewcommand{\thefigure}{S\arabic{figure}}
\renewcommand{\thesection}{S\arabic{section}}
\renewcommand{\thetheorem}{S\arabic{theorem}}
\renewcommand\rev[1]{{\color{black}{#1}}}

\newcommand{\TextUnderscore}{\rule{.4em}{.4pt}}

\section{Statistics of Fuzzy Seed Matching} \label{suppsec:fuzzy_statistics}
\subsection{Finding Fuzzy Matches Between Minimizers} \label{suppsec:min_fuzzy_statistics}
Supplementary Table~\ref{supptab:min_fuzzy_seed_matching} shows the overall statistics of the fuzzy seed matching we explain in the~\emph{\nameref{subsec:fuzzy_matching}} section. We find minimizers using 1)~the low collision hash function that minimap2 uses (i.e., \texttt{hash64}) and 2)~the SimHash technique~\protect\citesupp{supp_charikar_similarity_2002, supp_manku_detecting_2007} we use in \proposal. For \proposal, we use the \texttt{\proposal-I} technique to directly compare the minimizers found using \proposal and minimap2. We keep the seed length constant, 16. For \proposal, we use various numbers of immediately overlapping k-mers that \proposal extracts from seed sequences (i.e., \emph{neighbors}), as explained in the~\emph{\nameref{subsec:seed}} section. To keep the seed length ($|S|$) constant with a varying number of neighbors ($n$), we calculate the k-mer length ($k$) we extract from seeds as follows: $|S| = n+k-1$ where $|S|$ and $n$ are known. For each tool and configuration, we report the overall number of minimizers we find, the number of minimizer pairs that generate the same hash value (i.e., \emph{collision}), the ratio of collisions to all minimizers, and the average edit distance between the minimizer pairs that have the same hash value. We make our resulting dataset that includes the statistics shown in Figure~\ref{fig:fuzzy_seed_matching} and Supplementary Table~\ref{supptab:min_fuzzy_seed_matching} available at Zenodo\footnote{\url{https://doi.org/10.5281/zenodo.7317896}}.

\begin{table}[h]
\centering
\input{tables/supp_min_fuzzy_seed_matching}
\label{supptab:min_fuzzy_seed_matching}
\end{table}

\subsection{Finding Fuzzy Matches Between Similar Sequences} \label{suppsec:seq_fuzzy_statistics}
Our goal is to find the non-identical k-mers with the same hash value between similar sequences. To this end, we prepare a dataset that includes 25-character long sequences in four steps. First, we extract 25-character long non-overlapping sequences from the \emph{E. coli} reference genome~\protect\citesupp{supp_tvedte_comparison_2021} shown in Table~\ref{tab:dataset}, which we call \emph{sampled sequences} for simplicity. To evenly sample these sequences, each sampled sequence is separated by 75 characters from the previous sampled sequence. Second, our goal is to find \emph{all} sequences in the reference genome that are \emph{similar} and non-identical to the sampled sequences. To achieve this, we use bowtie~\protect\citesupp{supp_langmead_aligning_2010} and find \emph{all} sequences in the \emph{E. coli} reference genome that the sampled sequences align with at least one mismatch and, at most, three mismatches (i.e., at least $\sim 88\%$ similarity). Third, we extract the sequences from the reference regions that the sampled sequences align, which we call \emph{aligned sequences}. Fourth, we prepare our dataset that contains 1,077 FASTA files and 4,130 25-character long sequences overall. Each FASTA file includes 1)~a sampled sequence that has at least one alignment in the reference genome based on our mismatching criteria and 2)~all aligned sequences that the sampled sequence is aligned to.

To find the non-identical k-mers with the same hash value in each FASTA file, we generate the hash values of all overlapping 16-mers of all sequences in a FASTA file. We use the low-collision hash function that minimap2 uses (i.e., \texttt{hash64}) and the \texttt{\proposal-I} technique in \proposal to generate these hash values. For \proposal, we use various numbers of neighbors when generating the hash values of 16-mers (see Supplementary Section~\ref{suppsec:min_fuzzy_statistics} for the relation between the number of neighbors and the seed length, which is 16 in our evaluation). In Supplementary Table~\ref{supptab:seq_fuzzy_seed_matching}, we report the number of sequences in our dataset, the number of sequences that have at least one non-identical k-mer pair with the same hash value (i.e., collisions), the ratio of collisions to the overall number of sequences, and the average edit distance between k-mers with collision. We make our dataset available at Zenodo\footnote{\url{https://doi.org/10.5281/zenodo.7319786}}, which includes 1,077 FASTA files and the resulting files that we generate the numbers we show in Supplementary Table~\ref{supptab:seq_fuzzy_seed_matching}. These resulting files include the non-identical k-mers with the same hash value, the sequence pairs that we extract these k-mers from, and the edit distances with these k-mers.

\input{tables/supp_seq_fuzzy_seed_matching.tex}

\clearpage

\section{A Real Example of Generating the Hash Values of Seeds $S_k$ and $S_l$}\label{suppsec:real}
Our goal is to show how the k-mer length $k$ and the number of k-mers to include in a seed, $n$, affect the final hash value. To this end, we use the following two seeds as found in the Yeast reference genome: $S_k:$ \texttt{CGGATGCTACAGTATATACCA} and $S_l:$ \texttt{ATGCTACAGTATATACCATCT}. Both seeds are 21-character long. We use two different parameter settings when generating the hash values of these seeds. The first setting uses $k=7$ as the k-mer length and $n=15$ as the number of immediately overlapping k-mers to include in a seed so that we can generate the 21-character long seeds $S_l$ and $S_k$. The second setting uses $k=15$ as the k-mer length and $n=7$ as the number of k-mers to include in a seed. We use the  \texttt{hash64} hash function as provided in the minimap2 implementation to generate the hash values of the k-mers of seeds.

In Supplementary Tables~\ref{supptab:kmers7-15-1-32} - \ref{supptab:kmers15-7-2-16} we show k-mers, the hash values of the k-mers in their binary form, and the gradual change in the counter vectors used to calculate the hash values for seeds $S_k$ and $S_l$. We update the counter vectors based on the bits in the hash values of each k-mer. Finally, we show the hash values of $S_k$ and $S_l$ in the last rows of each table. In Supplementary Tables~\ref{supptab:kmers7-15-1-32}-~\ref{supptab:kmers7-15-2-16}, we use $k=7$ as the k-mer length and $n=15$ as the number of immediately overlapping k-mers to include in a seed. In Supplementary Tables~\ref{supptab:kmers15-7-1-32}-~\ref{supptab:kmers15-7-2-16}, we use $k=15$ as the k-mer length and $n=7$ as the number of k-mers to include in a seed.

We make two key observations. First, we observe that the hash values of $S_k$ and $S_l$ are equal ($B(S_k) = B(S_l) = $\texttt{0b11000100 01101100 11101001 10110100}) when we use a short k-mer with high number of neighbors even though these two seeds differ by 3 k-mers. Second, the hash values of these two seeds are not equal when we use fewer neighbors with larger k-mers. For $S_k$ we find the hash value $B(S_k) = $\texttt{0b01101000 01000001 01110100 11000000} and for $S_l$ we find $B(S_l) = $\texttt{0b00101101 10110000 01111100 01010011}. We note that the bit positions with large values in their corresponding counter vectors are less likely to differ between two seeds when the seeds have a large number of k-mers in common. This motivates as to design more intelligent hash functions that are aware of the values in the counter vectors to increase the chance of generating the hash value for similar seeds.

\input{tables/supp_hash_values_stepbystep.tex}

\clearpage

\section{SIMD Implementation of the SimHash Technique}\label{suppsec:simd_implementation}

\begin{figure}[htb!]
\centering
\includegraphics[width=0.9\linewidth]{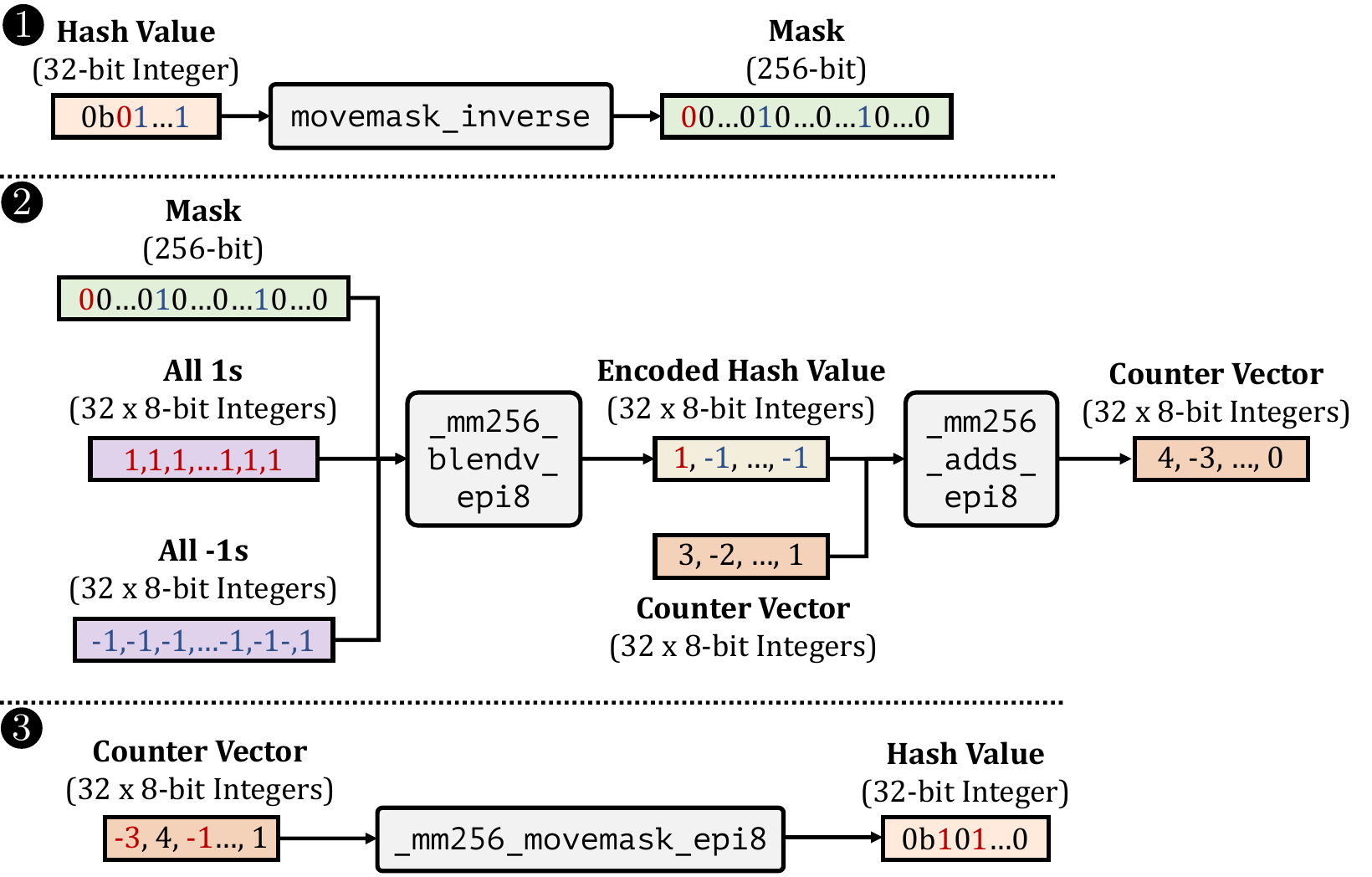}
\caption{SIMD execution flow when generating the hash value of a seed from its set items that \texttt{\proposal-I} or \texttt{\proposal-S} identifies. Colors highlight the propagation of bits and values to the outputs of functions.}
\label{suppfig:simd1}
\end{figure}

Supplementary Figure~\ref{suppfig:simd1} shows the high-level execution flow when calculating the hash value of a seed from its set items that \texttt{\proposal-I} or \texttt{\proposal-S} identifies, as explained in the~\emph{\nameref{subsec:seed}} and \emph{\nameref{subsec:seedhash}} sections. To efficiently perform the bitwise operations in the SimHash technique, \proposal utilizes the SIMD operations in three steps.

\begin{figure}[htb!]
\centering
\includegraphics[width=\linewidth]{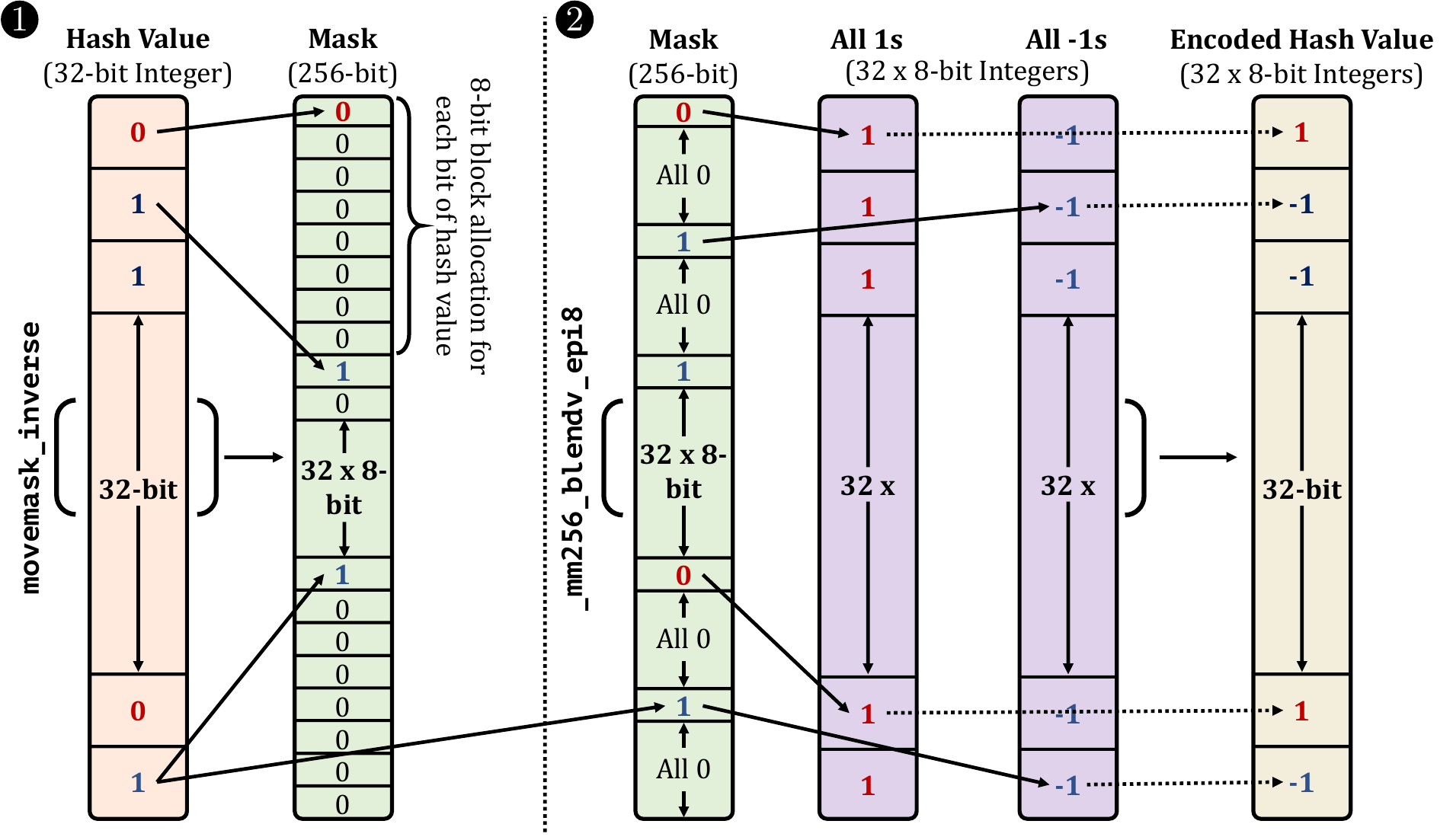}
\caption{Details of the \texttt{movemask\_inverse} and \texttt{\_mm256\_blendv\_epi8} executions. Colors and arrows highlight the propagation of bits and values to the outputs of functions.}
\label{suppfig:simd2}
\end{figure}

First, for each \emph{hash value} in set items, \proposal creates its corresponding \emph{mask} using the \texttt{movemask\_inverse} function, as shown in~\circlednumber{$1$}. For each bit position $t$ of the hash value, the \texttt{movemask\_inverse} function assigns the bit at position $t$ of the hash value to the bit position $t*8+7$ of a 256-bit SIMD register (i.e., the most significant bit positions of each 8-bit block), which \proposal uses it as a \emph{mask} in the next steps. We assume 0-based indexing and the mask register is initially 0. Supplementary Figure~\ref{suppfig:simd2} shows how each bit in hash value propagates to the mask register in~\circlednumber{$1$}. The \texttt{movemask\_inverse} is an in-house implementation that performs the reverse behavior of the \texttt{\_mm256\_movemask\_epi8}\footnote{\url{https://software.intel.com/sites/landingpage/IntrinsicsGuide/\#text=\_mm256\_movemask\_epi8&ig\_expand=4874}} SIMD function. Our function efficiently utilizes several other SIMD functions to perform the reverse behavior of \texttt{\_mm256\_movemask\_epi8}.

Second, for each mask created in the first step, \proposal updates the values in the counter vector (explained in the~\emph{\nameref{subsec:seedhash}} section), as shown in~\circlednumber{$2$}. To encode the hash value into its vector representation, \proposal uses the \texttt{\_mm256\_blendv\_epi8}\footnote{\url{https://software.intel.com/sites/landingpage/IntrinsicsGuide/\#text=\_mm256\_blendv\_epi8&ig\_expand=515}} SIMD function with 1)~the mask register \proposal creates in the first step, 2)~two 256-bit wide SIMD registers that include $32 \times$ 8-bit integers. For the first register, all 8-bit values are initialized to 1, and for the second register, all 8-bit values are initialized to -1. The \texttt{\_mm256\_blendv\_epi8} function generates a new 256-bit register with 8-bit integers where each 8-bit block is copied from either the first or the second register based on the most significant value in the mask register. If the most significant value in the mask register is 0, the corresponding 8-bit block in the first register is copied. Otherwise, the 8-bit block in the second register is copied. We show in detail how the values in these registers propagate to the resulting 256-bit register in Supplementary Figure~\ref{suppfig:simd2} in~\circlednumber{$2$}. \proposal, then, performs addition using the \texttt{\_mm256\_adds\_epi8}\footnote{\url{https://software.intel.com/sites/landingpage/IntrinsicsGuide/\#text=\_mm256\_adds\_epi8&ig\_expand=220}} function between the register that the \texttt{\_mm256\_blendv\_epi8} function generates and the 256-bit \emph{counter vector} that includes $32 \times$ 8-bit integers. We assume that all the 8-bit values in the counter vector are initially 0. \proposal keeps updating the counter vector as it iterates through the set items (i.e., hash values). The resulting value is written back to the counter vector to use it in the next iterations with the next set item. We note that the current design encodes 1 bits as -1 and 0 bits as 1, which is the opposite case of our explanation in the ~\emph{\nameref{subsec:seedhash}} section. Although we perform our encoding in an opposite way, \proposal generates the final hash values as originally explained. The reason for such a design change is due to the behavior of the function we use in the third step.

Third, \proposal converts the final result in the counter vector to its corresponding 32-bit hash value of a set (explained in the decoding step of the~\emph{\nameref{subsec:seedhash}} section), as shown in~\circlednumber{$3$}. \proposal uses the \texttt{\_mm256\_movemask\_epi8} function that takes a 256-bit register of 8-bit blocks and assigns the corresponding bit accordingly in a 32-bit value. The behavior of this function is essentially the reverse behavior of the \texttt{movemask\_inverse} function that we explain in the first step (i.e., reversing the arrows in Supplementary Figure~\ref{suppfig:simd2} \circlednumber{$1$} can simulate the \texttt{\_mm256\_movemask\_epi8} function). Thus, it assigns 1 to the bit position $t$ of a hash value if the bit at position $t*8+7$ (i.e., the most significant bit of an 8-bit integer) is 1. Since the most significant is 1 \emph{only} for the negative values according to the signed integer value convention, this function creates the opposite behavior of our decoding step we explain in the~\emph{\nameref{subsec:seedhash}} section. We resolve this issue by performing the encoding in an opposite way in the second step, where the resulting counter vector includes negative values when the majority of bits at a bit position is 1. Thus, the final hash value contains the same bits as explained in our main paper.

Although we omit the details here, \proposal avoids performing redundant computations when calculating the hash values of each input sequence, as the set items between each of these input sequences are likely to be shared. For example, minimap2 generates a hash value for each k-mer in a sequence and selects the k-mer with the minimum hash value in a window of k-mers as the minimizer, as shown in Figure~\ref{fig:seeding}. Assuming that the set items of each k-mer are l-mers, each k-mer differs by two l-mers at most with its next overlapping k-mer: one different l-mer contains the leading character of the first k-mer, and the other contains the trailing character of the next k-mer. Since there are two l-mer changes at most, \proposal calculates only the difference between subsequent k-mers. Thus, \proposal keeps a buffer in a first-in, first-out fashion such that the corresponding encoded hash value of the l-mer that is missing in the next k-mer is subtracted from the counter vector and popped from the queue while performing an addition \emph{only} for the l-mer that is missing from the previous k-mer and pushing it into the queue.

We perform our operations on 256-bit wide SIMD registers. \proposal works on 8-bit integer blocks assigned for each bit in a hash value. Since our registers are 256-bit wide, \proposal uses 32-bit hash values when calculating the SimHash value of a seed. Our implementation allows working on up to 64-bit hash values by dividing the most and least significant 32 bits into two 32-bit hash values. Each 32-bit hash value can independently follow the three steps we show in Supplementary Figure~\ref{suppfig:simd1}, and the final 64-bit value can be generated by the shift operations between the final 32-bit hash values. Although our approach is scalable to allow hash values with a larger number of bits, the current implementation does not support such flexible scaling and works on up to 64-bits.

\clearpage

\section{Parameter Exploration}
\subsection{The trade-off between \texttt{\proposal-I} and \texttt{\proposal-S}}\label{suppsec:tradeoff-blend}

\begin{figure}[htb!]
\centering
\includegraphics[width=\linewidth]{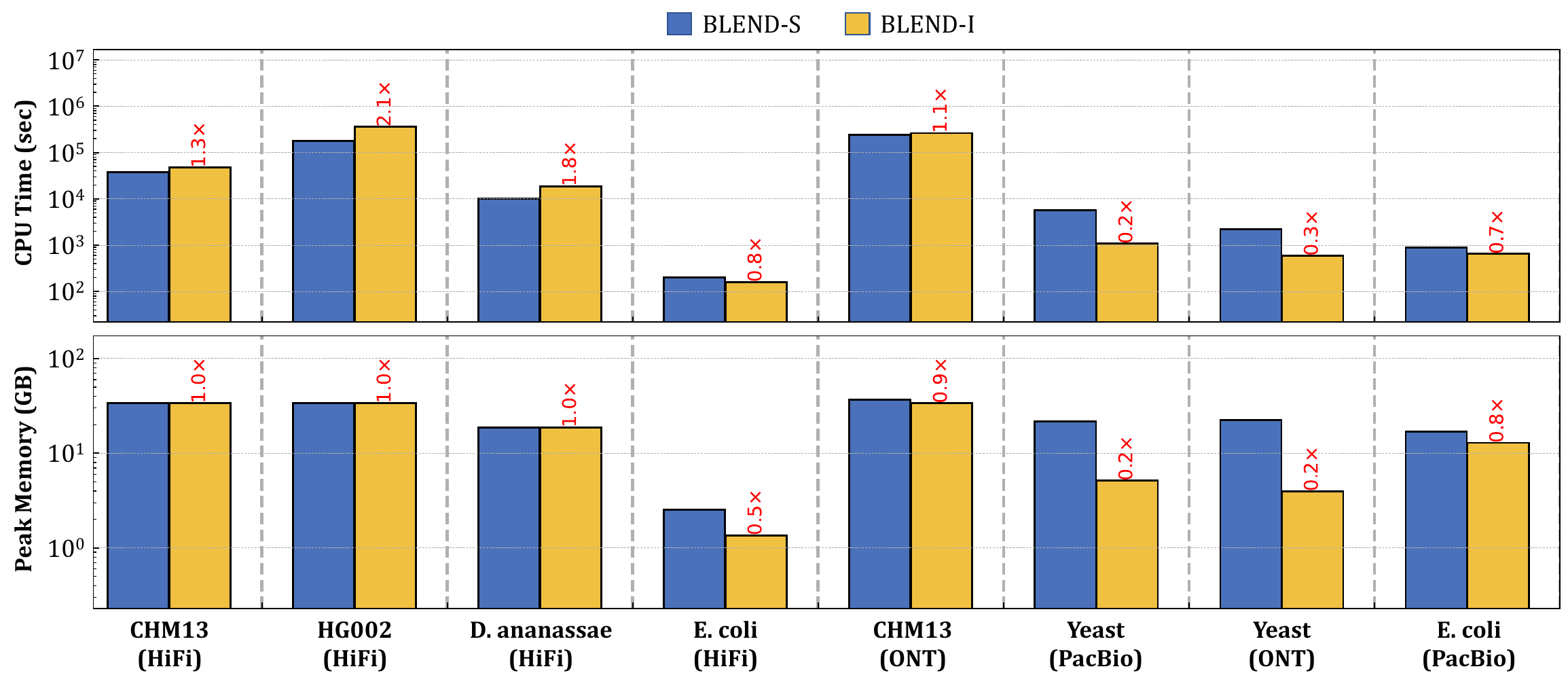}
\caption{CPU time and peak memory footprint comparisons of read mapping.}
\label{suppfig:read_mapping_perf-blend}
\end{figure}

Our goal is to show the performance and accuracy trade-offs between the seeding techniques that \proposal supports: \texttt{\proposal-I} and \texttt{\proposal-S}. In Supplementary Figures~\ref{suppfig:overlap_perf-blend} and ~\ref{suppfig:read_mapping_perf-blend}, we show the performance and peak memory usage comparisons when using \texttt{\proposal-I} and \texttt{\proposal-S} as the seeding technique by keeping all the other relevant parameters identical (e.g., number of k-mers to include in a seed $n$, window length $w$). In Supplementary Table~\ref{supptab:overlap_assembly-blend}, we show the assembly quality comparisons in terms of the accuracy and contiguity of the assemblies that we generate using the overlaps that \texttt{\proposal-I} and \texttt{\proposal-S} find. In Supplementary Tables~\ref{supptab:mapping_quality-blend} and ~\ref{supptab:mapping_accuracy-blend}, we show the read mapping quality and accuracy results using these two seeding techniques, respectively.

We also show the values for different parameters we test with \proposal in Supplementary Table~\ref{supptab:parameter_exploration}. We determine the default parameters of \proposal empirically based on the combination of best performance, memory overhead, and accuracy results.

\begin{figure}[htb!]
\centering
\includegraphics[width=\linewidth]{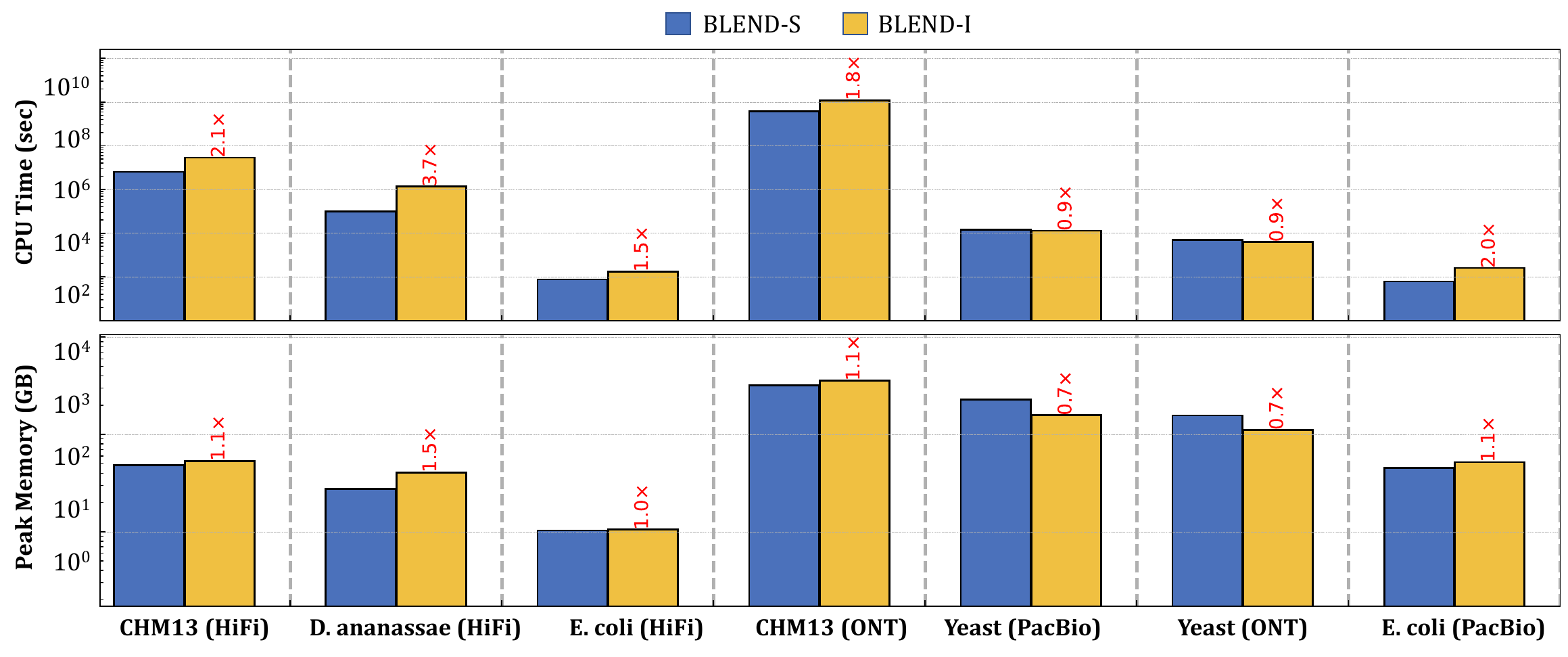}
\caption{CPU time and peak memory footprint comparisons of read overlapping.}
\label{suppfig:overlap_perf-blend}
\end{figure}

\clearpage

\input{tables/supp_assembly_quality_blendi_s}

\input{tables/supp_mapping_quality_blendi_s}

\input{tables/supp_mapping_accuracy-blend.tex}

\input{tables/supp_parameter_exploration.tex}

\clearpage

\subsection{The trade-off between \proposal and minimap2}\label{suppsec:tradeoff-eq}
Our goal is to compare \proposal and minimap2 using the same set of parameters that \proposal uses when generating its results. To achieve this, we control the following two conditions. First, we ensure that we use the same seeding technique that minimap2 uses. To this end, we use the \texttt{\proposal-I} seeding technique, which uses minimizers as seeds. We should note that \texttt{\proposal-I} does not always provide the best results in terms of performance or accuracy for the HiFi reads as the default seeding technique is \texttt{\proposal-S} for HiFi datasets in \proposal.

Second, we use the same seed length when we compare \proposal with minimap2. In minimap2, the seed length is the same as the k-mer length as minimap2 finds the minimizer k-mers from the hash values of k-mers. The seed length in \texttt{\proposal-I} is determined by \emph{both} the k-mer length and the number of k-mers that we include in a seed (i.e., $n$). For example, \proposal uses the \texttt{\proposal-I} seeding technique with the k-mer length $k=19$ and the number of neighbors $n=5$ for the PacBio reads. Combining immediately overlapping $5$-many $19$-mers generates seeds with length $19+5-1=23$. Thus, \texttt{\proposal-I} uses seeds of length $23$ based on these parameters. Supplementary Table~\ref{supptab:pardef} shows the seed length calculation for both \texttt{\proposal-I} and \texttt{\proposal-S}. We calculate the seed lengths for the datasets where \proposal uses \texttt{\proposal-I} as the default option (i.e., the PacBio and ONT datasets) in read overlapping. We note that \proposal uses the same seed length and window length as in minimap2 for mapping long reads. Thus, we do not report the read mapping results in this section, which are already reported in the main paper when comparing \proposal with minimap2. To run minimap2 with the same parameter conditions, we apply the same seed length and the window length that \proposal uses to minimap2 using the $k$ and $w$ parameters, respectively. We show these parameters in Supplementary Table~\ref{supptab:ovpars} (minimap-Eq). In the results we show below, minimap-Eq indicates the runs of minimap2 when using the same set of parameters that \proposal uses with the \texttt{\proposal-I} technique.

In Supplementary Figure~\ref{suppfig:overlap_perf-eq}, we show the performance and peak memory usage comparisons when using \proposal with the \texttt{\proposal-I} seeding technique, minimap2, and minimap2-Eq. In Supplementary Table~\ref{supptab:overlap_assembly-eq}, we show the assembly quality comparisons in terms of the accuracy and contiguity of the assemblies that we generate using the overlaps that each tool finds.

\begin{figure}[htb]
\centering
\includegraphics[width=0.5\linewidth]{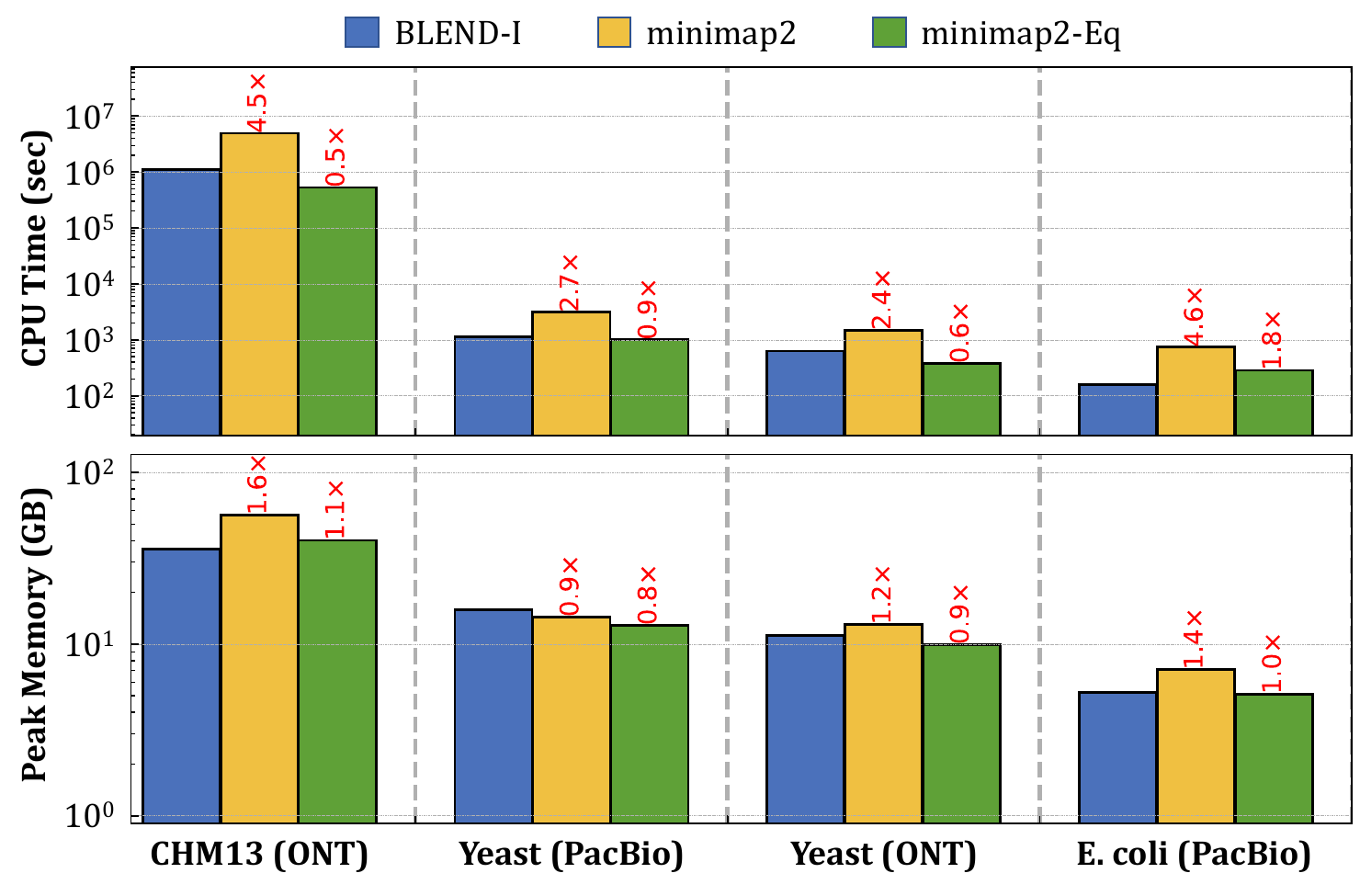}
\caption{CPU time and peak memory footprint comparisons of read overlapping.}
\label{suppfig:overlap_perf-eq}
\end{figure}

\input{tables/supp_overlap_assembly-eq.tex}

\vspace{-2cm}

\clearpage

\section{The Genome-wide Coverage Comparison}\label{suppsec:genomecov}

\begin{figure}[htb!]
\centering
\includegraphics[width=\linewidth]{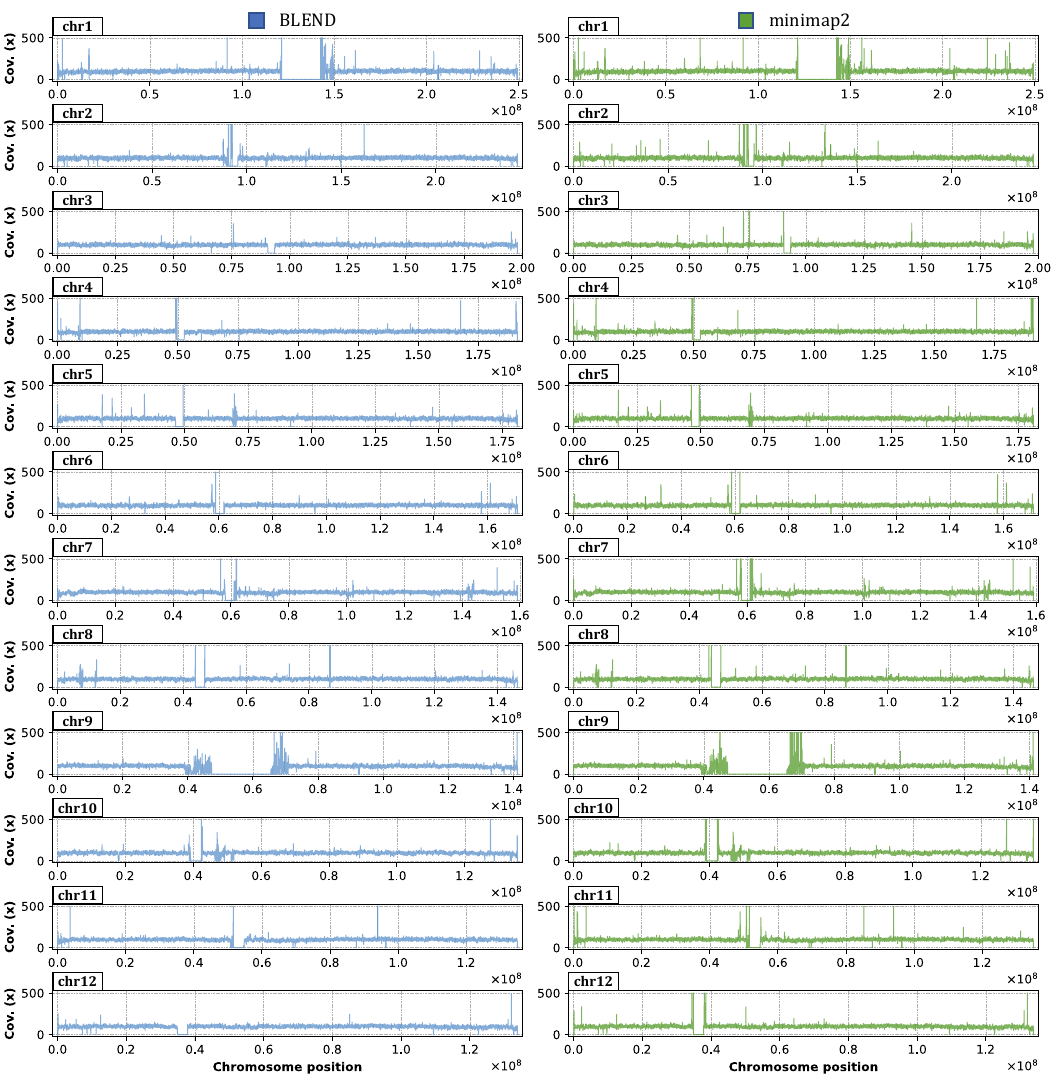}
\caption{Depth of coverage at each position (binned) of the GRCh37 reference genome (chromosomes 1 to 12) after mapping the HG002 reads using \proposal and minimap2. We label the chromosomes on the top left corner of each plot.}
\label{suppfig:genomedepth-1}
\end{figure}

\begin{figure}[htb!]
\centering
\includegraphics[width=\linewidth]{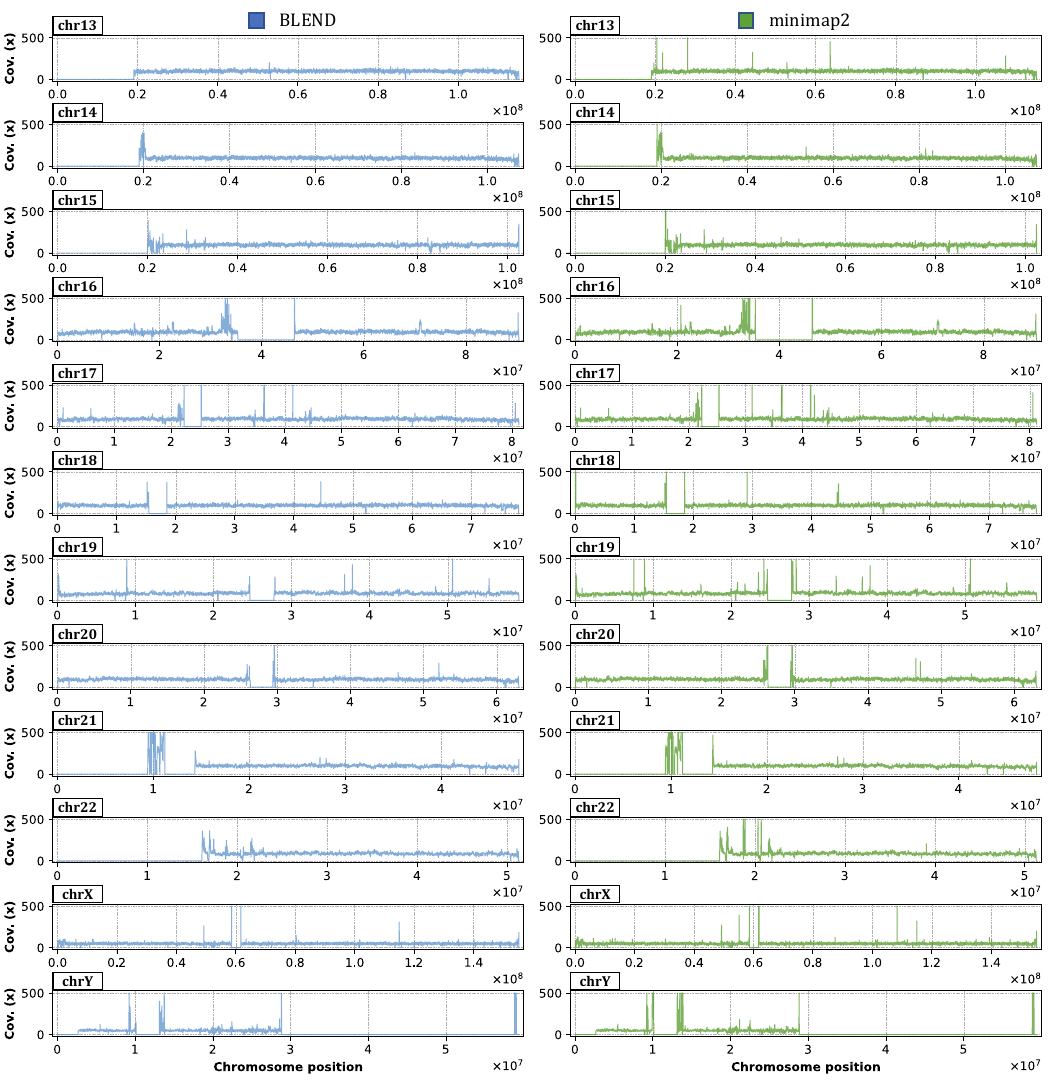}
\caption{Depth of coverage at each position (binned) of the GRCh37 reference genome (chromosomes 13 to Y) after mapping the HG002 reads using \proposal and minimap2. We label the chromosomes on the top left corner of each plot.}
\label{suppfig:genomedepth-13}
\end{figure}

We map the HG002 reads to the human reference genome (GRCh37) using \proposal and minimap2. Supplementary Figures~\ref{suppfig:genomedepth-1} and \ref{suppfig:genomedepth-13} show the depth of mapping coverage at each position of the reference genome chromosomes for \proposal and minimap2 on the left and right sides of the figures, respectively. To calculate the position-wise depth of coverage, we use the \texttt{multiBamSummary} tool from the deepTools2 package~\citesupp{supp_ramirez_deeptools2_2016}. The \texttt{multiBamSummary} tool divides the reference genome into consecutive bins of equal size (10,000 bases) to calculate the genome-wide coverage in fine granularity. For positions where the coverage is higher than $500\times$, we set the coverage to $500\times$ for visibility reasons as there are only a negligible amount of such regions where either \proposal or minimap2 exceeds this threshold without the other one exceeding it.

\clearpage

\begin{figure}[htb!]
\centering
\includegraphics[width=\linewidth]{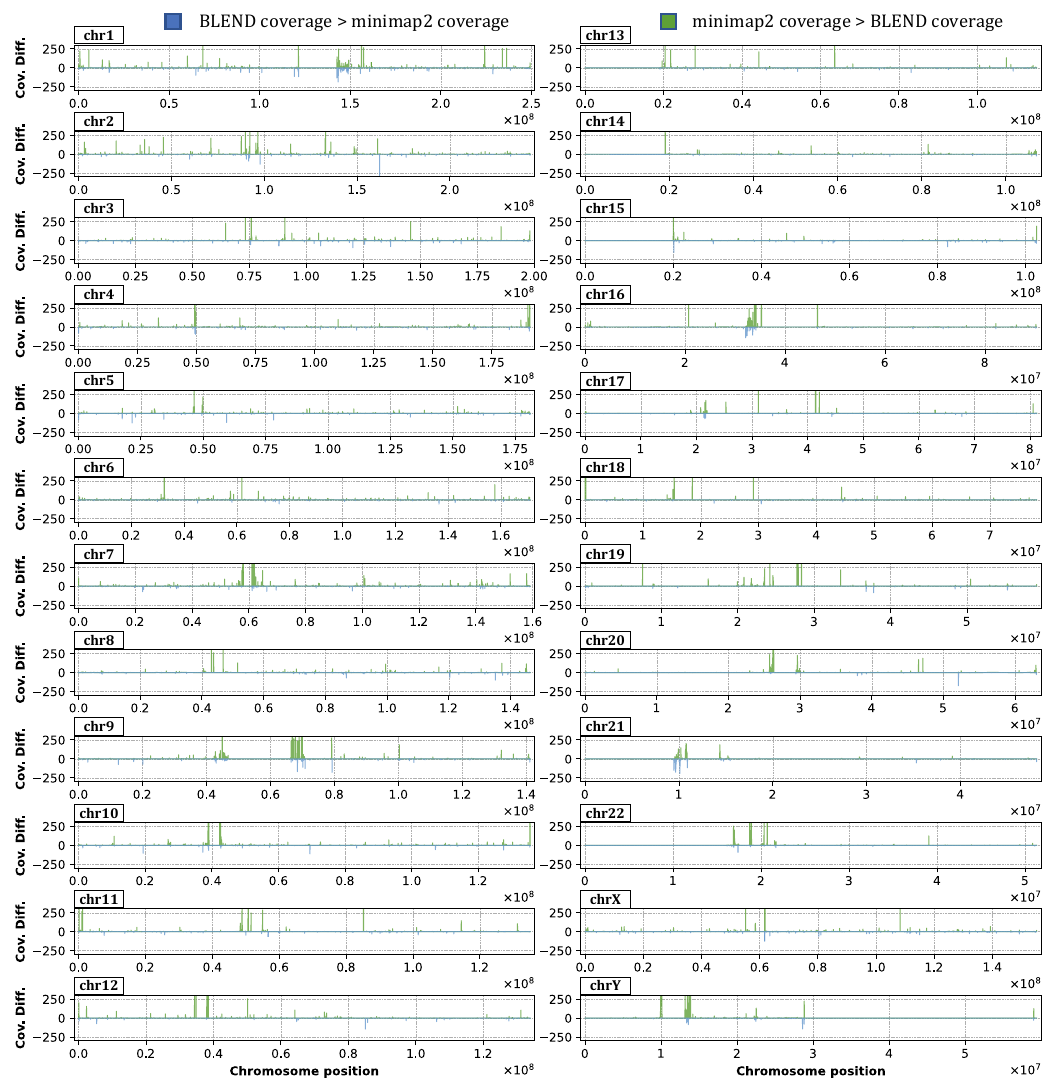}
\caption{Difference between the depth of coverage of minimap2 and \proposal. Positive values show the positions where minimap2 has higher coverage and negative values show the positions where \proposal has higher coverage. We label the chromosomes on the top left corner of each plot.}
\label{suppfig:genomedepth_diff}
\end{figure}

To find the positions where the depth of coverage significantly differs between \proposal and minimap2, we subtract the minimap2 coverage from the \proposal coverage for each chromosome position that we show in Figures~\ref{suppfig:genomedepth-1} and \ref{suppfig:genomedepth-13}. We show the coverage differences in Figure~\ref{suppfig:genomedepth_diff}, where the positive values show the positions that minimap2 has a higher depth of coverage than \proposal, and negative values show the positions that \proposal has a higher coverage.

\clearpage

\section{Parameters and Tool Versions}\label{suppsec:parameters}

Supplementary Table~\ref{supptab:pardef} shows the parameters we use in \proposal and their definition. Since \proposal uses the minimap2 implementation as a baseline, the rest of the parameters we do not show in Supplementary Table~\ref{supptab:pardef} can be found on the manual page of minimap2\footnote{\url{https://lh3.github.io/minimap2/minimap2.html}}. In Supplementary Table~\ref{supptab:ovpars}, we show the parameters we use with \proposal, minimap2, and MHAP~\protect\citesupp{supp_berlin_assembling_2015} for read overlapping. Since there are no default parameters for minimap2 and MHAP when using the HiFi reads, we used the parameters as suggested by the HiCanu tool~\protect\citesupp{supp_nurk_hicanu_2020}. We found these parameters in the source code of Canu. For minimap2 and MHAP, the HiFi parameters are found in the GitHub pages\footnote{\url{https://github.com/marbl/canu/blob/404540a944664cfab00617f4f4fa37be451b34e0/src/pipelines/canu/OverlapMMap.pm\#L63-L65}}$^,$\footnote{\url{https://github.com/marbl/canu/blob/404540a944664cfab00617f4f4fa37be451b34e0/src/pipelines/canu/OverlapMhap.pm\#L100-L131}}, respectively. In Supplementary Table~\ref{supptab:ovpars}, \emph{minimap-Eq} shows the parameters that are equivalent to the parameters we use with \proposal without the fuzzy seed matching capability.

In Supplementary Table~\ref{supptab:mappars}, we show the parameters we use with \proposal, minimap2~\protect\citesupp{supp_li_minimap2_2018}, LRA~\protect\citesupp{supp_ren_lra_2021}, Winnowmap2~\protect\citesupp{supp_jain_long-read_2022, supp_jain_weighted_2020}, S-conLSH~\protect\citesupp{supp_chakraborty_conlsh_2020, supp_chakraborty_s-conlsh_2021}, and Strobealign~\protect\citesupp{supp_sahlin_flexible_2022} for read mapping.

In Supplementary Table~\ref{supptab:version}, we show the version numbers of each tool. When calculating the performance and peak memory usage, we use the time command from Linux and append the following command to the beginning of each of our runs: \texttt{/usr/bin/time -vp}.

\input{tables/supp_pardef.tex}

\input{tables/supp_ovpars.tex}

\input{tables/supp_mappars.tex}

\input{tables/supp_version.tex}

\clearpage

\let\noopsort\undefined
\let\printfirst\undefined
\let\singleletter\undefined
\let\switchargs\undefined

\bibliographystylesupp{IEEEtran}
\bibliographysupp{main}

%% file: tables/supp_min_fuzzy_seed_matching.tex
\caption{Fuzzy seed matching statistics of minimizer seeds that we find using minimap2 and \proposal. The number of overlapping k-mers that \proposal extracts from seed sequences (i.e., neighbors or \emph{n}) are annotated as \proposal-n}
\begin{tabular}{@{}lrrrr@{}}\toprule
\textbf{Tool} & \textbf{Number of}  & \textbf{Number of} & \textbf{Collision/Minimizer} & \textbf{Avg. Edit Distance} \\	
 	          & \textbf{Minimizers} & \textbf{Collisions}       & \textbf{Ratio}               & \textbf{Between Minimizers} \\
 	          &                     &                               &                              & \textbf{With Collision}     \\\midrule
minimap2      & 903,043             & 15,306                        & 0.016949                     & 9.327061                    \\\midrule
\proposal-3   & 1,014,173           & 18,224                        & 0.017969                     & 9.393437                    \\
\proposal-5   & 1,090,468           & 20,659                        & 0.018945                     & 9.213660                    \\
\proposal-7   & 1,140,254           & 23,591                        & 0.020689                     & 8.874698                    \\
\proposal-9   & 1,173,198           & 28,411                        & 0.024217                     & 8.495301                    \\
\proposal-11  & 1,186,687           & 35,500                        & 0.029915                     & 8.067549                    \\
\proposal-13  & 1,197,966           & 72,078                        & 0.060167                     & 8.075918                    \\\bottomrule
\end{tabular}

%% file: tables/supp_seq_fuzzy_seed_matching.tex
\begin{table*}[h]
\centering
\caption{Fuzzy k-mer matching statistics of sequences that we find using minimap2 and \proposal. The number of overlapping k-mers that \proposal extracts from seed sequences (i.e., neighbors or \emph{n}) are annotated as \proposal-n}
\begin{tabular}{@{}lrrrr@{}}\toprule
\textbf{Tool} & \textbf{Number of}  & \textbf{Number of Sequences} & \textbf{Collision/Sequence} & \textbf{Avg. Edit Distance} \\	
 	          & \textbf{Sequences}  & \textbf{with Collision}   & \textbf{Ratio}              & \textbf{Between K-mers}  \\
 	          &                     &                           &                             & \textbf{With Collision}     \\\midrule
minimap2      & 4,130               & 0                         & 0                           & N/A                         \\\midrule
\proposal-3   & 4,130               & 0                         & 0                           & N/A                         \\
\proposal-5   & 4,130               & 11                        & 0.00263663                  & 1.45455                     \\
\proposal-7   & 4,130               & 50                        & 0.0119847                   & 1.5                         \\
\proposal-9   & 4,130               & 77                        & 0.0184564                   & 2.01299                     \\
\proposal-11  & 4,130               & 273                       & 0.0654362                   & 2.80952                     \\
\proposal-13  & 4,130               & 329                       & 0.0788591                   & 2.20669                     \\\bottomrule
\end{tabular}
\label{supptab:seq_fuzzy_seed_matching}
\end{table*}

%% file: tables/supp_hash_values_stepbystep.tex
\begin{table*}[hbt]
\centering
\caption{Hash Values of the k-mers of seed $S_k$: \texttt{CGGATGCTACAGTATATACCA} for $k=7$ and $n=15$. We show the most significant 16 bits of the counter vector $C(S_k)$. Last row shows the most significant 16 bits of the hash value of the seed.}
\vspace{0.2cm}
\resizebox{\linewidth}{!}{
\begin{tabular}{@{}llrrrrrrrrrrrrrrrr@{}}\toprule
\textbf{K-mer} & \textbf{Hash Value} & C[31] & C[30] & C[29] & C[28] & C[27] & C[26] & C[25] & C[24] & C[23] & C[22] & C[21] & C[20] & C[19] & C[18] & C[17] & C[16]\\\midrule
\texttt{CGGATGC} & \texttt{0b 10100000 01111111 10000110 10110101} & 1 & -1 & 1 & -1 & -1 & -1 & -1 & -1 & -1 & 1  & 1  & 1  & 1  & 1  & 1  & 1 \\
\texttt{GGATGCT} & \texttt{0b 10101101 11110000 01110100 11010000} & 2 & -2 & 2 & -2 & 0  & 0  & -2 & 0  & 0  & 2  & 2  & 2  & 0  & 0  & 0  & 0 \\
\texttt{GATGCTA} & \texttt{0b 01000010 01001011 11011001 10011011} & 1 & -1 & 1 & -3 & -1 & -1 & -1 & -1 & -1 & 3  & 1  & 1  & 1  & -1 & 1  & 1 \\
\texttt{ATGCTAC} & \texttt{0b 11001100 01110101 01010011 00100110} & 2 &  0 & 0 & -4 & 0  & 0  & -2 & -2 & -2 & 4  & 2  & 2  & 0  & 0  & 0  & 2 \\
\texttt{TGCTACA} & \texttt{0b 10110001 01101010 10101001 10100111} & 3 & -1 & 1 & -3 & -1 & -1 & -3 & -1 & -3 & 5  & 3  & 1  & 1  & -1 & 1  & 1 \\
\texttt{GCTACAG} & \texttt{0b 11000101 11010111 11010101 00100101} & 4 & 0 & 0  & -4 & -2 & 0  & -4 & 0  & -2 & 6  & 2  & 2  & 0  & 0  & 2  & 2 \\
\texttt{CTACAGT} & \texttt{0b 11001001 01111101 01001010 10110101} & 5 & 1 & -1 & -5 & -1 & -1 & -5 & 1  & -3 & 7  & 3  & 3  & 1  & 1  & 1  & 3 \\
\texttt{TACAGTA} & \texttt{0b 00101011 01101111 11111000 11111000} & 4 & 0 & 0  & -6 & 0  & -2 & -4 & 2  & -4 & 8  & 4  & 2  & 2  & 2  & 2  & 4 \\
\texttt{ACAGTAT} & \texttt{0b 11100100 01001110 01110101 00011010} & 5 & 1 & 1  & -7 & -1 & -1 & -5 & 1  & -5 & 9  & 3  & 1  & 3  & 3  & 3  & 3 \\
\texttt{CAGTATA} & \texttt{0b 10010010 00001101 00100011 10110100} & 6 & 0 & 0  & -6 & -2 & -2 & -4 & 0  & -6 & 8  & 2  & 0  & 4  & 4  & 2  & 4 \\
\texttt{AGTATAT} & \texttt{0b 01110100 00110000 10101100 00000000} & 5 & 1 & 1  & -5 & -3 & -1 & -5 & -1 & -7 & 7  & 3  & 1  & 3  & 3  & 1  & 3 \\
\texttt{GTATATA} & \texttt{0b 11001111 10110000 11001001 10010110} & 6 & 2 & 0  & -6 & -2 & 0  & -4 & 0  & -6 & 6  & 4  & 2  & 2  & 2  & 0  & 2 \\
\texttt{TATATAC} & \texttt{0b 10000001 00001000 00101111 01111111} & 7 & 1 & -1 & -7 & -3 & -1 & -5 & 1  & -7 & 5  & 3  & 1  & 3  & 1  & -1 & 1 \\
\texttt{ATATACC} & \texttt{0b 11001100 11100000 00101000 11011010} & 8 & 2 & -2 & -8 & -2 & 0  & -6 & 0  & -6 & 6  & 4  & 0  & 2  & 0  & -2 & 0 \\
\texttt{TATACCA} & \texttt{0b 00110100 00000100 11110100 10010100} & 7 & 1 & -1 & -7 & -3 & 1  & -7 & -1 & -7 & 5  & 3  & -1 & 1  & 1  & -3 & -1 \\\cmidrule{1-18}
& & B[31] & B[30] & B[29] & B[28] & B[27] & B[26] & B[25] & B[24] & B[23] & B[22] & B[21] & B[20] & B[19] & B[18] & B[17] & B[16]\\\midrule
& & 1 & 1 & 0 & 0 & 0 & 1 & 0 & 0 & 0 & 1 & 1 & 0 & 1 & 1 & 0 & 0 \\\bottomrule\\
\multicolumn{18}{l}{\footnotesize Best results are highlighted with \textbf{bold} text.} \\
\end{tabular}}
\label{supptab:kmers7-15-1-32}
\end{table*}

\begin{table*}[hbt]
\centering
\caption{Hash Values of the k-mers of seed $S_k$: \texttt{CGGATGCTACAGTATATACCA} for $k=7$ and $n=15$. We show the least significant 16 bits of the counter vector $C(S_k)$. Last row shows the least significant 16 bits of the hash value of the seed.}
\vspace{0.2cm}
\resizebox{\linewidth}{!}{
\begin{tabular}{@{}llrrrrrrrrrrrrrrrr@{}}\toprule
\textbf{K-mer} & \textbf{Hash Value} & C[15] & C[14] & C[13] & C[12] & C[11] & C[10] & C[9] & C[8] & C[7] & C[6] & C[5] & C[4] & C[3] & C[2] & C[1] & C[0]\\\midrule
\texttt{CGGATGC} & \texttt{0b 10100000 01111111 10000110 10110101} & 1  & -1 & -1 & -1 & -1 & 1  & 1  & -1 & 1  & -1 & 1  & 1  & -1 & 1  & -1 & 1 \\
\texttt{GGATGCT} & \texttt{0b 10101101 11110000 01110100 11010000} & 0  & 0  & 0  & 0  & -2 & 2  & 0  & -2 & 2  & 0  & 0  & 2  & -2 & 0  & -2 & 0 \\
\texttt{GATGCTA} & \texttt{0b 01000010 01001011 11011001 10011011} & 1  & 1  & -1 & 1  & -1 & 1  & -1 & -1 & 3  & -1 & -1 & 3  & -1 & -1 & -1 & 1 \\
\texttt{ATGCTAC} & \texttt{0b 11001100 01110101 01010011 00100110} & 0  & 2  & -2 & 2  & -2 & 0  & 0  & 0  & 2  & -2 & 0  & 2  & -2 & 0  & 0  & 0 \\
\texttt{TGCTACA} & \texttt{0b 10110001 01101010 10101001 10100111} & 1  & 1  & -1 & 1  & -1 & -1 & -1 & 1  & 3  & -3 & 1  & 1  & -3 & 1  & 1  & 1 \\
\texttt{GCTACAG} & \texttt{0b 11000101 11010111 11010101 00100101} & 2  & 2  & -2 & 2  & -2 & 0  & -2 & 2  & 2  & -4 & 2  & 0  & -4 & 2  & 0  & 2 \\
\texttt{CTACAGT} & \texttt{0b 11001001 01111101 01001010 10110101} & 1  & 3  & -3 & 1  & -1 & -1 & -1 & 1  & 3  & -5 & 3  & 1  & -5 & 3  & -1 & 3 \\
\texttt{TACAGTA} & \texttt{0b 00101011 01101111 11111000 11111000} & 2  & 4  & -2 & 2  & 0  & -2 & -2 & 0  & 4  & -4 & 4  & 2  & -4 & 2  & -2 & 2 \\
\texttt{ACAGTAT} & \texttt{0b 11100100 01001110 01110101 00011010} & 1  & 5  & -1 & 3  & -1 & -1 & -3 & 1  & 3  & -5 & 3  & 3  & -3 & 1  & -1 & 1 \\
\texttt{CAGTATA} & \texttt{0b 10010010 00001101 00100011 10110100} & 0  & 4  & 0  & 2  & -2 & -2 & -2 & 2  & 4  & -6 & 4  & 4  & -4 & 2  & -2 & 0 \\
\texttt{AGTATAT} & \texttt{0b 01110100 00110000 10101100 00000000} & 1  & 3  & 1  & 1  & -1 & -1 & -3 & 1  & 3  & -7 & 3  & 3  & -5 & 1  & -3 & -1 \\
\texttt{GTATATA} & \texttt{0b 11001111 10110000 11001001 10010110} & 2  & 4  & 0  & 0  & 0  & -2 & -4 & 2  & 4  & -8 & 2  & 4  & -6 & 2  & -2 & -2 \\
\texttt{TATATAC} & \texttt{0b 10000001 00001000 00101111 01111111} & 1  & 3  & 1  & -1 & 1  & -1 & -3 & 3  & 3  & -7 & 3  & 5  & -5 & 3  & -1 & -1 \\
\texttt{ATATACC} & \texttt{0b 11001100 11100000 00101000 11011010} & 0  & 2  & 2  & -2 & 2  & -2 & -4 & 2  & 4  & -6 & 2  & 6  & -4 & 2  & 0  & -2 \\
\texttt{TATACCA} & \texttt{0b 00110100 00000100 11110100 10010100} & 1  & 3  & 3  & -1 & 1  & -1 & -5 & 1  & 5  & -7 & 1  & 7  & -5 & 3  & -1 & -3 \\\cmidrule{1-18}
& & B[15] & B[14] & B[13] & B[12] & B[11] & B[10] & B[9] & B[8] & B[7] & B[6] & B[5] & B[4] & B[3] & B[2] & B[1] & B[0]\\\midrule
& & 1 & 1 & 1 & 0 & 1 & 0 & 0 & 1 & 1 & 0 & 1 & 1 & 0 & 1 & 0 & 0 \\\bottomrule\\
\multicolumn{18}{l}{\footnotesize Best results are highlighted with \textbf{bold} text.} \\
\end{tabular}}
\label{supptab:kmers7-15-1-16}
\end{table*}

\begin{table*}[hbt]
\centering
\caption{Hash Values of the k-mers of seed $S_l$: \texttt{ATGCTACAGTATATACCATCT} for $k=7$ and $n=15$. We show the most significant 16 bits of the counter vector $C(S_l)$. Last row shows the most significant 16 bits of the hash value of the seed.}
\vspace{0.2cm}
\resizebox{\linewidth}{!}{
\begin{tabular}{@{}llrrrrrrrrrrrrrrrr@{}}\toprule
\textbf{K-mer} & \textbf{Hash Value} & C[31] & C[30] & C[29] & C[28] & C[27] & C[26] & C[25] & C[24] & C[23] & C[22] & C[21] & C[20] & C[19] & C[18] & C[17] & C[16]\\\midrule
\texttt{ATGCTAC} & \texttt{0b 11001100 01110101 01010011 00100110} & 1  & 1  & -1 & -1 & 1  & 1  & -1 & -1 & -1 & 1  & 1  & 1  & -1 & 1  & -1 & 1 \\
\texttt{TGCTACA} & \texttt{0b 10110001 01101010 10101001 10100111} & 2  & 0  & 0  & 0  & 0  & 0  & -2 & 0  & -2 & 2  & 2  & 0  & 0  & 0  & 0  & 0 \\
\texttt{GCTACAG} & \texttt{0b 11000101 11010111 11010101 00100101} & 3  & 1  & -1 & -1 & -1 & 1  & -3 & 1  & -1 & 3  & 1  & 1  & -1 & 1  & 1  & 1 \\
\texttt{CTACAGT} & \texttt{0b 11001001 01111101 01001010 10110101} & 4  & 2  & -2 & -2 & 0  & 0  & -4 & 2  & -2 & 4  & 2  & 2  & 0  & 2  & 0  & 2 \\
\texttt{TACAGTA} & \texttt{0b 00101011 01101111 11111000 11111000} & 3  & 1  & -1 & -3 & 1  & -1 & -3 & 3  & -3 & 5  & 3  & 1  & 1  & 3  & 1  & 3 \\
\texttt{ACAGTAT} & \texttt{0b 11100100 01001110 01110101 00011010} & 4  & 2  & 0  & -4 & 0  & 0  & -4 & 2  & -4 & 6  & 2  & 0  & 2  & 4  & 2  & 2 \\
\texttt{CAGTATA} & \texttt{0b 10010010 00001101 00100011 10110100} & 5  & 1  & -1 & -3 & -1 & -1 & -3 & 1  & -5 & 5  & 1  & -1 & 3  & 5  & 1  & 3 \\
\texttt{AGTATAT} & \texttt{0b 01110100 00110000 10101100 00000000} & 4  & 2  & 0  & -2 & -2 & 0  & -4 & 0  & -6 & 4  & 2  & 0  & 2  & 4  & 0  & 2 \\
\texttt{GTATATA} & \texttt{0b 11001111 10110000 11001001 10010110} & 5  & 3  & -1 & -3 & -1 & 1  & -3 & 1  & -5 & 3  & 3  & 1  & 1  & 3  & -1 & 1 \\
\texttt{TATATAC} & \texttt{0b 10000001 00001000 00101111 01111111} & 6  & 2  & -2 & -4 & -2 & 0  & -4 & 2  & -6 & 2  & 2  & 0  & 2  & 2  & -2 & 0 \\
\texttt{ATATACC} & \texttt{0b 11001100 11100000 00101000 11011010} & 7  & 3  & -3 & -5 & -1 & 1  & -5 & 1  & -5 & 3  & 3  & -1 & 1  & 1  & -3 & -1 \\
\texttt{TATACCA} & \texttt{0b 00110100 00000100 11110100 10010100} & 6  & 2  & -2 & -4 & -2 & 2  & -6 & 0  & -6 & 2  & 2  & -2 & 0  & 2  & -4 & -2 \\
\texttt{ATACCAT} & \texttt{0b 00000111 10111111 11010101 01001100} & 5  & 1  & -3 & -5 & -3 & 3  & -5 & 1  & -5 & 1  & 3  & -1 & 1  & 3  & -3 & -1 \\
\texttt{TACCATC} & \texttt{0b 01010110 11100111 00100010 11001101} & 4  & 2  & -4 & -4 & -4 & 4  & -4 & 0  & -4 & 2  & 4  & -2 & 0  & 4  & -2 & 0 \\
\texttt{ACCATCT} & \texttt{0b 00010010 11001000 11001010 11100111} & 3  & 1  & -5 & -3 & -5 & 3  & -3 & -1 & -3 & 3  & 3  & -3 & 1  & 3  & -3 & -1 \\\cmidrule{1-18}
& & B[31] & B[30] & B[29] & B[28] & B[27] & B[26] & B[25] & B[24] & B[23] & B[22] & B[21] & B[20] & B[19] & B[18] & B[17] & B[16]\\\midrule
& & 1 & 1 & 0 & 0 & 0 & 1 & 0 & 0 & 0 & 1 & 1 & 0 & 1 & 1 & 0 & 0 \\\bottomrule\\
\multicolumn{18}{l}{\footnotesize Best results are highlighted with \textbf{bold} text.} \\
\end{tabular}}
\label{supptab:kmers7-15-2-32}
\end{table*}

\begin{table*}[hbt]
\centering
\caption{Hash Values of the k-mers of seed $S_l$: \texttt{ATGCTACAGTATATACCATCT} for $k=7$ and $n=15$. We show the least significant 16 bits of the counter vector $C(S_l)$. Last row shows the least significant 16 bits of the hash value of the seed.}
\vspace{0.2cm}
\resizebox{\linewidth}{!}{
\begin{tabular}{@{}llrrrrrrrrrrrrrrrr@{}}\toprule
\textbf{K-mer} & \textbf{Hash Value} & C[15] & C[14] & C[13] & C[12] & C[11] & C[10] & C[9] & C[8] & C[7] & C[6] & C[5] & C[4] & C[3] & C[2] & C[1] & C[0]\\\midrule
\texttt{ATGCTAC} & \texttt{0b 11001100 01110101 01010011 00100110} & -1 & 1  & -1 & 1  & -1 & -1 & 1  & 1  & -1 & -1 & 1  & -1 & -1 & 1  & 1  & -1  \\
\texttt{TGCTACA} & \texttt{0b 10110001 01101010 10101001 10100111} & 0  & 0  & 0  & 0  & 0  & -2 & 0  & 2  & 0  & -2 & 2  & -2 & -2 & 2  & 2  & 0  \\
\texttt{GCTACAG} & \texttt{0b 11000101 11010111 11010101 00100101} & 1  & 1  & -1 & 1  & -1 & -1 & -1 & 3  & -1 & -3 & 3  & -3 & -3 & 3  & 1  & 1  \\
\texttt{CTACAGT} & \texttt{0b 11001001 01111101 01001010 10110101} & 0  & 2  & -2 & 0  & 0  & -2 & 0  & 2  & 0  & -4 & 4  & -2 & -4 & 4  & 0  & 2  \\
\texttt{TACAGTA} & \texttt{0b 00101011 01101111 11111000 11111000} & 1  & 3  & -1 & 1  & 1  & -3 & -1 & 1  & 1  & -3 & 5  & -1 & -3 & 3  & -1 & 1  \\
\texttt{ACAGTAT} & \texttt{0b 11100100 01001110 01110101 00011010} & 0  & 4  & 0  & 2  & 0  & -2 & -2 & 2  & 0  & -4 & 4  & 0  & -2 & 2  & 0  & 0  \\
\texttt{CAGTATA} & \texttt{0b 10010010 00001101 00100011 10110100} & -1 & 3  & 1  & 1  & -1 & -3 & -1 & 3  & 1  & -5 & 5  & 1  & -3 & 3  & -1 & -1  \\
\texttt{AGTATAT} & \texttt{0b 01110100 00110000 10101100 00000000} & 0  & 2  & 2  & 0  & 0  & -2 & -2 & 2  & 0  & -6 & 4  & 0  & -4 & 2  & -2 & -2  \\
\texttt{GTATATA} & \texttt{0b 11001111 10110000 11001001 10010110} & 1  & 3  & 1  & -1 & 1  & -3 & -3 & 3  & 1  & -7 & 3  & 1  & -5 & 3  & -1 & -3 \\
\texttt{TATATAC} & \texttt{0b 10000001 00001000 00101111 01111111} & 0  & 2  & 2  & -2 & 2  & -2 & -2 & 4  & 0  & -6 & 4  & 2  & -4 & 4  & 0  & -2  \\
\texttt{ATATACC} & \texttt{0b 11001100 11100000 00101000 11011010} & -1 & 1  & 3  & -3 & 3  & -3 & -3 & 3  & 1  & -5 & 3  & 3  & -3 & 3  & 1  & -3  \\
\texttt{TATACCA} & \texttt{0b 00110100 00000100 11110100 10010100} & 0  & 2  & 4  & -2 & 2  & -2 & -4 & 2  & 2  & -6 & 2  & 4  & -4 & 4  & 0  & -4  \\
\texttt{ATACCAT} & \texttt{0b 00000111 10111111 11010101 01001100} & 1  & 3  & 3  & -1 & 1  & -1 & -5 & 3  & 1  & -5 & 1  & 3  & -3 & 5  & -1 & -5  \\
\texttt{TACCATC} & \texttt{0b 01010110 11100111 00100010 11001101} & 0  & 2  & 4  & -2 & 0  & -2 & -4 & 2  & 2  & -4 & 0  & 2  & -2 & 6  & -2 & -4 \\
\texttt{ACCATCT} & \texttt{0b 00010010 11001000 11001010 11100111} & 1  & 3  & 3  & -3 & 1  & -3 & -3 & 1  & 3  & -3 & 1  & 1  & -3 & 7  & -1 & -3 \\\cmidrule{1-18}
& & B[15] & B[14] & B[13] & B[12] & B[11] & B[10] & B[9] & B[8] & B[7] & B[6] & B[5] & B[4] & B[3] & B[2] & B[1] & B[0]\\\midrule
& & 1 & 1 & 1 & 0 & 1 & 0 & 0 & 1 & 1 & 0 & 1 & 1 & 0 & 1 & 0 & 0 \\\bottomrule\\
\multicolumn{18}{l}{\footnotesize Best results are highlighted with \textbf{bold} text.}
\end{tabular}}
\label{supptab:kmers7-15-2-16}
\end{table*}


\begin{table*}[hbt]
\centering
\caption{Hash Values of the k-mers of seed $S_k$: \texttt{CGGATGCTACAGTATATACCA} for $k=15$ and $n=7$. We show the most significant 16 bits of the counter vector $C(S_k)$. Last row shows the most significant 16 bits of the hash value of the seed.}
\vspace{0.2cm}
\resizebox{\linewidth}{!}{
\begin{tabular}{@{}llrrrrrrrrrrrrrrrr@{}}\toprule
\textbf{K-mer} & \textbf{Hash Value} & C[31] & C[30] & C[29] & C[28] & C[27] & C[26] & C[25] & C[24] & C[23] & C[22] & C[21] & C[20] & C[19] & C[18] & C[17] & C[16]\\\midrule
\texttt{CGGATGCTACAGTAT} & \texttt{0b 01001010 11101011 00100110 11001101} & -1 & 1  & -1 & -1 & 1  & -1 & 1  & -1 & 1  & 1  & 1  & -1 & 1  & -1 & 1  & 1  \\
\texttt{GGATGCTACAGTATA} & \texttt{0b 01101100 01000011 11111000 11000000} & -2 & 2  & 0  & -2 & 2  & 0  & 0  & -2 & 0  & 2  & 0  & -2 & 0  & -2 & 2  & 2  \\
\texttt{GATGCTACAGTATAT} & \texttt{0b 01011000 01000101 00110001 11011000} & -3 & 3  & -1 & -1 & 3  & -1 & -1 & -3 & -1 & 3  & -1 & -3 & -1 & -1 & 1  & 3  \\
\texttt{ATGCTACAGTATATA} & \texttt{0b 11100001 01110100 01100010 01000010} & -2 & 4  & 0  & -2 & 2  & -2 & -2 & -2 & -2 & 4  & 0  & -2 & -2 & 0  & 0  & 2  \\
\texttt{TGCTACAGTATATAC} & \texttt{0b 10111100 10011010 00111111 01011011} & -1 & 3  & 1  & -1 & 3  & -1 & -3 & -3 & -1 & 3  & -1 & -1 & -1 & -1 & 1  & 1  \\
\texttt{GCTACAGTATATACC} & \texttt{0b 11101010 01001100 01000100 11100001} & 0  & 2  & 2  & -2 & 4  & -2 & -2 & -4 & -2 & 4  & -2 & -2 & 0  & 0  & 0  & 0  \\
\texttt{CTACAGTATATACCA} & \texttt{0b 00101001 10010001 11111100 01010000} & -1 & 1  & 3  & -3 & 5  & -3 & -3 & -3 & -1 & 3  & -3 & -1 & -1 & -1 & -1 & 1  \\\cmidrule{1-18}
\multicolumn{2}{l}{Seed \texttt{CGGATGCTACAGTATATACCA}} & B[31] & B[30] & B[29] & B[28] & B[27] & B[26] & B[25] & B[24] & B[23] & B[22] & B[21] & B[20] & B[19] & B[18] & B[17] & B[16]\\\midrule
& & 0 & 1 & 1 & 0 & 1 & 0 & 0 & 0 & 0 & 1 & 0 & 0 & 0 & 0 & 0 & 1 \\\bottomrule\\
\multicolumn{18}{l}{\footnotesize Best results are highlighted with \textbf{bold} text.}
\end{tabular}}
\label{supptab:kmers15-7-1-32}
\end{table*}

\begin{table*}[hbt]
\centering
\caption{Hash Values of the k-mers of seed $S_k$: \texttt{CGGATGCTACAGTATATACCA} for $k=15$ and $n=7$. We show the least significant 16 bits of the counter vector $C(S_k)$. Last row shows the least significant 16 bits of the hash value of the seed.}
\vspace{0.2cm}
\resizebox{\linewidth}{!}{
\begin{tabular}{@{}llrrrrrrrrrrrrrrrr@{}}\toprule
\textbf{K-mer} & \textbf{Hash Value} & C[15] & C[14] & C[13] & C[12] & C[11] & C[10] & C[9] & C[8] & C[7] & C[6] & C[5] & C[4] & C[3] & C[2] & C[1] & C[0]\\\midrule
\texttt{CGGATGCTACAGTAT} & \texttt{0b 01001010 11101011 00100110 11001101} & -1 & -1 & 1  & -1 & -1 & 1  & 1  & -1 & 1  & 1  & -1 & -1 & 1  & 1  & -1 & 1  \\
\texttt{GGATGCTACAGTATA} & \texttt{0b 01101100 01000011 11111000 11000000} & 0  & 0  & 2  & 0  & 0  & 0  & 0  & -2 & 2  & 2  & -2 & -2 & 0  & 0  & -2 & 0  \\
\texttt{GATGCTACAGTATAT} & \texttt{0b 01011000 01000101 00110001 11011000} & -1 & -1 & 3  & 1  & -1 & -1 & -1 & -1 & 3  & 3  & -3 & -1 & 1  & -1 & -3 & -1 \\
\texttt{ATGCTACAGTATATA} & \texttt{0b 11100001 01110100 01100010 01000010} & -2 & 0  & 4  & 0  & -2 & -2 & 0  & -2 & 2  & 4  & -4 & -2 & 0  & -2 & -2 & -2 \\
\texttt{TGCTACAGTATATAC} & \texttt{0b 10111100 10011010 00111111 01011011} & -3 & -1 & 5  & 1  & -1 & -1 & 1  & -1 & 1  & 5  & -5 & -1 & 1  & -3 & -1 & -1 \\
\texttt{GCTACAGTATATACC} & \texttt{0b 11101010 01001100 01000100 11100001} & -4 & 0  & 4  & 0  & -2 & 0  & 0  & -2 & 2  & 6  & -4 & -2 & 0  & -4 & -2 & 0  \\
\texttt{CTACAGTATATACCA} & \texttt{0b 00101001 10010001 11111100 01010000} & -3 & 1  & 5  & 1  & -1 & 1  & -1 & -3 & 1  & 7  & -5 & -1 & -1 & -5 & -3 & -1 \\\cmidrule{1-18}
\multicolumn{2}{l}{Seed \texttt{CGGATGCTACAGTATATACCA}} & B[15] & B[14] & B[13] & B[12] & B[11] & B[10] & B[9] & B[8] & B[7] & B[6] & B[5] & B[4] & B[3] & B[2] & B[1] & B[0]\\\midrule
& & 0 & 1 & 1 & 1 & 0 & 1 & 0 & 0 & 1 & 1 & 0 & 0 & 0 & 0 & 0 & 0 \\\bottomrule\\
\multicolumn{18}{l}{\footnotesize Best results are highlighted with \textbf{bold} text.}
\end{tabular}}
\label{supptab:kmers15-7-1-16}
\end{table*}


\begin{table*}[hbt]
\centering
\caption{Hash Values of the k-mers of seed $S_l$: \texttt{ATGCTACAGTATATACCATCT} for $k=15$ and $n=7$. We show the most significant 16 bits of the counter vector $C(S_l)$. Last row shows the most significant 16 bits of the hash value of the seed.}
\vspace{0.2cm}
\resizebox{\linewidth}{!}{
\begin{tabular}{@{}llrrrrrrrrrrrrrrrr@{}}\toprule
\textbf{K-mer} & \textbf{Hash Value} & C[31] & C[30] & C[29] & C[28] & C[27] & C[26] & C[25] & C[24] & C[23] & C[22] & C[21] & C[20] & C[19] & C[18] & C[17] & C[16]\\\midrule
\texttt{ATGCTACAGTATATA} & \texttt{0b 11100001 01110100 01100010 01000010} & 1  & 1  & 1  & -1 & -1 & -1 & -1 & 1  & -1 & 1  & 1  & 1  & -1 & 1  & -1 & -1  \\
\texttt{TGCTACAGTATATAC} & \texttt{0b 10111100 10011010 00111111 01011011} & 2  & 0  & 2  & 0  & 0  & 0  & -2 & 0  & 0  & 0  & 0  & 2  & 0  & 0  & 0  & -2  \\
\texttt{GCTACAGTATATACC} & \texttt{0b 11101010 01001100 01000100 11100001} & 3  & 1  & 3  & -1 & 1  & -1 & -1 & -1 & -1 & 1  & -1 & 1  & 1  & 1  & -1 & -3  \\
\texttt{CTACAGTATATACCA} & \texttt{0b 00101001 10010001 11111100 01010000} & 2  & 0  & 4  & -2 & 2  & -2 & -2 & 0  & 0  & 0  & -2 & 2  & 0  & 0  & -2 & -2  \\
\texttt{TACAGTATATACCAT} & \texttt{0b 00001110 00100000 11011100 11110110} & 1  & -1 & 3  & -3 & 3  & -1 & -1 & -1 & -1 & -1 & -1 & 1  & -1 & -1 & -3 & -3  \\
\texttt{ACAGTATATACCATC} & \texttt{0b 00101111 10111010 00010000 11011111} & 0  & -2 & 4  & -4 & 4  & 0  & 0  & 0  & 0  & -2 & 0  & 2  & 0  & -2 & -2 & -4  \\
\texttt{CAGTATATACCATCT} & \texttt{0b 01100101 11100111 10111011 00111011} & -1 & -1 & 5  & -5 & 5  & 1  & -1 & 1  & 1  & -1 & 1  & 1  & -1 & -1 & -1 & -3 \\\cmidrule{1-18}
\multicolumn{2}{l}{Seed \texttt{ATGCTACAGTATATACCATCT}} & B[31] & B[30] & B[29] & B[28] & B[27] & B[26] & B[25] & B[24] & B[23] & B[22] & B[21] & B[20] & B[19] & B[18] & B[17] & B[16]\\\midrule
& & 0 & 0 & 1 & 0 & 1 & 1 & 0 & 1 & 1 & 0 & 1 & 1 & 0 & 0 & 0 & 0 \\\bottomrule\\
\multicolumn{18}{l}{\footnotesize Best results are highlighted with \textbf{bold} text.}
\end{tabular}}
\label{supptab:kmers15-7-2-32}
\end{table*}

\begin{table*}[hbt]
\centering
\caption{Hash Values of the k-mers of seed $S_l$: \texttt{ATGCTACAGTATATACCATCT} for $k=15$ and $n=7$. We show the least significant 16 bits of the counter vector $C(S_l)$. Last row shows the least significant 16 bits of the hash value of the seed.}
\vspace{0.2cm}
\resizebox{\linewidth}{!}{
\begin{tabular}{@{}llrrrrrrrrrrrrrrrr@{}}\toprule
\textbf{K-mer} & \textbf{Hash Value} & C[15] & C[14] & C[13] & C[12] & C[11] & C[10] & C[9] & C[8] & C[7] & C[6] & C[5] & C[4] & C[3] & C[2] & C[1] & C[0]\\\midrule
\texttt{ATGCTACAGTATATA} & \texttt{0b 11100001 01110100 01100010 01000010} & -1 & 1  & 1  & -1 & -1 & -1 & 1  & -1 & -1 & 1  & -1 & -1 & -1 & -1 & 1  & -1 \\
\texttt{TGCTACAGTATATAC} & \texttt{0b 10111100 10011010 00111111 01011011} & -2 & 0  & 2  & 0  & 0  & 0  & 2  & 0  & -2 & 2  & -2 & 0  & 0  & -2 & 2  & 0  \\
\texttt{GCTACAGTATATACC} & \texttt{0b 11101010 01001100 01000100 11100001} & -3 & 1  & 1  & -1 & -1 & 1  & 1  & -1 & -1 & 3  & -1 & -1 & -1 & -3 & 1  & 1  \\
\texttt{CTACAGTATATACCA} & \texttt{0b 00101001 10010001 11111100 01010000} & -2 & 2  & 2  & 0  & 0  & 2  & 0  & -2 & -2 & 4  & -2 & 0  & -2 & -4 & 0  & 0  \\
\texttt{TACAGTATATACCAT} & \texttt{0b 00001110 00100000 11011100 11110110} & -1 & 3  & 1  & 1  & 1  & 3  & -1 & -3 & -1 & 5  & -1 & 1  & -3 & -3 & 1  & -1  \\
\texttt{ACAGTATATACCATC} & \texttt{0b 00101111 10111010 00010000 11011111} & -2 & 2  & 0  & 2  & 0  & 2  & -2 & -4 & 0  & 6  & -2 & 2  & -2 & -2 & 2  & 0  \\
\texttt{CAGTATATACCATCT} & \texttt{0b 01100101 11100111 10111011 00111011} & -1 & 1  & 1  & 3  & 1  & 1  & -1 & -3 & -1 & 5  & -1 & 3  & -1 & -3 & 3  & 1 \\\cmidrule{1-18}
\multicolumn{2}{l}{Seed \texttt{ATGCTACAGTATATACCATCT}} & B[15] & B[14] & B[13] & B[12] & B[11] & B[10] & B[9] & B[8] & B[7] & B[6] & B[5] & B[4] & B[3] & B[2] & B[1] & B[0]\\\midrule
& & 0 & 1 & 1 & 1 & 1 & 1 & 0 & 0 & 0 & 1 & 0 & 1 & 0 & 0 & 1 & 1 \\\bottomrule\\
\multicolumn{18}{l}{\footnotesize Best results are highlighted with \textbf{bold} text.}
\end{tabular}}
\label{supptab:kmers15-7-2-16}
\end{table*}

%% file: tables/supp_assembly_quality_blendi_s.tex
\begin{table}[tb]
\centering
\caption{Assembly quality comparisons between \texttt{\proposal-I} and \texttt{\proposal-S}.}
\resizebox{\linewidth}{!}{
\begin{tabular}{@{}clrrrrrrrrrr@{}}\toprule
\textbf{Dataset} 	  & \textbf{Tool} 		 & \textbf{Average}	      & \textbf{Genome}		    & \textbf{K-mer}	   & \textbf{Aligned}	   & \textbf{Mismatch per} & \textbf{Average} & \textbf{Assembly}	  & \textbf{Largest} 	  & \textbf{NGA50}    & \textbf{NG50}	  \\
					  & 			  		 & \textbf{Identity (\%)} & \textbf{Fraction (\%)}  & \textbf{Compl. (\%)} & \textbf{Length (Mbp)} & \textbf{100Kbp (\#)}  & \textbf{GC (\%)} & \textbf{Length (Mbp)} & \textbf{Contig (Mbp)} & \textbf{(Kbp)}    & \textbf{(Kbp)}    \\\midrule
\emph{CHM13} 	  	  & \texttt{\proposal-I} & 99.7535	              & 96.7203		            & 83.65	               & 3,054.49 			   & 48.49	  	           & \textbf{40.79}	  & \textbf{3,059.29} 	  & \textbf{41.8342} 	  & \textbf{8,507.53} & \textbf{8,508.92} \\
(HiFi)				  & \texttt{\proposal-S} & \textbf{99.8526}	      & \textbf{98.4847}		& \textbf{90.15}	   & \textbf{3,092.54} 	   & \textbf{22.02}	  	   & 40.78	          & 3,095.21 	          & 22.8397 			  & 5,442.25 		  & 5,442.31 		  \\
					  & Reference 	  		 & 100 				      & 100 					& 100 				   & 3,054.83 			   & 0.00 				   & 40.85 			  & 3,054.83 			  & 248.387 			  & 154,260           & 154,260 	   	  \\\midrule
\emph{D. ananassae}   & \texttt{\proposal-I} & 99.6890	              & 97.2290		            & 77.85	               & \textbf{270.218} 	   & 233.18 	           & 41.95	          & 280.388 	          & 5.01099 	          & 356.745           & 356.745           \\
(HiFi)				  & \texttt{\proposal-S} & \textbf{99.7856}	      & \textbf{97.2308}		& \textbf{86.43}	   & 240.391 			   & \textbf{143.13} 	   & \textbf{41.75}	  & \textbf{247.153} 	  & \textbf{6.23256} 	  & \textbf{792.407}  & \textbf{798.913}  \\
					  & Reference 	  		 & 100 				      & 100 					& 100 				   & 213.805 			   & 0.00 				   & 41.81 			  & 213.818 			  & 30.6728 			  & 26,427.4 		  & 26,427.4 		  \\\midrule
\emph{E. coli} 		  & \texttt{\proposal-I} & 99.6902                & \textbf{99.8824} 		& 79.36	               & 5.04157 	           & 17.92	  	           & \textbf{50.52}	  & \textbf{5.04263} 	  & \textbf{4.94601}	  & \textbf{4,025.48} & \textbf{4,946.01} \\
(HiFi)				  & \texttt{\proposal-S} & \textbf{99.8320} 	  & 99.8801 		        & \textbf{87.91}	   & \textbf{5.12155} 	   & \textbf{3.77}	  	   & 50.53	          & 5.12155 			  & 3.41699	              & 3,416.99          & 3,416.99          \\
					  & Reference 	  		 & 100 				      & 100 					& 100 				   & 5.04628 			   & 0.00 				   & 50.52 			  & 5.04628 			  & 4.94446 			  & 4,944.46 		  & 4,944.46 		  \\\midrule
\emph{CHM13} 	  	  & \texttt{\proposal-I} & N/A	   			      & N/A					    & \textbf{29.26}	   & \textbf{2,891.28} 	   & 4,077.53	           & \textbf{41.32}	  & \textbf{2,897.87} 	  & \textbf{25.2071} 	  & \textbf{5,061.52} & \textbf{5,178.59} \\
(ONT)				  & \texttt{\proposal-S} & N/A	   			      & N/A					    & 0 			       & 0.010546 	   		   & \textbf{3,250.70} 	   & 51.30 			  & 0.010548 	          & 0.010548	          & 0                 & 0                 \\
					  & Reference 	  		 & 100 				      & 100 					& 100 			   	   & 3,117.29 			   & 0.00 				   & 40.75 			  & 3,117.29 			  & 248.387 			  & 150,617 	   	  & 150,617 	      \\\midrule
\emph{Yeast} 		  & \texttt{\proposal-I} & 89.1677 	   		      & \textbf{97.0854} 		& \textbf{33.81}	   & 12.3938 	           & \textbf{2,672.37} 	   & 38.84	  		  & \textbf{12.4176} 	  & \textbf{1.54807} 	  & \textbf{635.966}  & \textbf{636.669}  \\
(PacBio)			  & \texttt{\proposal-S} & \textbf{90.3347}       & 83.8814 				& 33.17 			   & \textbf{22.9473} 	   & 4,795.58 			   & \textbf{38.71}   & 22.9523 	          & 0.265118	          & 114.125           & 116.143           \\
					  & Reference 	  		 & 100 				      & 100 					& 100 				   & 12.1571 			   & 0.00 				   & 38.15 			  & 12.1571 			  & 1.53193 			  & 924.431 		  & 924.431 		  \\\midrule
\emph{Yeast} 	      & \texttt{\proposal-I} & 89.6889                & \textbf{99.2974} 	    & \textbf{35.95}	   & \textbf{12.3222} 	   & 2,529.47 	           & 38.64	          & \textbf{12.3225} 	  & \textbf{1.10582} 	  & \textbf{793.046}  & \textbf{793.046}  \\
(ONT)			      & \texttt{\proposal-S} & \textbf{91.0865} 	  & 7.9798		            & 4.90 			       & 0.898565 			   & \textbf{2,006.91} 	   & \textbf{38.35}   & 0.899654 			  & 0.043321	          & 0                 & 0                 \\
					  & Reference 	  		 & 100 				      & 100 					& 100 				   & 12.1571 			   & 0.00 				   & 38.15 			  & 12.1571 			  & 1.53193 			  & 924.431 		  & 924.431 		  \\\midrule
\emph{E. coli} 		  & \texttt{\proposal-I} & 88.5806                & \textbf{96.5238} 		& \textbf{32.32}	   & \textbf{5.90024} 	   & 1,857.56	           & \textbf{49.81}	  & \textbf{6.21598} 	  & \textbf{2.40671}	  & \textbf{769.981}  & \textbf{2,060.4}  \\
(PacBio)			  & \texttt{\proposal-S} & \textbf{90.3551} 	  & 36.6230 				& 17.07 			   & 2.10137 			   & \textbf{1,299.50} 	   & 48.91 			  & 2.10704 	          & 0.095505 	          & 0    	   	      & 0                 \\
					  & Reference 	  		 & 100 				      & 100 					& 100 				   & 5.6394 			   & 0.00 				   & 50.43 			  & 5.6394 				  & 5.54732 			  & 5,547.32 		  & 5,547.32 		  \\\bottomrule
\multicolumn{12}{l}{\footnotesize Best results are highlighted with \textbf{bold} text. For most metrics, the best results are the ones closest to the corresponding value of the reference genome.}\\
\multicolumn{12}{l}{\footnotesize The best results for \emph{Aligned Length} are determined by the highest number within each dataset. We do not highlight the reference results as the best results.}\\
\multicolumn{12}{l}{\footnotesize N/A indicates that we could not generate the corresponding result because tool, QUAST, or dnadiff failed to generate the statistic.} \\
\end{tabular}}
\label{supptab:overlap_assembly-blend}
\end{table}

%% file: tables/supp_mapping_quality_blendi_s.tex
\begin{table}[tb]
\centering
\caption{Read mapping quality comparisons between \texttt{\proposal-I} and \texttt{\proposal-S}.}
\begin{tabular}{@{}llrrrr@{}}\toprule
\textbf{Dataset} 		& \textbf{Tool} 		& \textbf{Average} 			 & \textbf{Breadth of} & \textbf{Aligned} 	 & \textbf{Properly}\\
				 		& 						& \textbf{Depth of} 		 & \textbf{Coverage}   & \textbf{Reads} 	 & \textbf{Paired}\\
				 		& 						& \textbf{Cov. (${\times}$)} & \textbf{(\%)} 	   & \textbf{(\#)} 		 & \textbf{(\%)}\\\midrule
\emph{CHM13} 	        & \texttt{\proposal-I} 	& 16.58 			 		 & 99.991 	   		   & \textbf{3,172,305}  & NA \\
(HiFi)					& \texttt{\proposal-S} 	& 16.58 			 		 & 99.991 	   		   & 3,171,916  		 & NA \\\midrule
\emph{HG002} 	        & \texttt{\proposal-I} 	& \textbf{51.25} 			 & \textbf{92.245} 	   & 6,813,886           & NA \\
(HiFi)					& \texttt{\proposal-S} 	& 11.24 			         & 13.860 	           & \textbf{11,424,762} & NA \\\midrule
\emph{D. ananassae} 	& \texttt{\proposal-I} 	& \textbf{57.51} 			 & 99.650 			   & \textbf{1,249,666}  & NA \\
(HiFi)					& \texttt{\proposal-S} 	& 57.37 					 & \textbf{99.662} 	   & 1,223,388 			 & NA \\\midrule
\emph{E. coli} 			& \texttt{\proposal-I} 	& 99.14 			 		 & 99.897 			   & \textbf{39,064} 	 & NA \\
(HiFi)					& \texttt{\proposal-S} 	& 99.14 			 		 & 99.897 			   & 39,048 			 & NA \\\midrule
\emph{CHM13} 	        & \texttt{\proposal-I} 	& \textbf{29.34} 			 & \textbf{99.999} 	   & \textbf{10,322,767} & NA \\
(ONT)			        & \texttt{\proposal-S} 	& 17.51 			 		 & 99.700 	   		   & 5,760,401 			 & NA \\\midrule
\emph{Yeast} 	        & \texttt{\proposal-I} 	& \textbf{195.87} 			 & \textbf{99.980} 	   & \textbf{270,064}    & NA \\
(PacBio)		        & \texttt{\proposal-S} 	& 142.31 			 		 & 99.975 	   		   & 179,039    		 & NA \\\midrule
\emph{Yeast} 		    & \texttt{\proposal-I} 	& \textbf{97.88} 			 & \textbf{99.964} 	   & \textbf{134,919} 	 & NA \\
(ONT)			        & \texttt{\proposal-S} 	& 59.57 			 		 & 99.906 	   		   & 75,110 	 		 & NA \\\midrule
\emph{E. coli} 			& \texttt{\proposal-I} 	& \textbf{97.51} 			 & 100 			       & \textbf{83,924}     & NA \\
(PacBio)			    & \texttt{\proposal-S} 	& 56.87 			 		 & 100 			       & 40,694 			 & NA \\\bottomrule
\multicolumn{6}{l}{\footnotesize Best results are highlighted with \textbf{bold} text.} \\
\multicolumn{6}{l}{\footnotesize Properly paired rate is only available for paired-end Illumina reads.} \\
\end{tabular}
\label{supptab:mapping_quality-blend}
\end{table}

%% file: tables/supp_mapping_accuracy-blend.tex
\begin{table}[tb]
\centering
\caption{Read mapping accuracy comparisons between \texttt{\proposal-I} and \texttt{\proposal-S}.}
\begin{tabular}{@{}lrrr@{}}\toprule
\textbf{Dataset} 		& \multicolumn{2}{c}{\textbf{Overall Error Rate (\%)}} \\\cmidrule{2-4}
				 		& \texttt{\proposal-I} & \texttt{\proposal-S} \\\midrule
\emph{CHM13} (ONT) 	    & \textbf{1.5168427}   & 5.996888    		  \\\midrule
\emph{Yeast} (PacBio) 	& \textbf{0.2403134}   & 0.6959378            \\\midrule
\emph{Yeast} (ONT) 		& \textbf{0.2386617}   & 0.6284117            \\\bottomrule
\multicolumn{4}{l}{\footnotesize Best results are highlighted with \textbf{bold} text.} \\
\end{tabular}
\label{supptab:mapping_accuracy-blend}
\end{table}

%% file: tables/supp_parameter_exploration.tex
\begin{table*}[tb]
\centering
\caption{Performance, memory, and accuracy comparisons using different parameter settings in \proposal.}
\resizebox{\linewidth}{!}{
\begin{tabular}{@{}lrrrrrrr@{}}\toprule
\textbf{Tool} 	   & \textbf{K-mer} 	   & \textbf{\# of k-mers} 	  & \textbf{Window} 	  & \textbf{CPU Time}  & \textbf{Peak}		  & \textbf{Average} 	   & \textbf{Genome}	    \\
 				   & \textbf{Length ($k$)} & \textbf{in a Seed ($n$)} & \textbf{Length ($w$)} & \textbf{(seconds)} & \textbf{Memory (KB)} & \textbf{Identity (\%)} & \textbf{Fraction (\%)} \\\midrule
\proposal 		   & 9					   & 11						  & 200					  &	62.38			   & 1,115,384			  & 99.7255			  	   & 99.8502	  			\\
\proposal 		   & 9					   & 13						  & 200					  &	58.13			   & 994,120				  & 99.7294			  	   & 99.7808	  			\\
\proposal 		   & 9					   & 15						  & 200					  &	49.79			   & 1,030,148			  & 99.7411			  	   & 99.7619	  			\\
\proposal 		   & 9					   & 17						  & 200					  &	45.03			   & 960,080				  & 99.7302			  	   & 99.7460	  			\\
\proposal 		   & 9					   & 21						  & 200					  &	36.84			   & 976,456				  & 99.7257			  	   & 99.6640	  			\\\cmidrule{1-8}
\proposal 		   & 15					   & 5						  & 200					  &	83.05			   & 1,168,612			  & 99.6735			  	   & 99.7625	  			\\
\proposal 		   & 15					   & 7						  & 200					  &	74.93			   & 1,137,360			  & 99.7009			  	   & 99.5874	  			\\
\proposal 		   & 15					   & 11						  & 200					  &	58.09			   & 1,051,912			  & 99.7149			  	   & 99.1166	  			\\\cmidrule{1-8}
\proposal 		   & 19					   & 5						  & 200					  &	77.16			   & 1,130,604			  & 99.7312			  	   & 99.8802	  			\\
\proposal 		   & 19					   & 7						  & 200					  &	50.50			   & 1,078,596			  & 99.7880			  	   & 99.8424	  			\\
\proposal 		   & 19					   & 11						  & 200					  &	46.26			   & 977,060				  & 99.8078			  	   & 99.6438	  			\\\cmidrule{1-8}
\proposal 		   & 21					   & 5						  & 200					  &	67.85			   & 1,116,684			  & 99.7472			  	   & 99.8835	  			\\
\proposal 		   & 21					   & 7						  & 200					  &	61.63			   & 1,042,724			  & 99.7969			  	   & 99.8605	  			\\
\proposal 		   & 21					   & 11						  & 200					  &	42.35			   & 969,184				  & 99.8340			  	   & 99.7515	  			\\\cmidrule{1-8}
\proposal 		   & 25					   & 5						  & 200					  &	65.61			   & 1,057,804			  & 99.7769			  	   & 99.8818	  			\\
\proposal 		   & 25					   & 7						  & 200					  &	54.88			   & 1,029,888			  & 99.8320			  	   & 99.8801	  			\\
\proposal 		   & 25					   & 11						  & 200					  &	37.01			   & 936,260				  & 99.8646			  	   & 99.8001	  			\\
\proposal 		   & 25					   & 15						  & 200					  &	29.83			   & 866,208				  & 99.8838			  	   & 99.7307	  			\\
\proposal 		   & 25					   & 17						  & 200					  &	29.59			   & 826,456				  & 99.8784			  	   & 99.7521	  			\\
\proposal 		   & 25					   & 21						  & 200					  &	26.09			   & 791,736				  & 99.8774			  	   & 99.6955	  			\\\cmidrule{1-8}
\proposal 		   & 9					   & 11						  & 50					  &	263.82			   & 1,786,516			  & 99.7013			  	   & 99.8612	  			\\
\proposal 		   & 9					   & 13						  & 50					  &	411.24			   & 1,805,800			  & 99.6995			  	   & 99.8573	  			\\
\proposal 		   & 9					   & 15						  & 50					  &	271.00			   & 1,729,784			  & 99.6798			  	   & 99.8517	  			\\
\proposal 		   & 9					   & 17						  & 50					  &	238.52			   & 1,690,912			  & 99.6690			  	   & 99.8083	  			\\
\proposal 		   & 9					   & 21						  & 50					  &	206.76			   & 1,725,168			  & 99.6496			  	   & 99.8150	  			\\\cmidrule{1-8}
\proposal 		   & 15					   & 5						  & 50					  &	330.84			   & 1,785,456			  & 99.6634			  	   & 99.8604	  			\\
\proposal 		   & 15					   & 7						  & 50					  &	337.95			   & 1,812,052			  & 99.6280			  	   & 99.8177	  			\\
\proposal 		   & 15					   & 11						  & 50					  &	236.82			   & 1,803,816			  & 99.5831			  	   & 99.6893	  			\\\cmidrule{1-8}
\proposal 		   & 19					   & 5						  & 50					  &	328.67			   & 1,692,248			  & 99.7077			  	   & 99.8794	  			\\
\proposal 		   & 19					   & 7						  & 50					  &	295.57			   & 1,713,940			  & 99.7188			  	   & 99.8579	  			\\
\proposal 		   & 19					   & 11						  & 50					  &	201.79			   & 1,700,412			  & 99.7015			  	   & 99.8578	  			\\\cmidrule{1-8}
\proposal 		   & 21					   & 5						  & 50					  &	378.58			   & 1,625,388			  & 99.7120			  	   & 99.8832	  			\\
\proposal 		   & 21					   & 7						  & 50					  &	278.56			   & 1,695,476			  & 99.7333			  	   & 99.8832	  			\\
\proposal 		   & 21					   & 11						  & 50					  &	189.33			   & 1,694,820			  & 99.7623			  	   & 99.8594	  			\\\cmidrule{1-8}
\proposal 		   & 25					   & 5						  & 50					  &	323.69			   & 1,685,304			  & 99.7272			  	   & 99.8831	  			\\
\proposal 		   & 25					   & 7						  & 50					  &	211.78			   & 1,647,984			  & 99.7722			  	   & 99.8831	  			\\
\proposal 		   & 25					   & 11						  & 50					  &	170.60			   & 1,683,736			  & 99.8094			  	   & 99.8866	  			\\
\proposal 		   & 25					   & 15						  & 50					  &	142.42			   & 1,622,452			  & 99.8170			  	   & 99.8576	  			\\
\proposal 		   & 25					   & 17						  & 50					  &	103.96			   & 1,590,776			  & 99.8073			  	   & 99.8206	  			\\
\proposal 		   & 25					   & 21						  & 50					  &	109.62			   & 1,548,228			  & 99.7792			  	   & 99.7880	  			\\\cmidrule{1-8}
\proposal 		   & 9					   & 11						  & 20					  &	837.50			   & 2,769,552			  & 99.6916			  	   & 99.8784	  			\\
\proposal 		   & 9					   & 13						  & 20					  &	813.50			   & 2,765,480			  & 99.6834			  	   & 99.8785	  			\\
\proposal 		   & 9					   & 15						  & 20					  &	764.91			   & 2,795,848			  & 99.6797			  	   & 99.8756	  			\\
\proposal 		   & 9					   & 17						  & 20					  &	739.52			   & 2,823,188			  & 99.6801			  	   & 99.8802	  			\\\cmidrule{1-8}
\multicolumn{8}{l}{\footnotesize We use the \emph{E.coli} dataset for all these runs} \\
\end{tabular}}
\label{supptab:parameter_exploration}
\end{table*}

%% file: tables/supp_overlap_assembly-eq.tex
\begin{table*}[h]
\centering
\caption{Assembly quality comparisons when using the parameters equivalent to \texttt{\proposal-I}.}
\resizebox{\linewidth}{!}{
\begin{tabular}{@{}clrrrrrrrrrr@{}}\toprule
\textbf{Dataset} 	  & \textbf{Tool} 		 & \textbf{Average}	      & \textbf{Genome}		    & \textbf{K-mer}	   & \textbf{Aligned}	   & \textbf{Mismatch per} & \textbf{Average} & \textbf{Assembly}	  & \textbf{Largest} 	  & \textbf{NGA50}    & \textbf{NG50}	  \\
					  & 			  		 & \textbf{Identity (\%)} & \textbf{Fraction (\%)}  & \textbf{Compl. (\%)} & \textbf{Length (Mbp)} & \textbf{100Kbp (\#)}  & \textbf{GC (\%)} & \textbf{Length (Mbp)} & \textbf{Contig (Mbp)} & \textbf{(Kbp)}    & \textbf{(Kbp)}    \\\midrule
\emph{CHM13} 	  	  & \texttt{\proposal-I} & N/A	   			      & N/A					    & 29.26	   			   & 2,891.28 	   		   & 4,077.53	           & \textbf{41.32}	  & 2,897.87 	  		  & 25.2071 	  		  & 5,061.52 		  & 5,178.59 		  \\
(ONT)				  & minimap2 	  		 & N/A	   				  & N/A						& 28.32 			   & 2,860.26 	   		   & 4,660.73 			   & 41.36 			  & \textbf{2,908.55} 	  & \textbf{66.7564}	  & \textbf{13,189.2} & \textbf{13,820.3} \\
					  & minimap2-Eq 		 & N/A	   			      & N/A					    & \textbf{29.32} 	   & \textbf{3,117.29} 	   & \textbf{4,025.22} 	   & \textbf{41.32}   & 2,882.94 	          & 24.6651	          	  & 3,634.05          & 3,653.47          \\
					  & Reference 	  		 & 100 				      & 100 					& 100 			   	   & 3,117.29 			   & 0.00 				   & 40.75 			  & 3,117.29 			  & 248.387 			  & 150,617 	   	  & 150,617 	      \\\cmidrule{1-12}
\emph{Yeast} 		  & \texttt{\proposal-I} & 89.1677 	   		      & 97.0854 				& 33.81	   			   & \textbf{12.3938} 	   & 2,672.37 	   		   & 38.84	  		  & 12.4176 	  		  & 1.54807 	  		  & 635.966  		  & 636.669  		  \\
(PacBio)			  & minimap2 	  		 & 88.9002 			   	  & 96.9709 				& 33.38 			   & 12.0128 			   & 2,684.38 			   & 38.85 			  & \textbf{12.3325} 	  & \textbf{1.56078}	  & \textbf{810.046}  & \textbf{828.212}  \\
					  & minimap2-Eq 		 & \textbf{89.2166}       & \textbf{97.2674} 		& \textbf{33.93} 	   & 12.3886 	   		   & \textbf{2,653.08} 	   & \textbf{38.82}   & 12.4241 	          & 1.53435	          	  & 643.136           & 781.136           \\
					  & Reference 	  		 & 100 				      & 100 					& 100 				   & 12.1571 			   & 0.00 				   & 38.15 			  & 12.1571 			  & 1.53193 			  & 924.431 		  & 924.431 		  \\\cmidrule{1-12}
\emph{Yeast} 	      & \texttt{\proposal-I} & \textbf{89.6889}       & 99.2974 	    		& \textbf{35.95}	   & \textbf{12.3222} 	   & 2,529.47 	           & \textbf{38.64}	  & \textbf{12.3225} 	  & 1.10582 	  		  & 793.046  		  & 793.046  		  \\
(ONT)			      & minimap2 	  		 & 88.9393 			   	  & \textbf{99.6878}		& 34.84 			   & 12.304 			   & 2,782.59 			   & 38.74 			  & 12.3725 			  & \textbf{1.56005}	  & \textbf{796.718}  & \textbf{941.588}  \\
					  & minimap2-Eq 		 & 89.6653 	  			  & 97.3273		            & 35.62 			   & 11.826 			   & \textbf{2,465.87} 	   & \textbf{38.64}   & 11.8282 			  & 1.07367	          	  & 605.201           & 677.415  		  \\
					  & Reference 	  		 & 100 				      & 100 					& 100 				   & 12.1571 			   & 0.00 				   & 38.15 			  & 12.1571 			  & 1.53193 			  & 924.431 		  & 924.431 		  \\\cmidrule{1-12}
\emph{E. coli} 		  & \texttt{\proposal-I} & 88.5806                & 96.5238 				& 32.32	   			   & \textbf{5.90024} 	   & 1,857.56	           & \textbf{49.81}	  & 6.21598 	  		  & 2.40671	  			  & 769.981  		  & 2,060.4  		  \\
(PacBio)			  & minimap2 	  		 & 88.1365 			   	  & 92.7603 				& 30.74 			   & 5.37728 			   & 2,005.72 			   & 49.66 			  & \textbf{6.02707} 	  & 3.77098 	  		  & 367.442    	   	  & 3,770.98 		  \\
					  & minimap2-Eq 		 & \textbf{88.6371} 	  & \textbf{96.8540} 		& \textbf{32.33} 	   & 5.82218 			   & \textbf{1,816.29} 	   & 49.76 			  & 6.05821 	          & \textbf{3.77318} 	  & \textbf{1,119.04} & \textbf{3,773.18} \\
					  & Reference 	  		 & 100 				      & 100 					& 100 				   & 5.6394 			   & 0.00 				   & 50.43 			  & 5.6394 				  & 5.54732 			  & 5,547.32 		  & 5,547.32 		  \\\bottomrule
\multicolumn{12}{l}{\footnotesize Best results are highlighted with \textbf{bold} text. For most metrics, the best results are the ones closest to the corresponding value of the reference genome.}\\
\multicolumn{12}{l}{\footnotesize The best results for \emph{Aligned Length} are determined by the highest number within each dataset. We do not highlight the reference results as the best results.}\\
\multicolumn{12}{l}{\footnotesize N/A indicates that we could not generate the corresponding result because tool, QUAST, or dnadiff failed to generate the statistic.} \\
\end{tabular}}
\label{supptab:overlap_assembly-eq}
\end{table*}

%% file: tables/supp_pardef.tex
\begin{table*}[tbh]
\centering
\caption{Definition of parameters in \proposal}\label{supptab:pardef}
\begin{tabular}{@{}ll@{}}\toprule
\textbf{Parameter} & \textbf{Definition} \\\midrule
--strobemers & Use the \texttt{\proposal-S} mechanism when generating the list of k-mers of a seed\\\midrule
--immediate & Use the \texttt{\proposal-I} mechanism when generating the list of k-mers of a seed\\\midrule
-H & Use homopolymer-compressed k-mers\\\midrule
-w INT & Window size used when finding minimizers.\\\midrule
-k INT & k-mer size used when generating the list of k-mers of a seed\\\midrule
--neighbors INT & Number of k-mers included in the list of seeds.\\
                & Combination of both -k ($k$) and --neighbors ($n$) determines the seed length.\\
                & Seed length in \texttt{\proposal-S} is calculated as: $k \times n$\\
                & Seed length in \texttt{\proposal-I} is calculated as: $k+(n-1)$\\\midrule
--fixed-bits INT & Bit length of hash values that \proposal generates for each seed.\\
                 & Setting it to $2 \times k$ is the default behavior.\\\midrule
-t INT & Number of CPU threads to use.\\\midrule
-x STR & Preset for setting the default parameters given the use case (STR)\\
-x map-ont & Preset for mapping ONT reads. It uses the following parameters:\\
           & --immediate -w 10 -k 9 --neighbors 7 --fixed-bits 30\\
-x map-pb & Preset for mapping erroneous PacBio reads. It uses the following parameters:\\
           & --immediate -H -w 10 -k 13 --neighbors 7 --fixed-bits 32\\
-x map-hifi & Preset for mapping accurate long (HiFi) reads. It uses the following parameters:\\
           & --strobemers -w 50 -k 19 --neighbors 5 --fixed-bits 38\\
-x sr & Preset for mapping short reads. It uses the following parameters:\\
           & --immediate -w 11 -k 21 --neighbors 5 --fixed-bits 32\\
-x ava-ont & Preset for overlapping ONT reads. It uses the following parameters:\\
           & --immediate -w 10 -k 15 --neighbors 5 --fixed-bits 30\\
-x ava-pb & Preset for overlapping erroneous PacBio reads. It uses the following parameters:\\
           & --immediate -H -w 10 -k 19 --neighbors 5 --fixed-bits 38\\
-x ava-hifi & Preset for overlapping accurate long (HiFi) reads. It uses the following parameters:\\
           & --strobemers -w 200 -k 25 --neighbors 7 --fixed-bits 50\\\bottomrule
\end{tabular}
\end{table*}

%% file: tables/supp_ovpars.tex
\begin{table*}[tbh]
\centering
\caption{Parameters$^{*}$ we use in our evaluation for each tool and dataset in read overlapping.}\label{supptab:ovpars}
\begin{tabular}{@{}lll@{}}\toprule
\textbf{Tool} & \textbf{Dataset} & \textbf{Parameters} \\\midrule
\proposal & \emph{CHM13 (HiFi)} & -x ava-hifi -t 32 \\
\proposal & \emph{D. ananassae (HiFi)} & -x ava-hifi -t 32\\
\proposal & \emph{E. coli (HiFi)} & -x ava-hifi -t 32\\
\proposal & \emph{CHM13 (ONT)} & -x ava-ont -t 32\\
\proposal & \emph{Yeast (PacBio)} & -x ava-pb -t 32\\
\proposal & \emph{Yeast (ONT)} & -x ava-ont -t 32\\
\proposal & \emph{E. coli (PacBio)} & -x ava-pb -t 32\\\midrule
minimap2 & \emph{CHM13 (HiFi)} & -x ava-pb -Hk21 -w14 -t 32 \\
minimap2 & \emph{D. ananassae (HiFi)} & -x ava-pb -Hk21 -w14 -t 32 \\
minimap2 & \emph{E. coli (HiFi)} & -x ava-pb -Hk21 -w14 -t 32\\
minimap2 & \emph{CHM13 (ONT)} & -x ava-ont -t 32\\
minimap2 & \emph{Yeast (PacBio)} & -x ava-pb -t 32\\
minimap2 & \emph{Yeast (ONT)} & -x ava-ont -t 32\\
minimap2 & \emph{E. coli (PacBio)} & -x ava-pb -t 32\\\midrule
minimap2-Eq & \emph{CHM13 (ONT)} & -x ava-ont -k19 -w10 -t 32\\
minimap2-Eq & \emph{Yeast (PacBio)} & -x ava-pb -k23 -w10 -t 32\\
minimap2-Eq & \emph{Yeast (ONT)} & -x ava-ont -k19 -w10 -t 32\\
minimap2-Eq & \emph{E. coli (PacBio)} & -x ava-pb -k23 -w10 -t 32\\\midrule
MHAP & \emph{CHM13 (HiFi)} & --store-full-id --ordered-kmer-size 18 --num-hashes 128 --num-min-matches 5\\
& & --ordered-sketch-size 1000 --threshold 0.95 --num-threads 32\\
MHAP & \emph{D. ananassae (HiFi)} & --store-full-id --ordered-kmer-size 18 --num-hashes 128 --num-min-matches 5\\
& & --ordered-sketch-size 1000 --threshold 0.95 --num-threads 32\\
MHAP & \emph{E. coli (HiFi)} & --store-full-id --ordered-kmer-size 18 --num-hashes 128 --num-min-matches 5\\
& & --ordered-sketch-size 1000 --threshold 0.95 --num-threads 32\\
MHAP & \emph{Yeast (PacBio)} & --store-full-id --num-threads 32\\
MHAP & \emph{Yeast (ONT)} & --store-full-id --num-threads 32\\
MHAP & \emph{E. coli (PacBio)} & --store-full-id --num-threads 32\\\bottomrule
\multicolumn{3}{l}{\footnotesize $^{*}$ For the definitions of the parameters we use in \proposal, please see Supplementary Table~\ref{supptab:pardef}}\\
\end{tabular}
\end{table*}

%% file: tables/supp_mappars.tex
\begin{table*}[tbh]
\centering
\caption{Parameters we use in our evaluation for each tool and dataset in read mapping.}\label{supptab:mappars}
\resizebox{0.6\linewidth}{!}{
\begin{tabular}{@{}lll@{}}\toprule
\textbf{Tool} & \textbf{Dataset} & \textbf{Parameters} \\\midrule
\proposal & \emph{CHM13 (HiFi)} & -ax map-hifi -t 32 --secondary=no\\
\proposal & \emph{HG002 (HiFi)} & -ax map-hifi -t 32 --secondary=no\\
\proposal & \emph{D. ananassae (HiFi)} & -ax map-hifi -t 32 --secondary=no \\
\proposal & \emph{E. coli (HiFi)} & -ax map-hifi -t 32 --secondary=no \\
\proposal & \emph{CHM13 (ONT)} & -ax map-ont -t 32 --secondary=no\\
\proposal & \emph{Yeast (PacBio)} & -ax map-pb -t 32 --secondary=no\\
\proposal & \emph{Yeast (ONT)} & -ax map-ont -t 32 --secondary=no\\
\proposal & \emph{Yeast (Illumina)} & -ax sr -t 32 \\
\proposal & \emph{E. coli (PacBio)} & -ax map-pb -t 32 --secondary=no\\\midrule
minimap2 & \emph{CHM13 (HiFi)} & -ax map-hifi -t 32 --secondary=no\\
minimap2 & \emph{HG002 (HiFi)} & -ax map-hifi -t 32 --secondary=no\\
minimap2 & \emph{D. ananassae (HiFi)} & -ax map-hifi -t 32 --secondary=no\\
minimap2 & \emph{E. coli (HiFi)} & -ax map-hifi -t 32 --secondary=no\\
minimap2 & \emph{CHM13 (ONT)} & -ax map-ont -t 32 --secondary=no\\
minimap2 & \emph{Yeast (PacBio)} & -ax map-pb -t 32 --secondary=no\\
minimap2 & \emph{Yeast (ONT)} & -ax map-ont -t 32 --secondary=no\\
minimap2 & \emph{Yeast (Illumina)} & -ax sr -t 32 \\
minimap2 & \emph{E. coli (PacBio)} & -ax map-pb -t 32 --secondary=no\\\midrule
Winnowmap2 & \emph{CHM13 (HiFi)} & meryl count k=15\\
 & & meryl print greater-than distinct=0.9998\\
 & & -ax map-pb -t 32\\
Winnowmap2 & \emph{HG002 (HiFi)} & meryl count k=15\\
 & & meryl print greater-than distinct=0.9998\\
 & & -ax map-pb -t 32\\
Winnowmap2 & \emph{D. ananassae (HiFi)} & meryl count k=15\\
& &  meryl print greater-than distinct=0.9998\\
& & -ax map-pb -t 32\\
Winnowmap2 & \emph{E. coli (HiFi)} & meryl count k=15\\
& &  meryl print greater-than distinct=0.9998\\
& & -ax map-pb -t 32\\
Winnowmap2 & \emph{CHM13 (ONT)} & meryl count k=15\\
& &  meryl print greater-than distinct=0.9998\\
& & -ax map-ont -t 32\\
Winnowmap2 & \emph{Yeast (PacBio)} & meryl count k=15\\
& &  meryl print greater-than distinct=0.9998\\
& & -ax map-pb-clr -t 32\\
Winnowmap2 & \emph{Yeast (ONT)} & meryl count k=15\\
& &  meryl print greater-than distinct=0.9998\\
& & -ax map-ont -t 32\\
Winnowmap2 & \emph{E. coli (PacBio)} & meryl count k=15\\
& &  meryl print greater-than distinct=0.9998\\
& & -ax map-pb-clr -t 32\\\midrule
LRA & \emph{CHM13 (HiFi)} & align -CCS -t 32 -p s \\
LRA & \emph{HG002 (HiFi)} & align -CCS -t 32 -p s \\
LRA & \emph{D. ananassae (HiFi)} & align -CCS -t 32 -p s \\
LRA & \emph{E. coli (HiFi)} & align -CCS -t 32 -p s \\
LRA & \emph{CHM13 (ONT)} & align -ONT -t 32 -p s \\
LRA & \emph{Yeast (PacBio)} & align -CLR -t 32 -p s \\
LRA & \emph{Yeast (ONT)} & align -ONT -t 32 -p s \\
LRA & \emph{E. coli (PacBio)} & align -CLR -t 32 -p s \\\midrule
S-conLSH & \emph{CHM13 (HiFi)} & --threads 32 --align 1\\
S-conLSH & \emph{E. coli (HiFi)} & --threads 32 --align 1\\
S-conLSH & \emph{CHM13 (ONT)} & --threads 32 --align 1\\
S-conLSH & \emph{Yeast (PacBio)} & --threads 32 --align 1\\
S-conLSH & \emph{Yeast (ONT)} & --threads 32 --align 1\\
S-conLSH & \emph{E. coli (PacBio)} & --threads 32 --align 1\\\midrule
Strobealign & \emph{Yeast (Illumina)} & -t 32\\\bottomrule
\multicolumn{3}{l}{\footnotesize $^{*}$ For the definitions of the parameters we use in \proposal, please see Supplementary Table~\ref{supptab:pardef}}\\
\end{tabular}}
\end{table*}

%% file: tables/supp_version.tex
\begin{table*}[tbh]
\centering
\caption{Versions of each tool.}\label{supptab:version}
\resizebox{\linewidth}{!}{
\begin{tabular}{@{}lll@{}}\toprule
\textbf{Tool} & \textbf{Version} & \textbf{GitHub or Conda Link to the Version} \\\midrule
\proposal & 1.0 & \url{https://github.com/CMU-SAFARI/BLEND}\\\midrule
minimap2 & 2.24 & \url{https://github.com/lh3/minimap2/releases/tag/v2.24}\\\midrule
MHAP & 2.1.3 & \url{https://anaconda.org/bioconda/mhap/2.1.3/download/noarch/mhap-2.1.3-hdfd78af_1.tar.bz2}\\\midrule
LRA & 1.3.2 & \url{https://anaconda.org/bioconda/lra/1.3.2/download/linux-64/lra-1.3.2-ha140323_0.tar.bz2}\\\midrule
Winnowmap2 & 2.03 & \url{https://anaconda.org/bioconda/Winnowmap/2.03/download/linux-64/Winnowmap2-2.03-h2e03b76_0.tar.bz2}\\\midrule
S-conLSH & 2.0 & \url{https://github.com/anganachakraborty/S-conLSH-2.0/tree/292fbe0405f10b3ab63fc3a86cba2807597b582e} \\\midrule
Strobealign & 0.7.1 & \url{https://anaconda.org/bioconda/strobealign/0.7.1/download/linux-64/strobealign-0.7.1-hd03093a_1.tar.bz2} \\\bottomrule
\end{tabular}}
\end{table*}